\documentclass[sigconf,10pt,nonacm,authorversion]{acmart}
\renewcommand\footnotetextcopyrightpermission[1]{} 
\usepackage{pifont}

\usepackage{tabularx}
\usepackage[english]{babel}
\usepackage{blindtext}
\usepackage{color}
\usepackage{appendix}
\usepackage{url}
\usepackage{amsmath}
\usepackage{mathrsfs}
\usepackage{caption}
\usepackage{graphicx}
\usepackage{pifont}
\usepackage{flushend}
\usepackage{multirow}
\usepackage{enumitem}
\usepackage{array}
\usepackage[normalem]{ulem}
\usepackage{amsfonts}
\usepackage{algorithm}
\usepackage{algorithmic}
\usepackage{booktabs}
\usepackage{wrapfig}
\usepackage{xspace}
\usepackage{colortbl}
\usepackage{cancel}
\usepackage{textcase}
\usepackage{titlesec}

\usepackage{subcaption}
\usepackage{hyperref}
\definecolor{myLightGreen}{RGB}{200, 255, 200}

\makeatletter
\def\@seccntformat#1{\csname the#1\endcsname\hspace{0.45em}} % Reduce spacing after number

\renewcommand{\subsubsection}{\@startsection{subsubsection}{2}{\z@}{-.4ex\@plus -.2ex \@minus -.2ex}{.2ex \@plus .2ex \@minus .2ex}{\normalfont\large\slshape}}
\makeatother
\renewcommand{\paragraph}[1]{\vspace{2pt plus 0pt minus 2pt}\noindent{\bfseries #1}}

\def\sys{\textsf{Gandalf}\xspace}
\def\tool{\textsf{CellNinjia}\xspace}

\begin{document}
\title[]{\huge Seeing Through the Fog: Empowering Mobile Devices to Expose and Mitigate RAN Buffer Effects on Delay-Sensitive Protocols}
% \author{paperID: 503}
\author{Yuxin Liu}
\affiliation{
  \institution{University at Buffalo, SUNY}
  \country{}
}
% \email{yuxinliu@buffalo.edu}

\author{Tianyang Zhang}
\affiliation{
  \institution{University at Buffalo, SUNY}
  \country{}
}
% \email{tzhang52@buffalo.edu}

% \author{Qiang Wu}
% \affiliation{
%   \institution{Peking University}
%   \country{}
% }
% \email{2300022799@stu.pku.edu.cn}

% \author{Ju Ren}
% \affiliation{
%   \country{}
% }
% \email{renju@tsinghua.edu.cn}

\author{Kyle Jamieson}
\affiliation{
  \institution{Princeton University}
  \country{}
}
% \email{kylej@princeton.edu}

\author{Yaxiong Xie\textsuperscript{*}}
\affiliation{
  \institution{University at Buffalo, SUNY}
  \country{}
}
% \email{yaxiongx@buffalo.edu}
\thanks{\textsuperscript{*}Yaxiong Xie is the corresponding author.}

\def\UrlBreaks{\do\A\do\B\do\C\do\D\do\E\do\F\do\G\do\H\do\I\do\J
\do\K\do\L\do\M\do\N\do\O\do\P\do\Q\do\R\do\S\do\T\do\U\do\V
\do\W\do\X\do\Y\do\Z\do\[\do\\\do\]\do\^\do\_\do\`\do\a\do\b
\do\c\do\d\do\e\do\f\do\g\do\h\do\i\do\j\do\k\do\l\do\m\do\n
\do\o\do\p\do\q\do\r\do\s\do\t\do\u\do\v\do\w\do\x\do\y\do\z
\do\.\do\@\do\\\do\/\do\!\do\_\do\|\do\;\do\>\do\]\do\)\do\,
\do\?\do\'\do+\do\=\do\#}

\begin{abstract}
Delay-based protocols rely on end-to-end delay measurements to detect network congestion. 
However, in cellular networks, Radio Access Network (RAN) buffers introduce significant delays unrelated to congestion, 
fundamentally challenging these protocols' assumptions. 
We identify two major types of RAN buffers - retransmission buffers and uplink scheduling buffers - that can 
introduce delays comparable to congestion-induced delays, 
severely degrading protocol performance. 
We present \tool, a software-based system providing real-time visibility into RAN operations, 
and \sys, which leverages this visibility to systematically handle RAN-induced delays. 
Unlike existing approaches that treat these delays as random noise, 
\sys identifies specific RAN operations and compensates for their effects. 
Our evaluation in commercial 4G LTE and 5G networks 
shows that \sys enables substantial performance improvements - up to 7.49$\times$ for Copa and 9.53$\times$ for PCC Vivace - without 
modifying the protocols' core algorithms, 
demonstrating that delay-based protocols can realize their full potential in cellular networks.

%Delay-based protocols rely on end-to-end delay measurements to 
%detect network congestion and adapt their behavior. 
%However, in cellular networks, Radio Access Network (RAN) buffers introduce significant delays unrelated to congestion, 
%fundamentally challenging these protocols' core assumptions. 
%Through extensive measurement, we identify two major types of RAN buffers---retransmission buffers and uplink scheduling buffers---that 
%introduce delays comparable to or exceeding typical congestion-induced delays,
%These RAN-induced delays severely impact protocol performance, 
%causing throughput degradation of up to 89\% in protocols like Copa and PCC Vivace. 
%We present \tool, a software-based system that provides 
%real-time visibility into RAN operations on mobile devices, 
%and \sys, which leverages this visibility to systematically 
%handle RAN-induced delays. 
%Unlike existing approaches that treat these delays as random noise, 
%\sys identifies specific RAN operations and compensates for their effects. 
%Our evaluation in commercial 4G LTE and 5G SA networks shows that 
%\sys enables dramatic performance improvements---up to 7.49$\times$ for Copa and 9.53$\times$ for PCC Vivace---without 
%modifying the protocols' core algorithms. 
%These results demonstrate that by providing protocols with the ability to distinguish 
%between RAN-induced delays and true congestion, 
%\sys allows delay-based protocols to realize their full potential in cellular networks.
\end{abstract}

\maketitle
\pagenumbering{arabic}

\section{Introduction}
Efficient congestion control is paramount for application performance, 
and timely detection of congestion is crucial for minimizing latency. 
End-to-end delay serves as a critical early indicator: 
as congestion builds, packets experience increased buffering, 
leading to delay increases that precede packet loss. 
Consequently, protocols leveraging end-to-end delay can react proactively, 
maximizing network utilization while minimizing buffering and latency. 
This principle underpins numerous transport protocols and real-time applications, 
from classic TCP variants like TCP Vegas~\cite{brakmo_tcp_1994, mishra_keeping_2024, ware_ccanalyzer_2024} to modern video conferencing systems~\cite{dhawaskar_sathyanarayana_converge_2023}. 
Even machine learning-based congestion control algorithms~\cite{winstein_tcp_nodate, sivaraman_experimental_2014, dong_pcc_2015, dong_pcc_2018, yan_pantheon_2018, winstein_stochastic_2013, yen_computers_2023}, integrate delay signals 
into their optimization objectives, 
recognizing delay's value as an early congestion indicator.

For delay to serve as an accurate and timely indicator of network congestion, 
delay variations should primarily reflect packet queuing at bottleneck links. 
The underlying principle is straightforward: 
when network resources become scarce, packets queue up at bottleneck points, 
causing measurable increases in end-to-end delay. 
By monitoring these delay increases, protocols can detect emerging congestion 
before packet loss occurs, enabling them to proactively adjust their sending rates. 
This relationship between queuing and congestion forms a fundamental assumption 
for modern delay-based protocols---that increased delays directly signal network resource exhaustion.

However, our investigation reveals a critical challenge in cellular networks: 
the RAN contains multiple \textit{internal buffers} where packets 
are frequently queued for reasons entirely unrelated to network congestion. 
We identify two major RAN buffers that significantly impact end-to-end delay:

\paragraph{1) Retransmission Buffer (\S\ref{s:reTx_buf}).} 
To ensure reliable transmission over error-prone wireless links, 
the cellular network implements retransmissions at two protocol layers. 
When transmission errors occur, subsequent packets must wait in retransmission buffers until recovery succeeds. 
The first layer handles quick recovery of wireless errors with relatively short delays (around 8ms), 
while the second layer provides additional reliability with longer delays (60-120ms). 
This two-layer approach and the associated buffering are fundamental to 
reliable wireless communication, operating independently of application traffic patterns.

\paragraph{2) Uplink Scheduling Buffer (\S\ref{sec:uplink}).} 
Before any device can upload data, 
it must receive permission from the base station. 
This coordination requires devices to regularly report their buffer status to the base station, 
which then schedules transmission opportunities. 
During this process, packets wait in an uplink buffer between status reporting and receiving permission to transmit. 
This scheduling mechanism is essential for managing the shared wireless medium among multiple devices, 
and its operation is determined by cellular protocols rather than application behavior.

These buffering mechanisms are not optional features---they are 
fundamental to how cellular networks operate, 
ensuring reliable communication and efficient resource sharing among multiple devices. 
However, they introduce substantial delay variations that can reach tens of milliseconds, 
comparable to or even exceeding typical congestion-induced delays. 
Because these delays arise from essential cellular operations rather than network congestion, 
they fundamentally challenge the assumptions of delay-based protocols.

Today's delay-sensitive protocols lack visibility into these RAN internal operations, 
leaving them unable to differentiate between congestion-induced delays and RAN buffering delays. 
This visibility gap exists because cellular networks are largely closed systems---chipset manufacturers and 
network operators keep their implementations proprietary, 
and RAN operations are abstracted away from higher protocol layers. 
Without access to RAN operational information, protocols can only treat these RAN-induced delays as random noise~\cite{zaki_adaptive_2015, arun_copa_2018, ni_polycorn_nodate, wang_active-passive_2019}, 
applying various filtering techniques in an attempt to extract true congestion signals. 
However, our analysis reveals that this approach fundamentally falls short. 
The delay patterns introduced by RAN buffers are neither random nor noise---they 
are systematic effects tied to specific RAN operations. 
Simply filtering these delays fails to address their root cause and 
leaves protocols struggling to distinguish between actual congestion and RAN effects.

To address this visibility challenge, we design \tool, 
a software-based monitoring system that provides real-time visibility 
into cellular RAN operations directly on mobile devices. 
\tool efficiently collects and processes diagnostic messages from cellular modems, 
enabling detailed tracking of various RAN behaviors including retransmission events, 
scheduling decisions, and buffer status reports (\S\ref{sec:cellninja}). 
Unlike existing cellular monitoring approaches, 
\tool achieves this without requiring additional hardware and with minimal overhead. 
Previous tools like NG-Scope~\cite{xie_ng-scope_2022} and NR-Scope~\cite{wan_nr-scope_2024} 
rely on expensive software-defined radios (USRPs) 
for passive monitoring and can only decode unencrypted control messages, 
severely limiting their utility. 
Other diagnostic tools, like MobileInsight~\cite{li_mobileinsight_2016} and QC-Super~\cite{noauthor_p1secqcsuper_2025}, 
lack support for 5G Standalone networks~\cite{ye_closer_2023} and 
cannot provide real-time monitoring capabilities. 
\tool overcomes these limitations through novel 
buffer draining mechanisms and real-time message processing (\S\ref{sec:cellninjia_design}), 
decoding over 60 types of 5G messages and providing comprehensive and immediate feedback
needed for delay-sensitive protocols to effectively react to RAN conditions.

Building on \tool's visibility into RAN operations, 
we develop \sys, a system that fundamentally 
changes how protocols handle RAN-induced delays. 
Instead of attempting to filter out these delays as noise, 
\sys enables protocols to understand and systematically account for them. 
Using the real-time RAN information from \tool, 
\sys identifies specific RAN operations and compensates for their delay effects (\S\ref{sec:gandalf}). 
For retransmission buffers, \sys leverages a key insight: 
retransmissions are discrete, independent events with well-defined temporal boundaries. 
By tracking the start and end of each retransmission event through RAN messages, 
\sys can precisely identify affected packets and subtract the corresponding delay inflation~(\S\ref{sec:retx}). 
For uplink scheduling buffers, 
\sys exploits its understanding of the buffer status reporting mechanism to perform informed filtering---rather 
than applying generic smoothing techniques, it uses the actual reporting interval from RAN to 
determine the precise frequency of delay variations these buffers introduce~(\S\ref{sec:uplink}). 
This RAN-aware filtering ensures we remove only the delay components 
that match known RAN behaviors while preserving true congestion signals. 
This systematic approach allows delay-sensitive protocols to operate in cellular networks as 
effectively as they do in traditional networks, 
maintaining their core congestion control logic 
while gaining awareness of RAN-specific behaviors.

% \begin{table}[tb]
%     \centering
%     \begin{tabular}{c|c|c|c|c|c|c|c}
%         \toprule
%         \multicolumn{2}{c|}{Copa thput} & \multicolumn{2}{c|}{COPA delay} &  \multicolumn{2}{c|}{PCC thput} & \multicolumn{2}{c}{PCC delay }\\
%         \hline
%          dl             &ul             &dl           & ul                  & dl        & ul            & dl            & ul \\
%         \hline
%         7.4 $\times$   &2.7$\times$   &1.1$\times$   &0.9$\times$      &9.5$\times$     &2.2$\times$     &0.9$\times$     &0.9$\times$ \\
%         \bottomrule
%     \end{tabular}
%     \caption{Comparison table for Copa and PCC}
%     \label{tab:copa_pcc}
% \end{table}

\begingroup
\renewcommand{\arraystretch}{0.95}
\setlength{\tabcolsep}{2pt}

\begin{table}[]
\centering
\begin{tabular}{c|cc|cc|cc|cc}
\toprule
\multirow{2}{*}{\centering \shortstack{Network \\ Type}} & \multicolumn{2}{c|}{\begin{tabular}[c]{@{}c@{}}COPA \\ thput\end{tabular}}   & \multicolumn{2}{c|}{\begin{tabular}[c]{@{}c@{}}COPA \\ delay\end{tabular}}   & \multicolumn{2}{c|}{\begin{tabular}[c]{@{}c@{}}PCC \\ thput\end{tabular}}    & \multicolumn{2}{c}{\begin{tabular}[c]{@{}c@{}}PCC \\ delay\end{tabular}}     \\ 
\cline{2-9} 
& \multicolumn{1}{c|}{{\cellcolor{myLightGreen} DL}}  & { UL}  
& \multicolumn{1}{c|}{{\cellcolor{myLightGreen} DL}}  & { UL}  
& \multicolumn{1}{c|}{{\cellcolor{myLightGreen} DL}}  & { UL}  
& \multicolumn{1}{c|}{{\cellcolor{myLightGreen} DL}}  & { UL}  \\ 
\hline
5G  & \multicolumn{1}{c|}{{\cellcolor{myLightGreen} 7.5$\times$}} & { 2.8$\times$} 
    & \multicolumn{1}{c|}{{\cellcolor{myLightGreen} 1.1$\times$}} & { 1.0$\times$} 
    & \multicolumn{1}{c|}{{\cellcolor{myLightGreen} 9.5$\times$}} & { 2.2$\times$} 
    & \multicolumn{1}{c|}{{\cellcolor{myLightGreen} 1.0$\times$}} & { 1.0$\times$} \\ 
\hline
LTE & \multicolumn{1}{c|}{{\cellcolor{myLightGreen} 6.3$\times$}} & { 3.8$\times$} 
    & \multicolumn{1}{c|}{{\cellcolor{myLightGreen} 1.1$\times$}} & { 1.0$\times$} 
    & \multicolumn{1}{c|}{{\cellcolor{myLightGreen} 5.3$\times$}} & { 2.9$\times$} 
    & \multicolumn{1}{c|}{{\cellcolor{myLightGreen} 1.0$\times$}} & { 1.0$\times$} \\ 
\bottomrule
\end{tabular}
\caption{Performance improvements with \sys integration: 
throughput (thput) and delay ratios for COPA and PCC across 5G SA and LTE networks.}
\label{tab:perf_improvement}
\vspace{-0.7cm}
\end{table}

\endgroup
Our extensive evaluation reveals that RAN-induced delays
severely degrade the performance of three representative delay-sensitive protocols.
COPA's~\cite{arun_copa_2018} rate control becomes excessively conservative,
as it misinterprets normal RAN buffering as severe network congestion.
PCC Vivace~\cite{dong_pcc_2018} experiences erratic rate adjustments due to RAN-induced delay variations,
leading to significant throughput degradation.
WebRTC's video quality suffers from frequent,
unnecessary quality reductions when its congestion control mechanism mistakes RAN delays for network congestion.
By integrating \sys with these protocols, we achieve dramatic performance improvements
without modifying their core algorithms, as shown in Table~\ref{tab:perf_improvement}.
In commercial 4G LTE and 5G networks, \sys improves throughput by up to 7.5$\times$ for COPA
and 9.5$\times$ for PCC while maintaining similar latency.
These results demonstrate that by providing protocols with the ability to distinguish
between RAN-induced delays and true congestion,
\sys enables existing delay-based algorithms to realize their full potential in cellular networks.
\section{Cellular Primer}\label{s:primer}

%\subsection{Physical Layer Architecture}
\paragraph{Frame Structure.}
The system frame (SFN) lasts 10ms in both LTE and 5G networks.
LTE~\cite{3gpp_release_2017} divides each SFN into 10 subframes of 1ms,
while 5G uses variable-duration slots determined by the subcarrier spacing configuration~\cite{3gpp_ts_nodate}.
%(Table\ref{tab:numerology}).

%The system frame (SFN) is the fundamental time unit in cellular,
%with a fixed duration of 10ms. 
%In LTE~\cite{3gpp_release_2017}, each SFN is divided into 10 subframes of 1ms each.
%In 5G, each SFN is divided into a variable number of slots,
%where the slot duration is determined by the
%\textit{subcarrier spacing configuration} parameter~\cite{3gpp_ts_nodate} (Table\ref{tab:numerology} of Appendix).

\paragraph{Transmission Time Interval.}
A \textit{transmission time interval} (TTI) defines the minimum time unit
during which a cellular base station can schedule radio resources
for user uplink or downlink transmissions.
In LTE, a TTI equals one subframe (1ms),
while in 5G SA, a TTI corresponds to one slot.

%\subsection{Cellular Protocol Layers and Retransmissions}

\begin{figure}[htb]
    \centering
    \includegraphics[width=0.98\linewidth]{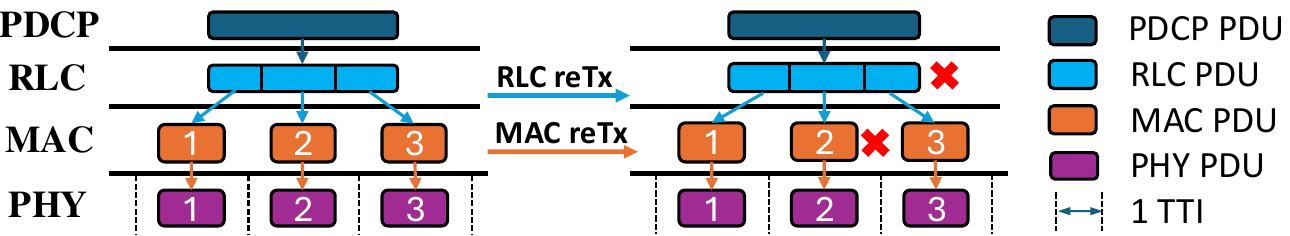}
    \caption{Illustration of cellular network's two-level retransmission hierarchy and 
    protocol layer interactions, showing PDU relationships across PDCP, RLC, MAC, and PHY layers.}
    \label{fig:rlc_reTx}
\end{figure}

\paragraph{PDCP, RLC, MAC, and PHY Layer Interactions.}
The packet data convergence protocol (PDCP) layer passes
its protocol data units (PDUs) to the radio link control (RLC) layer,
which then forwards its PDUs
to the medium access control (MAC) layer.
As shown in Figure~\ref{fig:rlc_reTx},
the MAC layer adaptively processes RLC PDUs based on
available PHY radio resources:
it may segment one RLC PDU into multiple MAC PDUs,
or alternatively combine multiple RLC PDUs into a single MAC PDU.
Each MAC PDU is then transmitted through the physical (PHY) layer
within one TTI.

\paragraph{Two-level Retransmission for Reliability.}
Cellular networks implement a two-level retransmission mechanism to maximize data reliability.
When a MAC PDU is received with errors,
the MAC layer attempts retransmission to recover the lost data.
If MAC layer retransmissions fail to recover the data,
resulting in a corrupted or partially received RLC PDU,
the RLC layer initiates its own retransmission process.
Both RLC and MAC layers employ
\textit{automatic repeat request} (ARQ) schemes for retransmission:
the receiver signals successful reception with an acknowledgment (ACK)
or failure with a negative acknowledgment (NACK),
prompting the sender to retransmit the corresponding data as needed.

%Cellular networks implement a two-level retransmission mechanism to maximize data reliability.
%When a MAC PDU is received with errors,
%the MAC layer attempts retransmission up to three times in LTE
%and \textit{maxHARQ-Tx}\footnote{The value of \textit{maxHARQ-Tx} ranges from 1 to 16.} times in 5G networks.
%If MAC layer retransmissions fail to recover the data,
%resulting in a corrupted or partially received RLC PDU,
%the RLC layer initiates its own retransmission process.
%The RLC layer attempts retransmission up to \textit{maxRetxThreshold}\footnote{The value of \textit{maxRetxThreshold} ranges from 1 to 64.} times.
%Both RLC and MAC layers employ
%\textit{automatic repeat request} (ARQ) schemes for retransmission:
%the receiver signals successful reception with an acknowledgment (ACK)
%or failure with a negative acknowledgment (NACK),
%prompting the sender to retransmit the corresponding data as needed.
\section{Internal Buffers inside RAN}\label{sec:ran_delay}
In this section, we analyze internal buffers inside RAN.
%cellular networks.
%that play crucial roles in both reliable transmission and resource management.

\subsection{Retransmission Buffer}\label{s:reTx_buf}
In this section, we examine cellular networks' retransmission mechanisms, 
specifically focusing on retransmission buffer management and their effects on packet transmission.

\subsubsection{Retransmission Buffer and its Impact}
\begin{figure}[htb]
    \centering
    \includegraphics[width=0.98\linewidth]{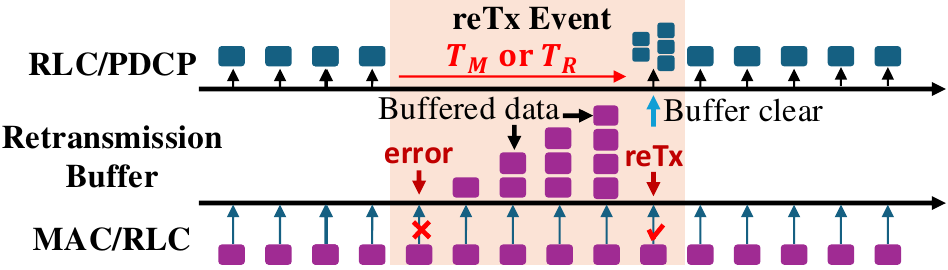}
    \caption{\textbf{Retransmission buffer}: 
    packets received between the original transmission and retransmission 
    are buffered and released together upon successful retransmission.}
    \label{fig:reorder_buf}
\end{figure}

\paragraph{In-order Delivery and Retransmission Buffer.}
The ARQ scheme introduces a delay
between the original transmission and retransmission,
as illustrated in Figure~\ref{fig:reorder_buf}.
We denote these retransmission delays as $T_{M}$ and $T_{R}$
for MAC and RLC layers respectively.
During the retransmission period,
data transmission continues uninterrupted.
To maintain in-order delivery, all data received during
this interval is stored in a \textit{retransmission buffer},
which resides at the UE for downlink traffic and
at the base station for uplink traffic,
as shown in Figure~\ref{fig:reorder_buf}.
Upon successful retransmission, the entire contents of the
buffer are delivered simultaneously to the next protocol layer:
to the RLC layer for MAC layer retransmissions,
and to the PDCP layer for RLC layer retransmissions.

\paragraph{Retransmission Event.}
We define a \textit{retransmission event} (or \textit{reTx event})
as the complete process of recovering from a single data error.
Each retransmission event spans a duration of $T_{M}$ or $T_{R}$ for MAC and RLC retransmissions respectively,
as illustrated in Figure~\ref{fig:reorder_buf}.
The retransmission buffer remains non-empty exclusively during these events.

%Multiple reTx events can overlap and merge into a longer composite event 
%encompassing multiple retransmissions. 
%Such overlapping events can originate from either 
%a single base station or multiple base stations when carrier aggregation is enabled.

\begin{figure}[tb]
    \begin{minipage}[t]{0.55\linewidth}
        \centering
        \begin{subfigure}[b]{0.49\linewidth}
            \centering
            \includegraphics[width=0.99\textwidth]{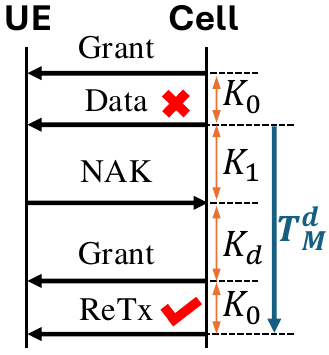}
            \caption{Downlink}
            \label{fig:dl_retx_time}
        \end{subfigure}
        \hfill
        \begin{subfigure}[b]{0.49\linewidth}
            \centering
            \includegraphics[width=0.99\textwidth]{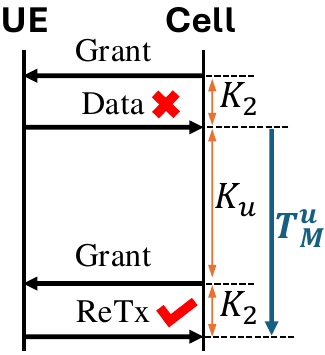}
            \caption{Uplink}
            \label{fig:ul_retx_time}
        \end{subfigure}
        \caption{Time sequence of MAC layer retransmission operations:
        (\textbf{a}) downlink and (\textbf{b}) uplink.}
        %for: (\textbf{a}) downlink and (\textbf{b}) uplink transmissions.}
        \label{fig:reTx_time}
    \end{minipage}
    \hfill
    \begin{minipage}[t]{0.43\linewidth}
        \centering
        \includegraphics[width=0.98\linewidth]{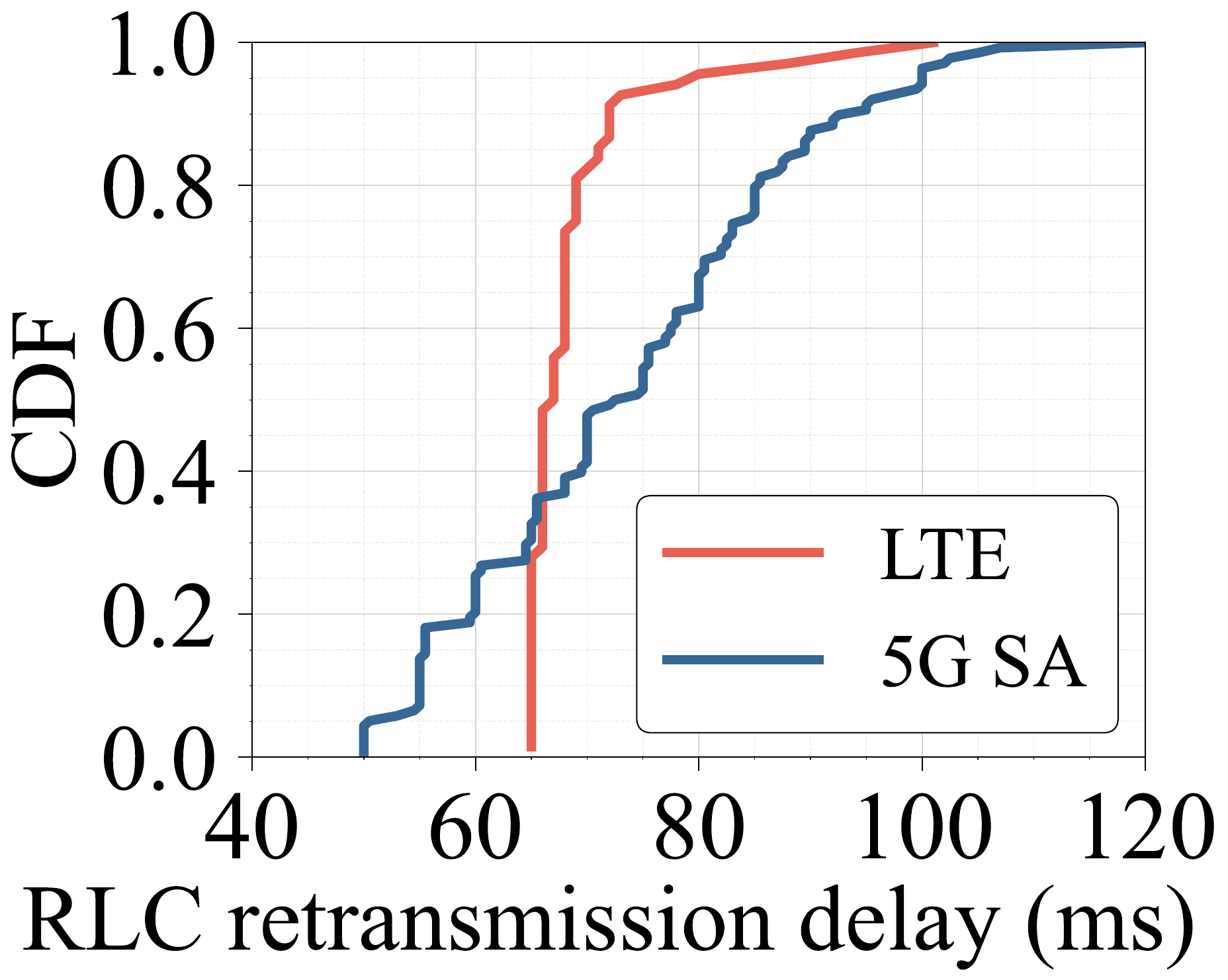}
        \caption{CDF of measured RLC retransmission delay $T_R$ from LTE and 5G SA.}
        \label{fig:rlc_retx_delay}
    \end{minipage}
\end{figure}

\paragraph{Impact on Packet Delivery and Delay.}
During each reTx event,
all incoming packets are held in the retransmission buffer,
completely suspending all packet delivery to higher protocol layers
and creating an \textit{idle period} equal to the retransmission delay duration,
as shown in Figure~\ref{fig:reorder_buf}.
This buffering mechanism introduces additional packet delay
to all packets currently stored in the buffer.
For packets experiencing transmission errors, this delay reaches a maximum of
$T_{M}$ for MAC layer traffic and $T_{R}$ for RLC layer traffic.
%
%\begin{figure}[!t]
%    \centering
%    \begin{subfigure}[b]{0.49\linewidth}
%        \centering
%        \includegraphics[width=0.99\textwidth]{figure/dl_retx_time.pdf}
%        \caption{Downlink reTx.}
%        \label{fig:dl_retx_time}
%    \end{subfigure}
%    \hfill
%    \begin{subfigure}[b]{0.49\linewidth}
%        \centering
%        \includegraphics[width=0.99\textwidth]{figure/ul_retx_time.pdf}
%        \caption{Uplink reTx.}
%        \label{fig:ul_retx_time}
%    \end{subfigure}
%    \caption{Time sequence of retransmission operations in cellular networks for: (\textbf{a}) downlink and (\textbf{b}) uplink transmissions.}
%    \label{fig:reTx_time}
%\end{figure}

\subsubsection{Quantifying Retransmission Delays}
%This section analyzes the retransmission delay values
%for both RLC and MAC layers in LTE and 5G networks.

\paragraph{MAC Layer.}
The MAC layer retransmission follows a predefined procedure regulated by the 3GPP standard,
making $T_{M}$ deterministic.
For downlink transmission, the base station first sends a grant to UE,
%the UE indicating the data location, 
followed by the actual data transmission,
with these events separated by $K_0$ TTIs,
as in Figure~\ref{fig:dl_retx_time}.
Upon detecting an error, the UE sends a NAK after $K_1$ TTIs,
and the base station issues a new grant after $K_d$ TTIs.
Therefore, the downlink retransmission delay $T^d_{M}$ is:
\begin{equation}
T^d_{M} = K_0 + K_1 + K_d
\end{equation}
For uplink transmission, as shown in Figure~\ref{fig:ul_retx_time},
the UE must receive a grant before transmitting data,
with an interval of $K_2$ TTIs between grant and data.
When an error occurs, the base station schedules another grant
after $K_u$ TTIs, resulting in an uplink retransmission delay $T^u_{M}$ of:
\begin{equation}
T^u_{M} = K_2 + K_u.
\end{equation}
The values of $T^d_M$ and $T^u_M$ differ between LTE and 5G networks. 
In LTE, both delays are fixed at 8~ms. 
In 5G SA networks, however, both $T^d_M$ and $T^u_M$ vary 
with the slot index $i$ ($i \in [0,19]$) within a system frame. 
We detail the derivation of these delays for both LTE and 5G in \S\ref{app:t_m} using \tool. 

%The values of $T^d_{M}$ and $T^u_{M}$ differ between LTE and 5G networks.
%In LTE, both delays are fixed at 8 ms.
%In 5G SA networks, however, $T_M$ varies with the TTI index 
%of the data transmission.
%We detail the derivation of $T_M$ for both LTE and 5G 
%in \S\ref{app:t_m} using \tool.

%these delays become time-varying
%as $K_1$, $K_2$, $K_u$, and $K_d$ depend on
%the TTI position within the system frame
%(detailed values for different configurations provided in
%Tables~\ref{tab:dl_reTx_config1}, \ref{tab:dl_reTx_config2}, and \ref{tab:ul_reTx_config} of Appendix).
%While variable, both $T^d_{M}$ and $T^u_{M}$ remain deterministic for a given TTI index
%and range from 4.5 ms to 8.5 ms.

\paragraph{RLC Layer.}
RLC retransmission relies on a \textit{polling} mechanism:
the sender periodically sets a polling bit in RLC PDUs 
to request ACK/NAK feedback,
and initiates retransmission upon receiving NAK.
Polling frequency depends on two configurable thresholds:
after every \textit{pollPDU} PDUs or \textit{pollByte} bytes transmitted.
The actual timing of these polling events varies significantly
because PDU transmission pace depends on MAC layer scheduling opportunities—
MAC may combine multiple RLC PDUs into one TTI
or segment a single RLC PDU across multiple TTIs
based on available radio resources.
As a result, both polling bit delivery and subsequent retransmissions
experience variable delays due to this MAC layer dependency.
This is evident in our measurements shown in Figure~\ref{fig:rlc_retx_delay},
where RLC retransmission delays $T_R$ range from 60-100ms in LTE and 60-120ms in 5G SA networks.

\subsubsection{Characterization of Retransmission Frequency}
To understand retransmission patterns in real-world scenarios,
we conducted experiments with a mobile user carrying a Xiaomi 10 Lite
while walking along city streets.
We measured both MAC PDU Error Rate (MPER) and RLC retransmission events
during continuous data transfer over 5G SA network.

\begin{figure}[tb]
    \setlength{\abovecaptionskip}{2pt}
    \setlength{\belowcaptionskip}{-2pt}
    \centering
    \begin{subfigure}[b]{0.28\linewidth}
        \centering
        \includegraphics[width=\textwidth]{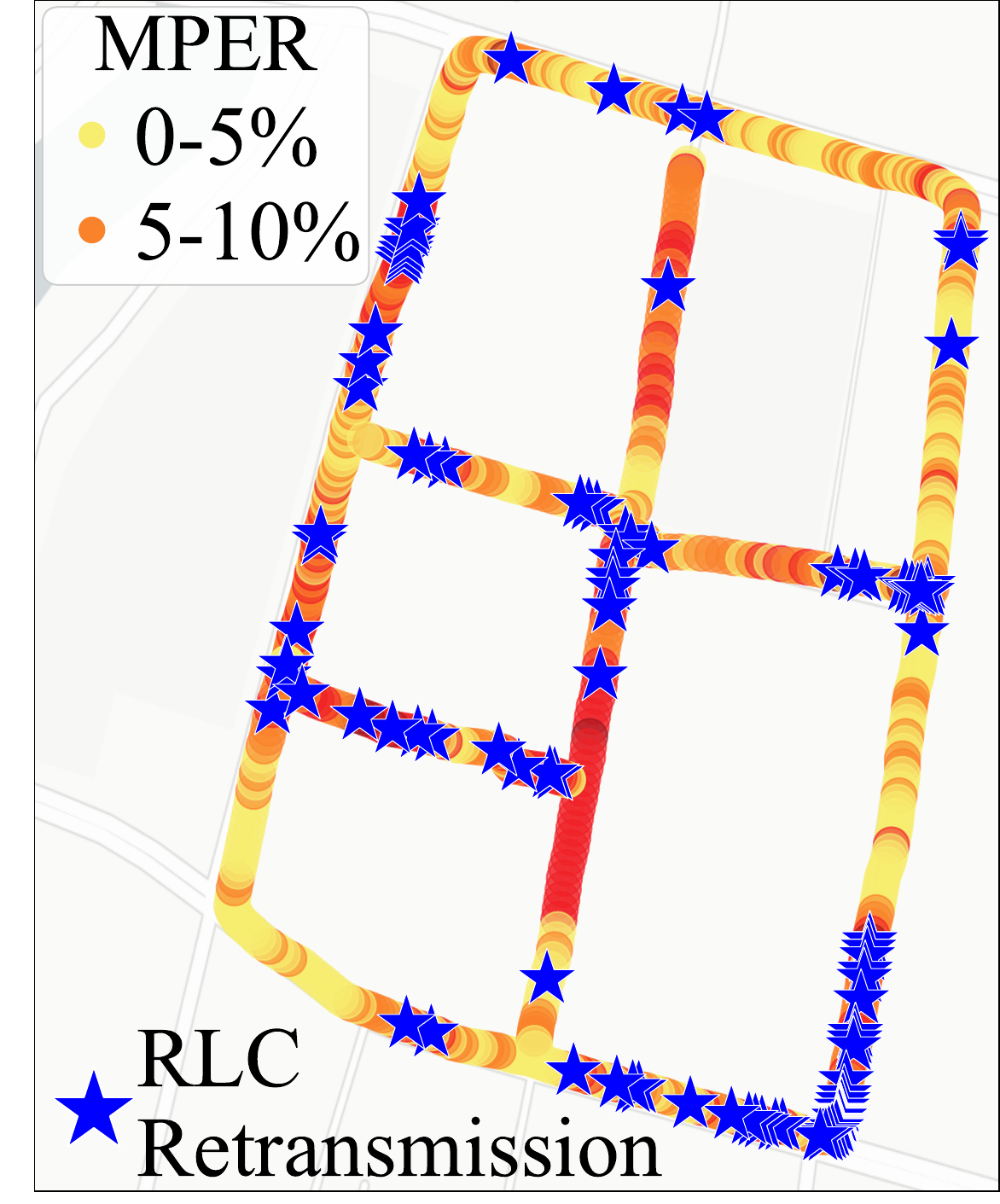}
        \caption{Downlink}
        \label{fig:5g_dl_mper_rlc}
    \end{subfigure}
    \hfill
    \begin{subfigure}[b]{0.28\linewidth}
        \centering
        \includegraphics[width=\textwidth]{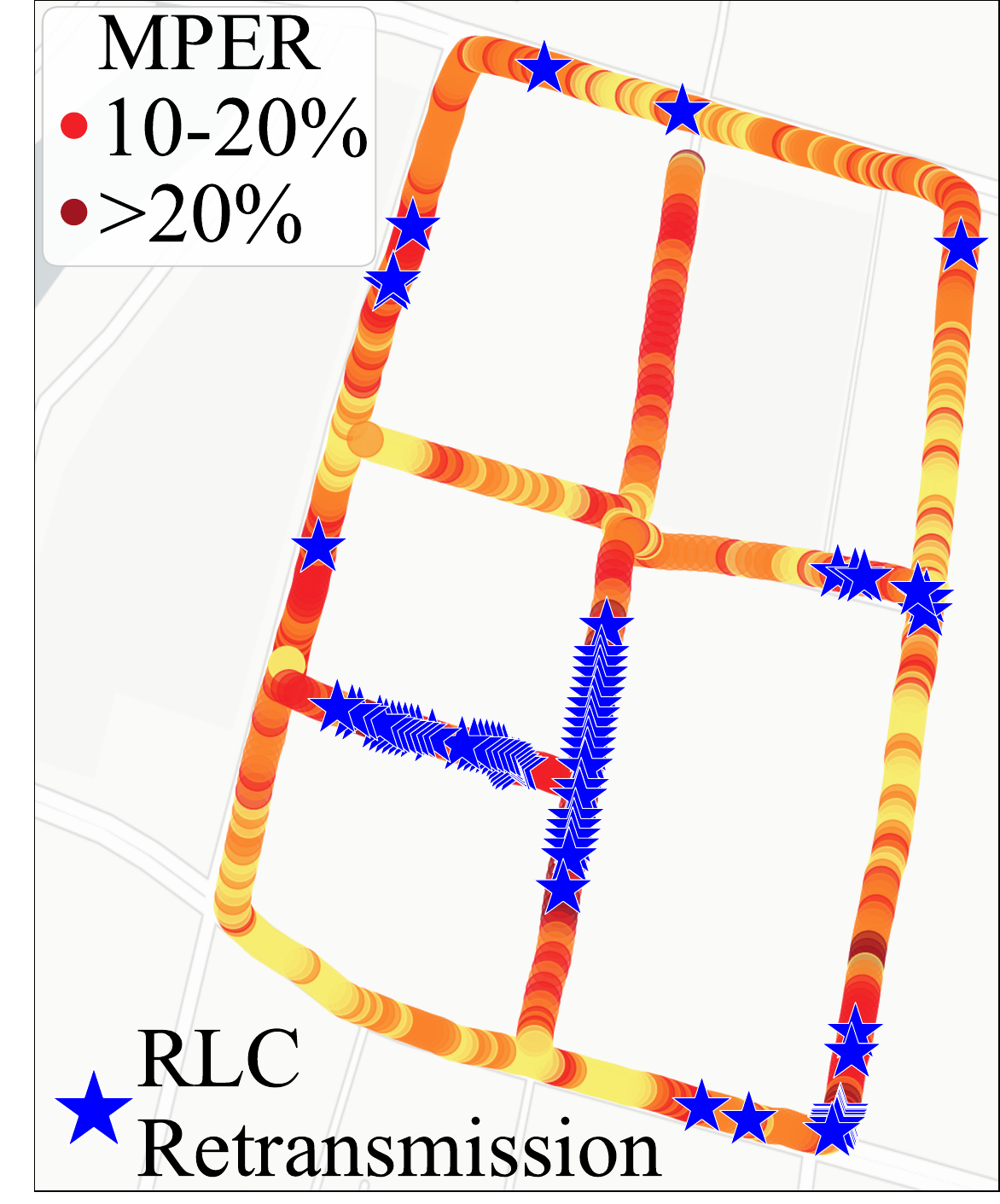}
        \caption{Uplink}
        \label{fig:5g_ul_mper_rlc}
    \end{subfigure}
    \hfill
    \begin{subfigure}[b]{0.41\linewidth}
        \centering
        \includegraphics[width=\textwidth]{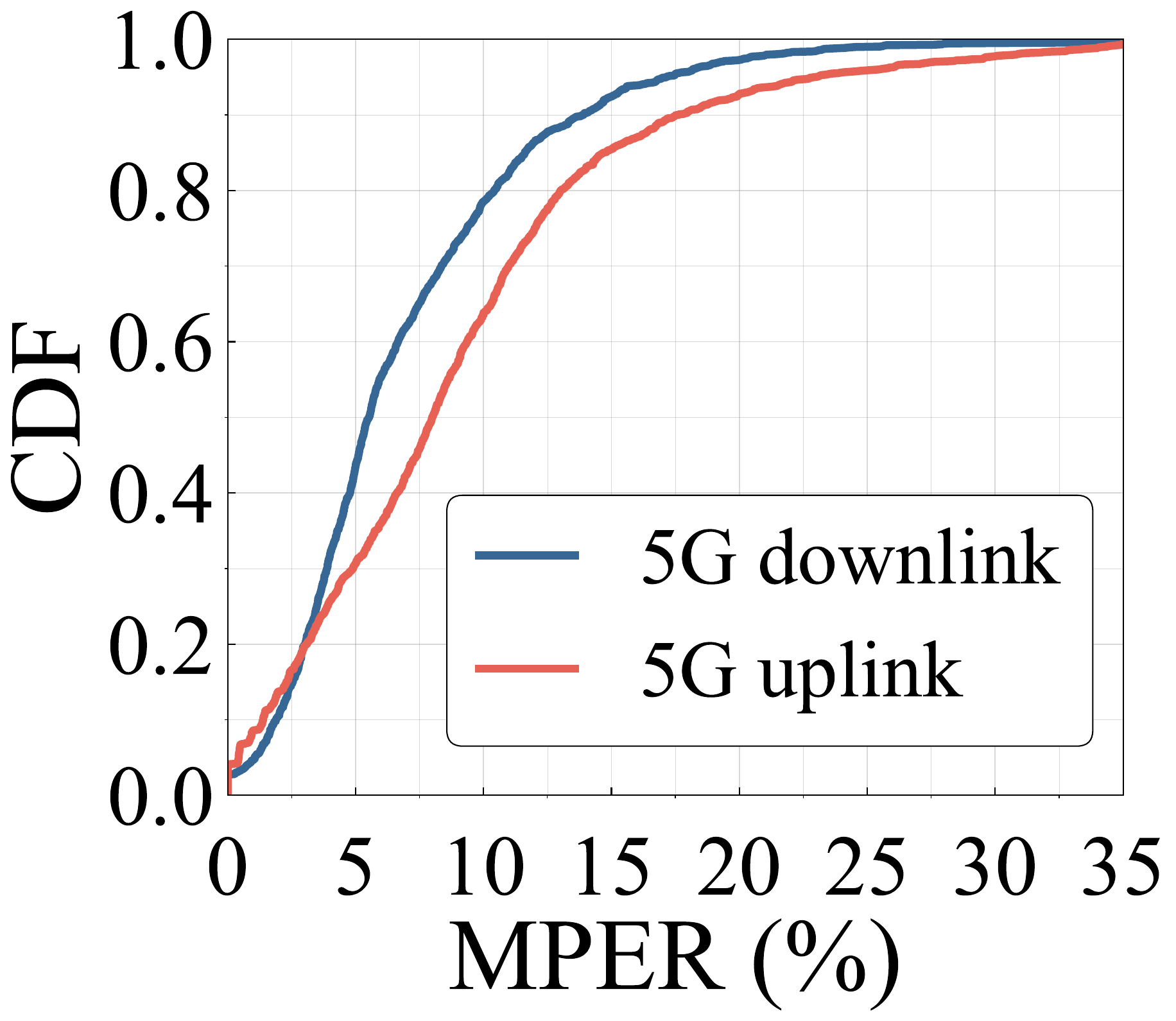}
        \caption{CDF of MPER}
        \label{fig:mper_cdf}
    \end{subfigure}
    \caption{\textbf{Retransmission patterns in 5G SA networks:} (\textbf{a}) downlink and (\textbf{b}) uplink MPER across locations, 
    with stars indicating RLC retransmission events. 
    Colors represent different MPER ranges. (\textbf{c}) CDF of per-second MPER. }
    %\textbf{MPER and RLC retransmission in 5G}:
    %(a) downlink and (b) uplink show the spatial distribution of MPER and RLC retransmission on a map, (c) presents the CDF of MPER in 5G derived from the spatial data in (a) and (b).}
    \vspace{-0.5cm}
    \label{fig:retx_freq}
\end{figure}

Figure~\ref{fig:retx_freq} presents our findings.
Specifically,
Figure~\ref{fig:5g_dl_mper_rlc} and \ref{fig:5g_ul_mper_rlc}
show downlink and uplink MPER distributions across different locations,
with RLC retransmission events marked as stars.
Figure~\ref{fig:mper_cdf} shows the CDF of per-second MPER.
Our measurements reveal that MAC layer retransmissions are pervasive:
the median MPER is 5.54\% in downlink and 8.00\% in uplink, 
indicating that a significant portion of MAC PDUs
require retransmission.
While RLC layer retransmissions occur less frequently,
the scattered stars across both uplink and downlink 
show they remain a consistent feature of cellular transmission.
Importantly, these retransmissions are inherent behaviors of cellular RAN
caused by wireless channel errors,
occurring independently of traffic patterns. 

%\paragraph{Takeaway.}
%MAC layer retransmissions occur frequently,
%thus introducing frequent but relatively small delay inflations.
%In contrast, RLC layer retransmissions occur less frequently
%but introduce significantly larger delays.

\subsection{Uplink Scheduling and Uplink Buffer}

\begin{figure}[t]
    \centering
    \begin{subfigure}[b]{0.5\linewidth}
        \centering
        \includegraphics[width=0.99\textwidth]{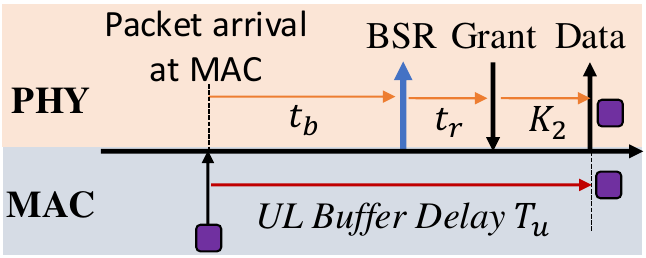}
        \caption{Single packet timing.}
        \label{fig:ul_sch_delay_a}
    \end{subfigure}
    \hfill
    \begin{subfigure}[b]{0.48\linewidth}
        \centering
        \includegraphics[width=0.99\textwidth]{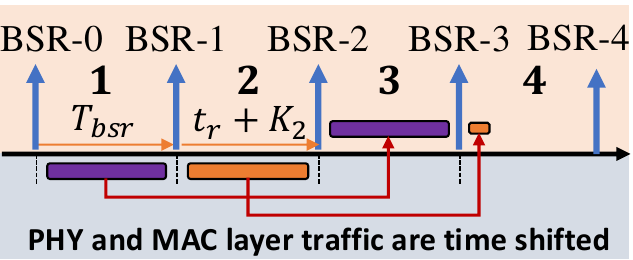}
        \caption{Periodic BSR reporting.}
        \label{fig:ul_sch_delay_b}
    \end{subfigure}
    \caption{\textbf{MAC-to-PHY layer packet progression}: 
    (\textbf{a}) decomposition of MAC-to-PHY layer buffer delay for a single packet and 
    (\textbf{b}) time shift between PHY and MAC layer traffic due to periodic BSR reporting.}
    \label{fig:ul_sch_delay}
\end{figure}

\paragraph{Uplink Scheduling.}
The MAC layer of the UE cellular modem cannot transmit packets directly to the PHY layer
because uplink transmissions are controlled by the base station through grants.
This creates a fundamental challenge:
the base station is unaware of packet arrivals at the UE
and cannot proactively schedule resources.
To bridge this gap, the UE sends a \textit{buffer status report} (BSR)
to inform the base station of its pending upload data volume.
Upon receiving the BSR,
the base station allocates appropriate resources and issues a grant to the UE,
as illustrated in Figure~\ref{fig:ul_sch_delay_a}.
Following grant reception, the UE must wait for $K_2$ TTIs
before initiating the uplink transmission (\S\ref{s:reTx_buf}).
Once transmission begins, the UE periodically reports BSR with an interval of $T_{bsr}$,
enabling the base station to track ongoing buffer status.

\subsubsection{Uplink Buffer Delay.}\label{sec:buffer_acu}
Our analysis shows that each packet experiences a delay $T_u$ in the UE's MAC layer \textit{uplink buffer}
before transmission to the PHY layer,
as illustrated in Figure~\ref{fig:ul_sch_delay_a}.
This total delay consists of three components:
\begin{equation}
T_U = t_b + t_r + K_2
\label{eqn:ul_buf_delay}
\end{equation}
In Equ.~\ref{eqn:ul_buf_delay},
the value $t_b$ represents the interval between packet arrival and BSR reporting.
Due to the periodic nature of BSR transmission,
$t_b \in [0, T_{bsr}]$.
The parameter $K_2$ is a constant system parameter.
%whose values are
%detailed in Table~\ref{tab:ul_reTx_config}
%of Appendix.
The delay $t_r$ between BSR reception and grant transmission
is determined by the base station's uplink scheduling algorithm.
Our empirical measurements across four operators
(two in the US and two in China)
demonstrate that in both LTE and 5G SA networks,
$t_r$ is consistently configured to maintain:
\begin{equation}
T_{bsr} \leq t_r + K_2 \leq 2T_{bsr}
\end{equation}
This scheduling configuration ensures that traffic arriving during the $i$-$th$ BSR interval
is transmitted during the $(i+2)$-$th$ interval,
as shown in Figure~\ref{fig:ul_sch_delay_b}.
Consequently, MAC layer and PHY layer traffic patterns exhibit a time shift
of two BSR intervals.
When sufficient uplink resources are available,
the base station typically schedules all uplink traffic,
including both data and BSR, within a single TTI.
Figure~\ref{fig:ul_sch_delay_b} illustrates this behavior,
where data accumulated during the second BSR interval
is transmitted together with the third BSR.
This combined transmission of BSR and user data
reduces signaling overhead and simplifies the uplink process,
as BSR transmission also requires grants from the base station.
\begin{figure*}
    \centering
    \begin{minipage}{0.235\linewidth}
        \centering 
        \includegraphics[width=0.95\textwidth]{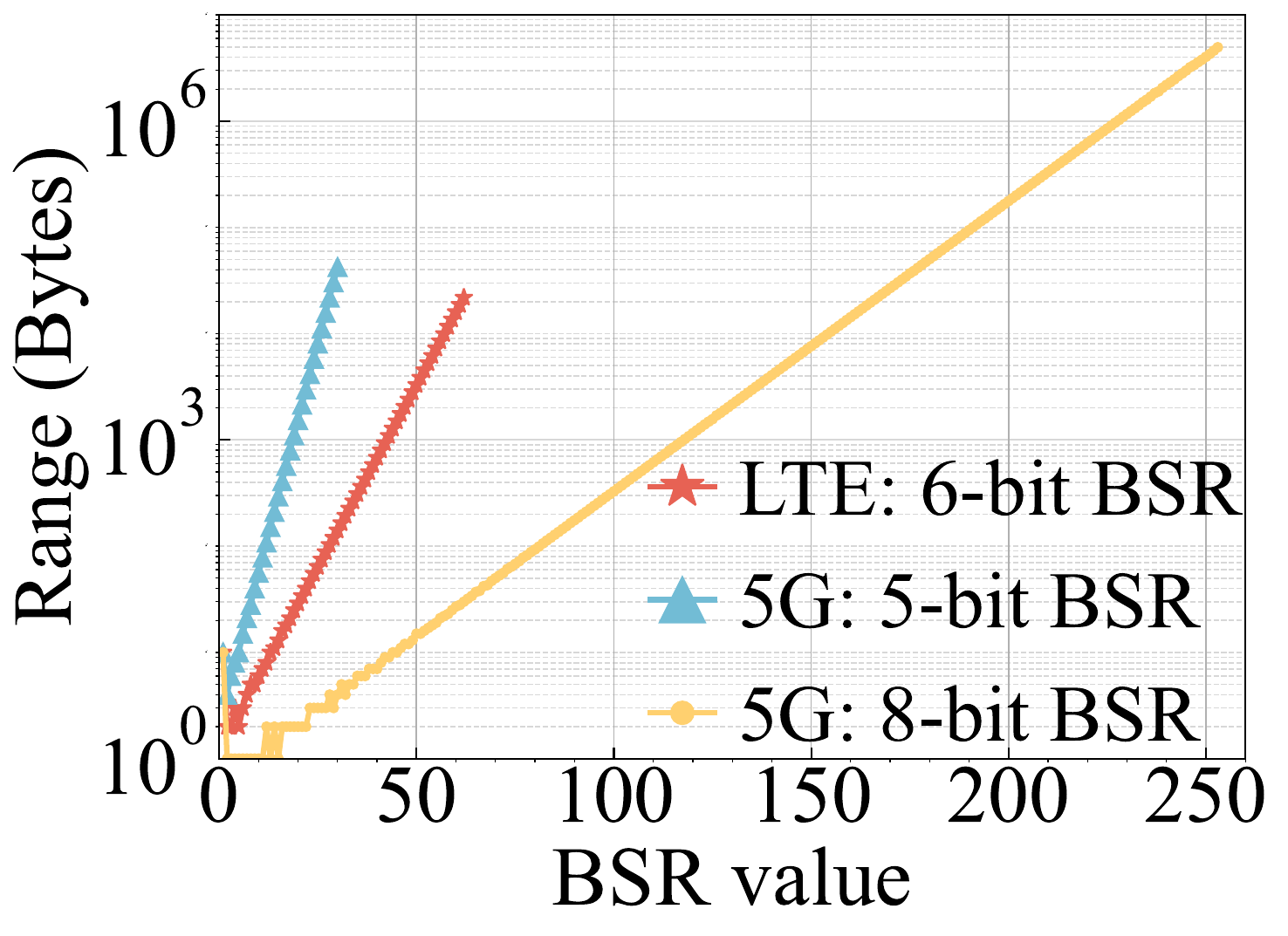}
        \caption{The range of bits BSR represents in LTE and 5G.}
        % The buffer size field in 5G can be either 5 bits or 8 bits in length, depending on the BSR format.
        \label{fig:bsr_range}      
    \end{minipage}
    \hfill
    \begin{minipage}{0.285\linewidth}
        \centering 
        \includegraphics[width=0.95\textwidth]{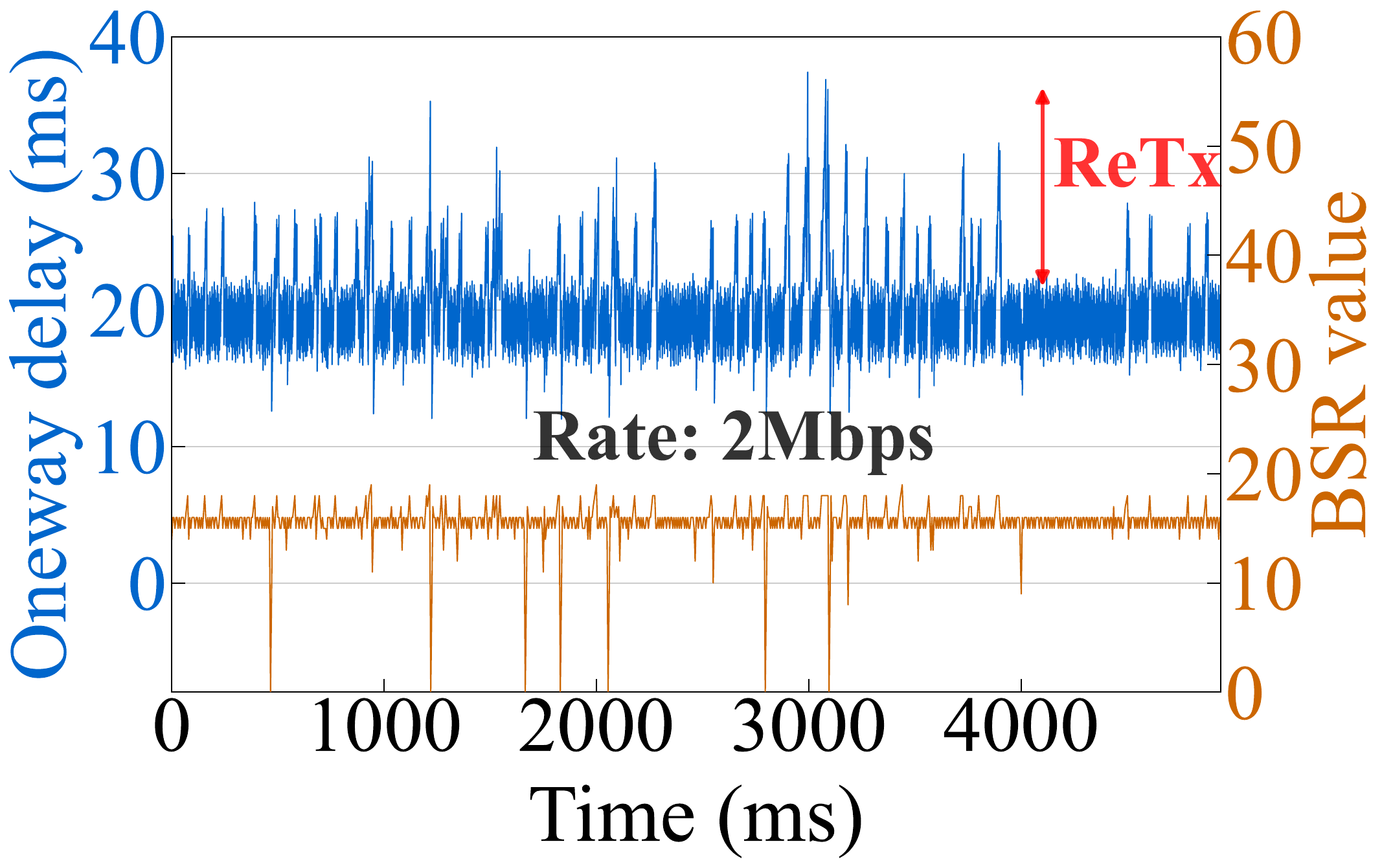}
        \caption{The 5G uplink oneway delay and corresponding BSR value.}
        \label{fig:bsr_value}      
    \end{minipage}
    \hfill
    \begin{minipage}{0.465\linewidth}
        \centering 
        \includegraphics[width=0.95\textwidth]{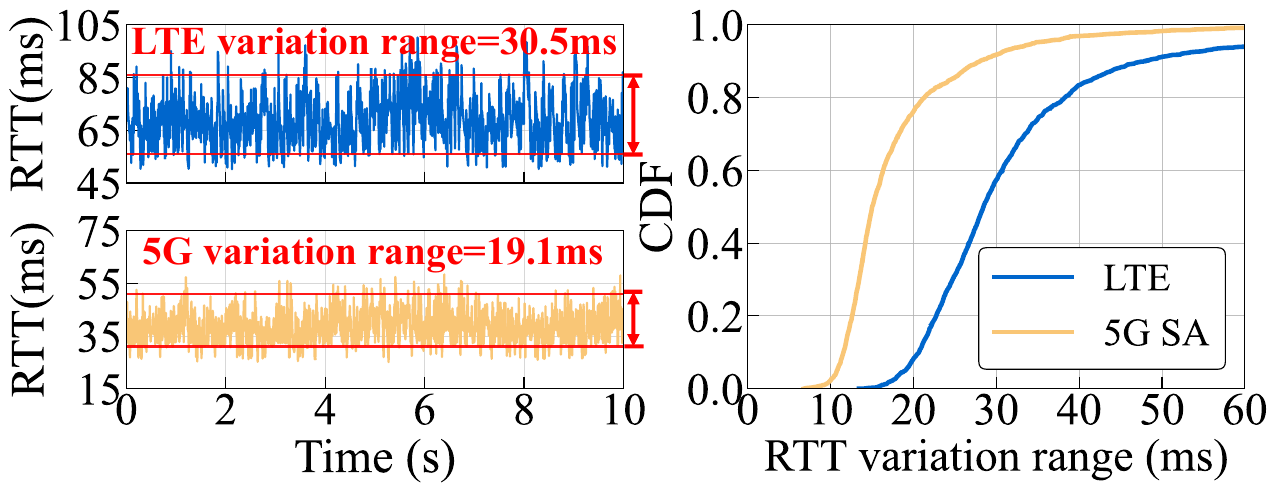}
        \caption{Collective impact of RAN-induced delay on end-to-end packet delivery.}
        \label{fig:delay_variance}      
    \end{minipage}
\end{figure*}

\paragraph{Takeaway.}
Each uplink packet experiences a buffer delay of one to two BSR intervals
in the uplink buffer before PHY layer transmission,
with the exact duration depending on packet arrival time and BSR reporting schedule.

% \begin{figure}[htb]
%     \centering
%     \includegraphics[width=0.49\linewidth]{figure/bsr_index.pdf}
%     \caption{The range of buffered bits represented by each BSR value of LTE and 5G. The buffer size field in 5G can be either 5 bits or 8 bits in length, depending on the BSR format.}
%     \label{fig:bsr_range}
% \end{figure}

% \begin{figure}[htb]
%     \centering
%     \includegraphics[width=0.49\linewidth]{figure/bsr_value.pdf}
%     \caption{The 5G uplink oneway delay and corresponding BSR value.}
%     \label{fig:bsr_value}
% \end{figure}

\subsubsection{Practical Buffer Estimation Challenges.}\label{sec:buffer_clear}
Theoretical analysis in \S\ref{sec:buffer_acu} suggests that during continuous transmission,
the uplink buffer should never be empty, with all packets experiencing
a minimum buffering delay of one BSR interval.
This fixed delay could be considered equivalent to propagation delay,
as it affects only absolute delay values rather than variance.
However, practical observations reveal packets occasionally experience
less than one BSR interval of buffering, or even no buffering delay.

This discrepancy stems from the base station's inability to accurately estimate
the UE's buffered data volume, due to two fundamental challenges.
First, the BSR format has limited precision:
rather than conveying exact buffer sizes, each BSR value represents a range
that increases exponentially, as shown in Figure~\ref{fig:bsr_range}
for both LTE and 5G SA networks.
Second, data may be reported multiple times in successive BSRs.
For instance, as shown in Figure~\ref{fig:ul_sch_delay_b},
the second BSR reports both newly accumulated data from its interval
and previously reported but untransmitted data from the first interval.
If the base station were to allocate resources solely based on BSR reports,
it might redundantly allocate resources for the same data multiple times.
To avoid this inefficiency, the base station must estimate 
actual buffer status
by correlating scheduled resources, transmitted data, and BSR reports
to distinguish new data from previously reported content.

These two challenges make the base station's estimation of buffered data
imprecise and frequently result in resource over-allocation.
The UE utilizes any excess granted resources to transmit additional packets,
effectively allowing some packets to bypass the expected buffering delay.
This behavior is evident in Figure~\ref{fig:bsr_value},
which depicts the relationship between oneway delay and BSR value.
With the UE strictly pacing at 2Mbps,
the BSR value should remain stable.
However, we observe sporadic drops of BSR to zero,
indicating that the base station's over-allocation allows the UE
to transmit more data than its steady arrival rate.
Note that if the UE did not strictly pace its sending rate at 2Mbps,
we would observe more frequent BSR drops since varying transmission rates
make it more challenging for the base station to estimate buffer status accurately.
%These occasional buffer depletions and the resulting opportunistic transmissions
%explain the observed variations in actual buffering delays.

\paragraph{Takeaway.}
In practice, packets deviate from the theoretical delay pattern
that combines a fixed one BSR interval with an additional variable delay of up to one BSR interval.
Instead, uplink packets experience variable uplink buffering delay $T_U$
ranging from zero to two BSR intervals.

% \begin{figure}[htb]
%     \centering
%     \includegraphics[width=0.98\linewidth]{figure/delay_variance.pdf}
%     \caption{Collective impact of RAN-induced delay on end-to-end packet delivery.}
%     \label{fig:delay_variance}
% \end{figure}
\subsection{Collective Impact of RAN Delays}
The RAN introduces multiple sources of delay through MAC retransmissions,
RLC retransmissions, and uplink buffering,
which can affect packets either individually or in combination.
The impact becomes particularly significant for RTT-based applications,
as RTT measurements encompass both paths:
downlink packets may experience both MAC and RLC retransmissions and thus get delayed,
while uplink packets potentially face all three delay sources.

To quantify this impact on RTT,
we conducted extensive experiments between an AWS server and a UE.
The server transmitted packets at 2Mbps with random inter-packet intervals,
while the UE responded with 40-byte ACKs for RTT measurement.
We maintained this low data rate to minimize congestion effects.
The experiment comprised 400 30-second trials across one week.
Figure~\ref{fig:delay_variance}(a) and (b) show representative RTT traces
for LTE and 5G networks respectively,
while Figure~\ref{fig:delay_variance}(c) presents the CDF of RTT variation range (95th percentile minus 5th percentile RTT)
across all traces.
Even without network congestion,
we observe significant RTT variations:
individual traces show RTT variation ranges 
of 30.5ms in LTE and 19.1ms in 5G.
These variations, stemming from RAN mechanisms rather than network congestion,
can mislead applications about network conditions.

%The RAN introduces multiple sources of delay through MAC retransmissions,
%RLC retransmissions, and uplink buffering,
%which can affect packets either individually or in combination.
%The impact becomes particularly significant for RTT-based applications,
%as RTT measurements encompass both paths:
%downlink packets may experience both MAC and RLC retransmissions and thus get delayed,
%while uplink packets potentially face all three delay sources.
%
%To quantify this impact on RTT,
%we conducted extensive experiments between an AWS server and a UE.
%The server transmitted packets at 2Mbps with random inter-packet intervals,
%while the UE responded with 40-byte ACKs for RTT measurement.
%We maintained this low data rate to minimize congestion effects.
%The experiment comprised 400 30-second trials across one week.
%Figure~\ref{fig:delay_variance}(a) and (b) show representative RTT traces
%for LTE and 5G networks respectively,
%while Figure~\ref{fig:delay_variance}(c) presents the CDF of RTT variance
%across all traces.
%Even without network congestion,
%we observe significant RTT variations:
%individual traces show RTT spreads of 30.5ms in LTE and 19.1ms in 5G.
%These variations, stemming from RAN mechanisms rather than network congestion,
%can mislead applications about network conditions.
%
%

% \begin{figure}[htb]
%     \centering
%     \includegraphics[width=0.99\linewidth]{figure/delay_variance.pdf}
%     \caption{Collective impact of RAN-induced delay on end-to-end packet delivery.}
%     \label{fig:delay_variance}
% \end{figure}

\section{Mitigating Impact of Buffers inside RAN}
We present our solution for mitigating RAN buffer impacts
on end-to-end protocols and applications.
We first provide a system overview in §\ref{sec:overview},
then present \tool (§\ref{sec:cellninja}) for real-time RAN visibility
and \sys (§\ref{sec:gandalf}) for eliminating RAN buffer impacts.

%We present our solution for mitigating RAN buffer impacts 
%on end-to-end protocols and applications.
%First, we provide a system overview (§\ref{sec:overview}) 
%that introduces our architecture
%for enabling cross-layer RAN-aware optimization.
%We then present \tool  (§\ref{sec:cellninja}), which provides instantaneous visibility into RAN operations,
%followed by \sys (§\ref{sec:gandalf}), which leverages this visibility to eliminate RAN buffer impacts on end-to-end performance.

\subsection{Overview of Solution}\label{sec:overview}

To mitigate the impact of RAN internal buffers,
we require both instantaneous visibility into cellular RAN operations
and the ability to correlate this information with data transmission.
Our solution leverages the dual-subsystem architecture of cellular modems:
a data subsystem responsible for user data delivery,
and a diagnostic subsystem that collects and reports RAN statistics
for debugging purposes.
As shown in Figure~\ref{fig:system_overview},
our solution comprises two key components that integrate these subsystems,
both implemented purely in software and running directly on mobile devices.
\tool provides comprehensive, low-latency visibility into RAN operations
by efficiently collecting and processing diagnostic messages.
\sys correlates this RAN information with network data
to identify and compensate for RAN-induced effects on data delivery in real time.
The diagnostic subsystem reports RAN statistics via diag messages,
while the data subsystem handles actual data transmission through the network protocol stack.

This architecture establishes a novel framework that fundamentally transforms
how protocols and applications can interact with cellular networks.
By providing unprecedented, instantaneous visibility into RAN operations through \tool,
this architecture enables a new generation of cross-layer cellular-aware applications
that can dynamically adapt to fine-grained network characteristics.
The purely software-based implementation, requiring no hardware modifications,
makes this powerful capability immediately accessible to any mobile device.
This novel framework opens possibilities for innovative optimization techniques
that were previously impossible without detailed RAN insights,
potentially revolutionizing mobile network performance optimization.

%This software-based framework opens possibilities for various optimization techniques
%$that leverage fine-grained RAN information for enhanced mobile performance.
%\subsection{CellNinjia: Real-time, Comprehensive and Cross-Layer Visibility into the RAN}

\subsection{CellNinjia: Visibility into the RAN}\label{sec:cellninja}
%Real-Time Mobile Cellular Network Diagnostics and Analysis}
\paragraph{Design Goals.}
To enable real-time optimization of cellular network performance,
we need comprehensive and instantaneous visibility into RAN operations.
Our system must meet the following requirements:
\noindent
\textit{1) Software-based Implementation.}
    The system must be directly implementable on mobile devices
    without requiring additional hardware or external tools.
\noindent
\textit{2) Comprehensive Parameter Access.}
    The tool must provide access to a wide range of internal RAN parameters
    shared between base stations and UEs.
    %such as BSR intervals, DL-UL configurations, and timing parameters
    %($K_0$, $K_1$, $K_2$, $K_d$, and $K_u$), to name a few.
\noindent
\textit{3) Fine-grained Monitoring.}
    The system must monitor data transmission of the cellular modem 
    at TTI-level granularity
    to accurately track fine-grained RAN activities,
    particularly retransmission events.
\noindent
\textit{4) Real-time Operation.}
    The system must provide instantaneous visibility into RAN behaviors
    directly on mobile devices,
    enabling real-time correlation with data transmission
    and supporting immediate protocol adaptation
    without external processing or additional hardware.
\noindent
\textit{5) Multi-generation Support.}
    The system must support both LTE and 5G SA networks.
    %\vspace{-0.3cm}

\begin{figure}
    \centering
    \includegraphics[width=0.85\linewidth]{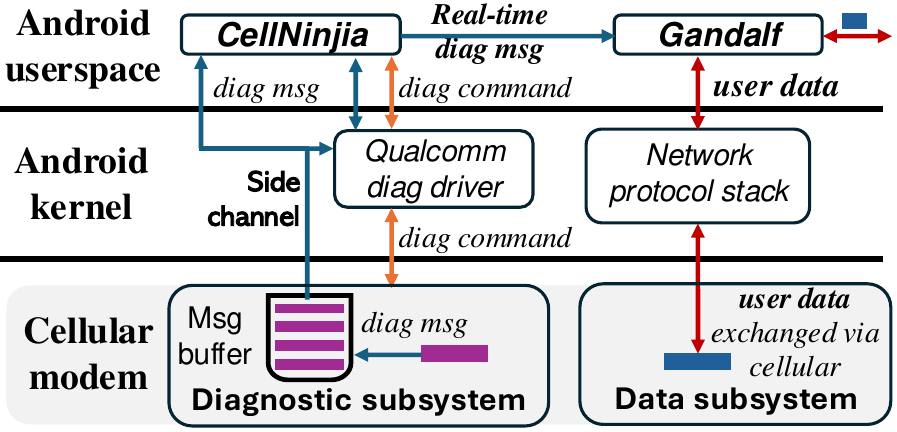}
    \caption{\textbf{System architecture:} 
    \tool provides instantaneous visibility into RAN through the diagnostic subsystem, 
    while \sys correlates this information with user data 
    from the data subsystem to optimize cellular network performance. 
    Both components run directly on the mobile device, leveraging existing cellular modem capabilities.}
    \label{fig:system_overview}
\end{figure}

\noindent
\textbf{Qualcomm Diagnostic Subsystem.}
Qualcomm provides access to RAN internal states
through a \textit{diagnostic subsystem} (diag),
whose architecture is shown in Figure~\ref{fig:system_overview}.
The cellular modem hardware collects RAN
status and statistics, encodes this information
into \textit{diag messages},
and stores them in a \textit{diag message buffer}.
Upon buffer saturation,
the modem aggregates all buffered messages into a single
\textit{diag message group}, transfers this group to the Android kernel,
and clears the buffer.
A Qualcomm-implemented diag driver in the kernel
receives these message groups and buffers them in the kernel.
The buffered message groups are reported to userspace applications upon query;
otherwise, they are discarded.

\noindent
\textbf{Challenge.}
While the diag system provides valuable access to RAN status,
it presents two significant challenges.
First, the modem's buffering mechanism introduces substantial delay
in diag message collection.
Second, while Qualcomm's diag driver is open-source,
the modem firmware and its interaction protocols remain closed-source,
making both communication with the modem and interpretation of its diagnostic messages challenging.
This limited transparency particularly impacts the research community,
requiring significant reverse engineering effort to achieve reliable system operation.

\subsubsection{Design of CellNinjia}\label{sec:cellninjia_design}
We design and implement \tool as a mobile phone software solution
that leverages the Qualcomm diagnostic system
to achieve all the above goals.

\begin{figure}[t]
    \centering
        \begin{minipage}[t]{0.49\linewidth}
        \centering
        \includegraphics[width=0.99\linewidth]{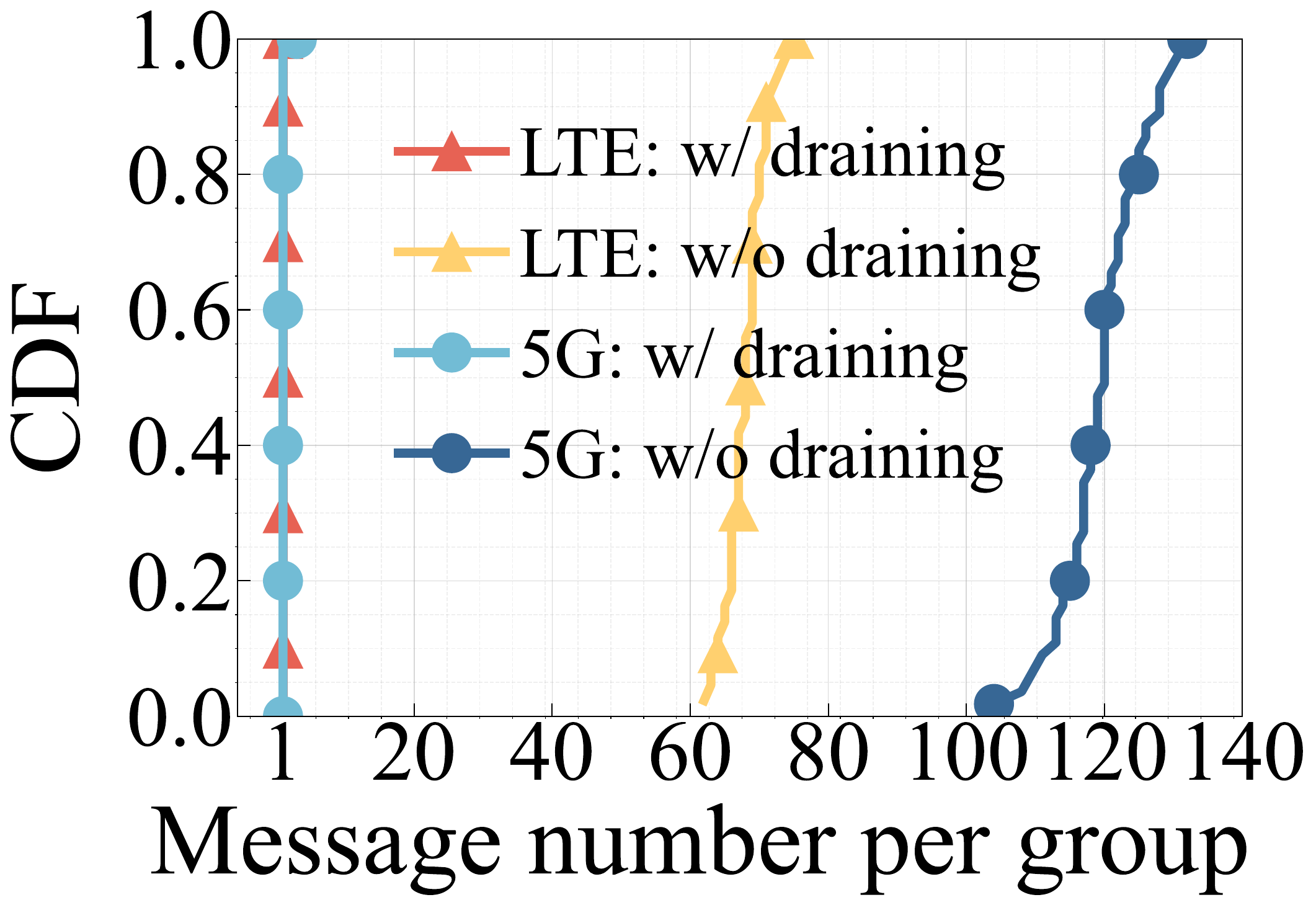}
        \caption{CDF of message number per group with and without buffer draining.}
        \label{fig:single_msggroup_size}
    \end{minipage}
    \hfill
    \begin{minipage}[t]{0.49\linewidth}
        \centering
        \includegraphics[width=0.99\linewidth]{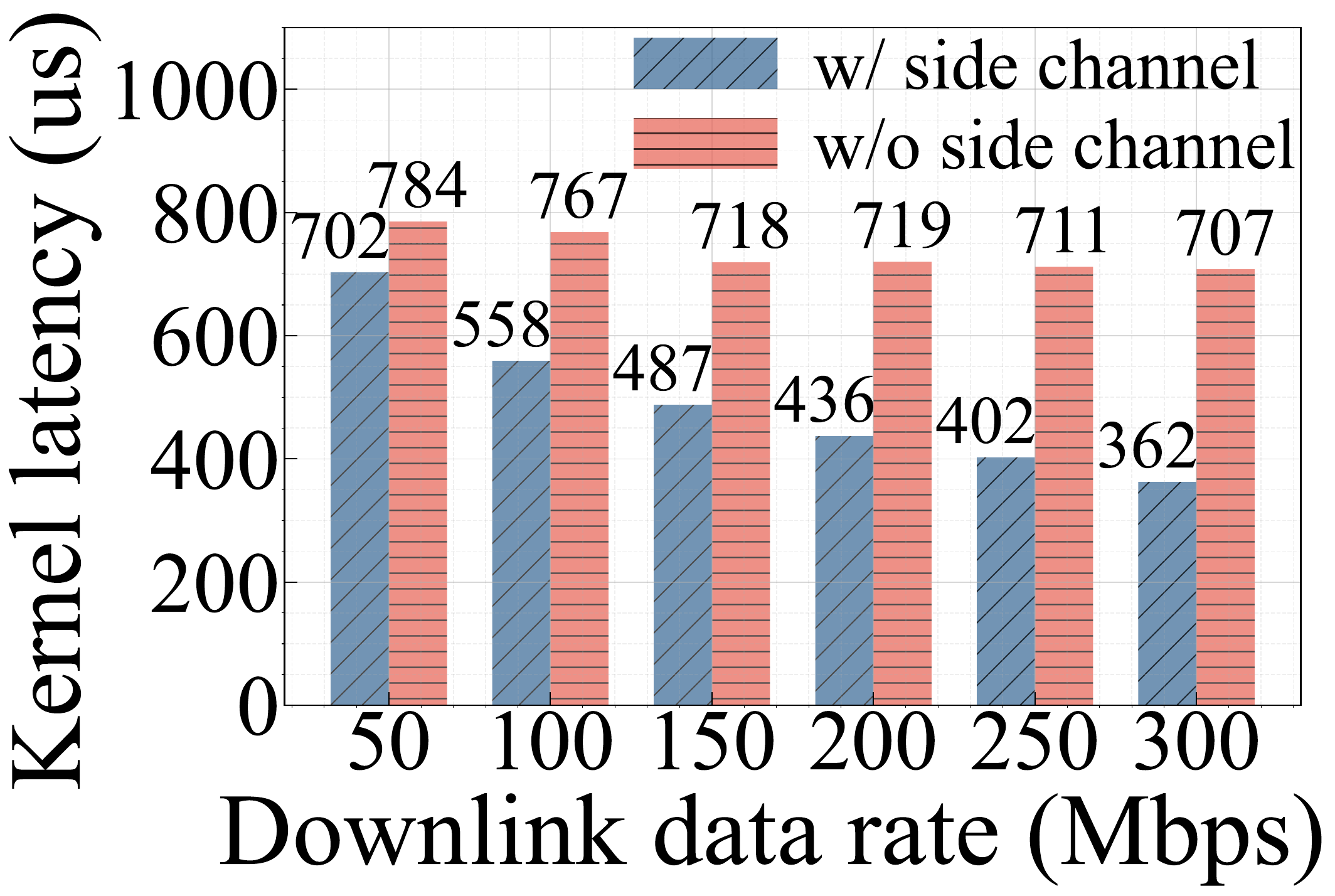}
        \caption{The median kernel latency with and without side channel.}
        \label{fig:kernel_latency}
    \end{minipage}
\end{figure}

\paragraph{Buffer Draining.}
To achieve instantaneous RAN message collection,
we leverage the diag system's support for manually draining the diag message
buffer in the modem through commands sent via the diag driver.
To validate this approach,
we configure the modem to collect only one message type:
\textsf{LTE\_LL1\_PUSCH\_Tx\_Report} for LTE
and \textsf{NR5G\_MAC\_UL\_\\PHY\_Channel\_Schedule\_Report} for 5G SA,
each describing UE uplink status within one TTI.
Since these messages are generated only during active transmission,
we saturated the uplink to ensure theoretical generation of one message per TTI.
Figure~\ref{fig:single_msggroup_size} shows our measurements of message group sizes
under continuous querying of the diag interface.
With manual buffer draining,
we reduced the message group size to one,
indicating immediate message reporting upon generation.
In contrast, without draining,
the modem accumulates approximately 70 messages for LTE
and 120 messages for 5G SA before kernel transfer.
We note that existing open-source tools
including MobileInsight~\cite{li_mobileinsight_2016},
QCSuper~\cite{noauthor_p1secqcsuper_2025},
and SCAT~\cite{noauthor_fgsectscat_2025}
do not implement this buffer draining mechanism
and 
thus cannot achieve instantaneous message reporting.

\paragraph{Side Channel.}
The Qualcomm diag driver handles multiple functionalities
with a complex processing chain for each message
received from the modem,
for example, modem handles both diag messages and control messages.
This extensive processing introduces significant delay between
a message's arrival at the Android kernel and its delivery to userspace.
To minimize this processing latency,
we implement a kernel-level side channel that
immediately forwards diag messages to \tool
upon their arrival in the Android kernel.
This side channel also provides an interface for other kernel modules
to access diag messages directly.
To evaluate our optimization,
we measured the end-to-end delay from message arrival at the kernel
to delivery in userspace,
comparing our side channel against the standard diag driver path.
As shown in Figure~\ref{fig:kernel_latency},
our side channel reduces this delay by approximately 50\%,
with particularly significant improvements under high data rates.

\paragraph{Comprehensive Visibility.}
\tool provides comprehensive visibility into cellular RAN operations
through extensive message monitoring capabilities.
%We summarize the message types that \tool collects
%from the cellular modem via the Qualcomm \textit{diag} interface
%in Table~\ref{tab:cellninjia}.
The current LTE and 5G SA versions of \tool
decode 99 and 60 message types respectively,
providing detailed insights into various aspects of cellular network operation.
%\footnote{\tool is still under active developing and more messages are being added.}.
In comparison, MobileInsight~\cite{li_mobileinsight_2016}
decodes only 57 LTE message types and eight 5G message types,
with 5G support limited to NSA networks~\cite{li_mobileinsight_2016}.
Alternative sniffing approaches like NG-Scope~\cite{xie_ng-scope_2022} and NR-Scope~\cite{wan_nr-scope_2024}
rely on sniffing and can only decode unencrypted broadcast physical layer messages.
At last, \tool enables precise configuration of both 
which message types to collect and how many message types to monitor.
A comprehensive comparison with existing tools is provided in Appendix~\ref{appendix:tool_comparison}.
%demonstrating \tool's enhanced monitoring capabilities.

%\begin{figure}[h]
%    \centering
%    \includegraphics[width=0.49\linewidth]{figure/enable_all.pdf}
%    \caption{CellNinja and MobileInsight kernel latency under various iperf dl data rates.}
%    \label{fig:enable_all}
%\end{figure}

\subsubsection{System Overhead Analysis of \tool}
We quantify \tool's operational overhead through two critical metrics:
the data rate of diagnostic messages from the diag subsystem
and the CPU utilization.
These measurements are particularly important as \tool operates continuously
as a background service on mobile devices.

\begin{figure}[t]
    \centering
        \begin{minipage}[t]{0.49\linewidth}
        \centering
        \includegraphics[width=0.99\linewidth]{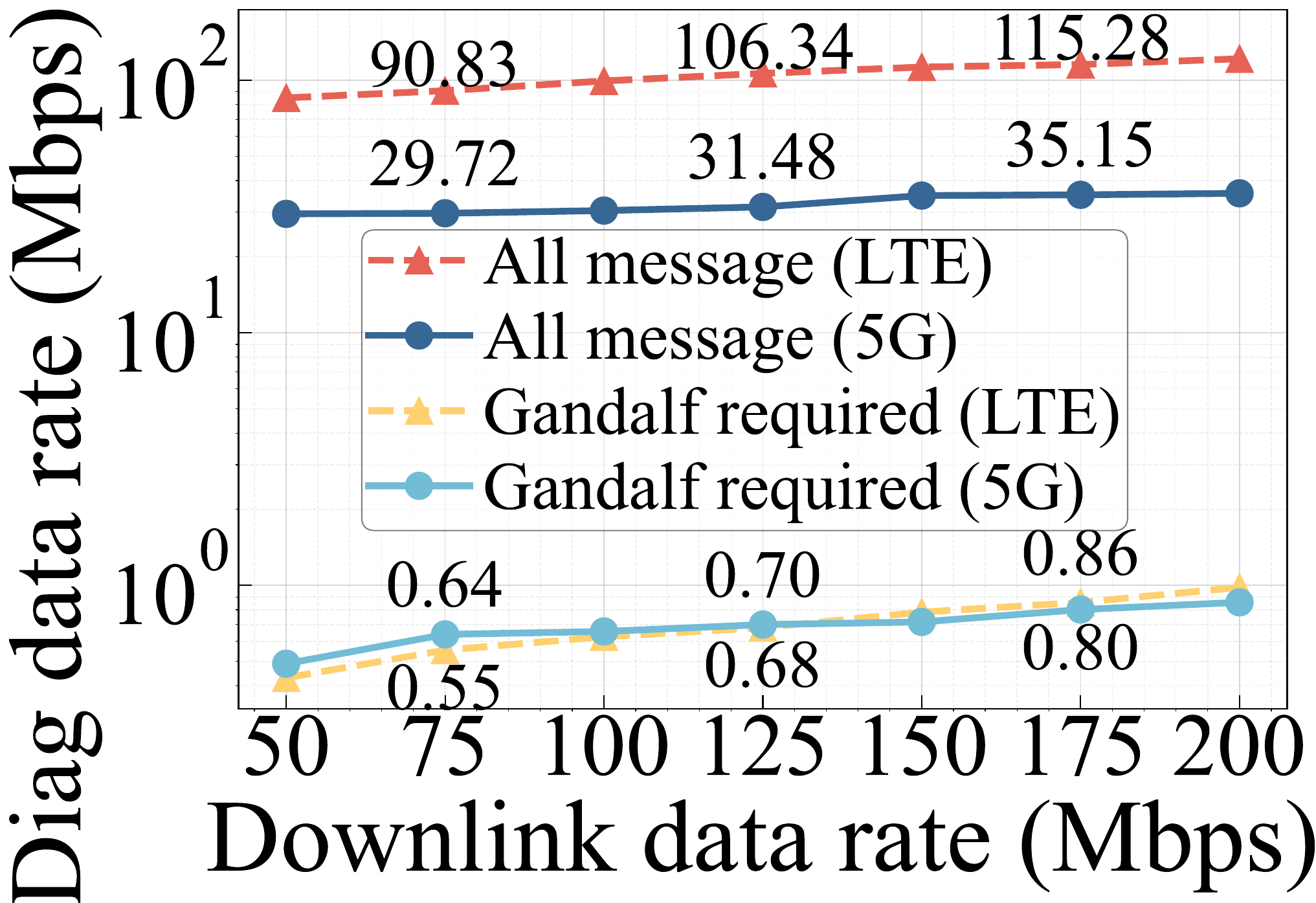}
        \caption{Comparison of diagnostic data rate for all messages and \sys required message under LTE and 5G.}
        \label{fig:diag_data_rate}
    \end{minipage}
    \hfill
    \begin{minipage}[t]{0.49\linewidth}
        \centering
        \includegraphics[width=0.99\linewidth]{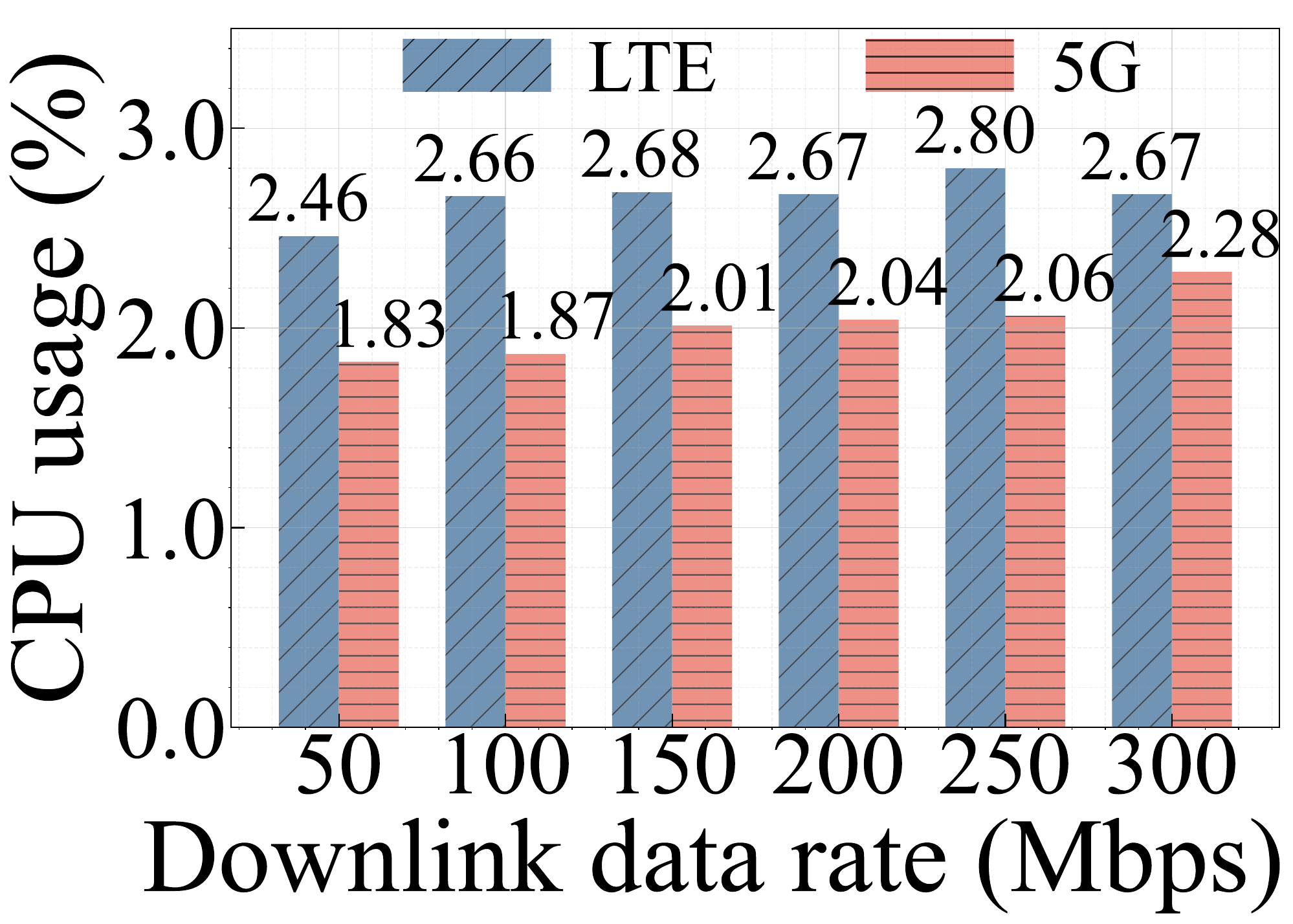}
        \caption{The average CPU usage by \tool in LTE and 5G with only \sys required message enabled.}
        \label{fig:cpu_usage}
    \end{minipage}
\end{figure}

\paragraph{Diagnostic Message Data Rate.}
The volume of diagnostic messages generated depends on the number of message types enabled in \tool.
To quantify this data rate, we evaluated two configurations:
one with all supported message types enabled,
and another with only the messages required by \sys.
We measured diagnostic message data rates while varying the device's download speed
from 50 Mbps to 200 Mbps, testing both LTE and 5G SA networks.
As shown in Figure~\ref{fig:diag_data_rate},
when all message types are enabled,
LTE generates approximately 100 Mbps of diagnostic data,
with a slight increase corresponding to higher download rates.
5G SA produces a lower volume at around 30 Mbps.
In contrast, the minimal configuration required by \sys 
(all the message types \sys used are listed in Appendix~\ref{app:msg_list})
generates less than 1 Mbps of diagnostic data
for both LTE and 5G SA.

These results demonstrate two key points.
First, \sys's minimal overhead
makes it practical for continuous operation on mobile devices.
Second, since \sys uses only a subset of \tool's diagnostic messages,
the rich RAN information captured by \tool
offers significant potential for other cellular network optimizations.

\paragraph{CPU Utilization.}
We evaluate \tool's CPU overhead on a Xiaomi 10 Lite
while varying the device's downlink data rate from 50 Mbps to 300 Mbps,
testing both LTE and 5G SA networks.
As shown in Figure~\ref{fig:cpu_usage},
\tool maintains modest CPU utilization across all test conditions.
For LTE networks, CPU usage remains stable at approximately 2.8%,
while 5G SA networks show even lower utilization at around 2.0%.
These results demonstrate that \tool's CPU overhead
is both minimal and consistent,
making it suitable for continuous background operation
on mobile devices.

\subsection{\sys: Cross-layer Protocol Adaptation}\label{sec:gandalf}
This section presents Gandalf's mechanisms for protocol adaptation
using CellNinja's RAN visibility.
We detail our algorithms for handling retransmission delays (§\ref{sec:retx})
and uplink buffering delays (§\ref{sec:uplink}) separately,
enabling real-time compensation for RAN-induced effects.

\subsubsection{Retransmission Buffer induced Delay}\label{sec:retx}
\begin{figure}[htb]
    \centering
    \begin{subfigure}[b]{0.51\linewidth}
        \centering
        \includegraphics[width=0.99\textwidth]{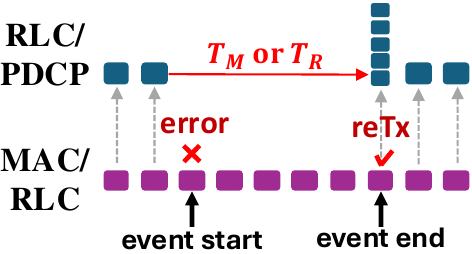}
        \caption{reTx-induced delay inflation.}
        \label{fig:retx_revert_a}
    \end{subfigure}
    \hfill
    \begin{subfigure}[b]{0.48\linewidth}
        \centering
        \includegraphics[width=0.99\textwidth]{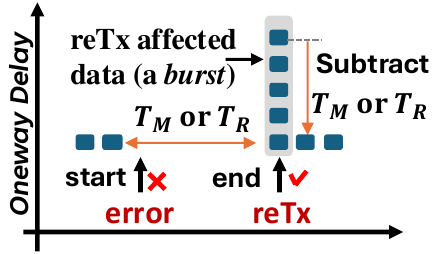}
        \caption{Delay compensation scheme.}
        \label{fig:retx_revert_b}
    \end{subfigure}
    \caption{\textbf{Retransmission delay compensation}: 
    (\textbf{a}) delay inflation of $T_M$ or $T_R$ for packets between error and retransmission,
    (\textbf{b}) subtraction of added delay from affected packets.}
    \label{fig:reTx_delay_revert}
\end{figure}

\noindent
\textbf{Key Observation.}
The retransmission buffer exhibits several key properties
that enable systematic delay mitigation.
First, buffer occupancy occurs exclusively during retransmission events,
having no impact on packet delays during normal operation.
Second, each retransmission event occurs independently,
with well-defined starting and ending points in time,
just as shown in Figure~\ref{fig:retx_revert_a}.
Third, the impact of each event is deterministic,
introducing a delay inflation of $T_M$ for MAC layer
and $T_R$ for RLC layer retransmissions.

\paragraph{Delay Compensation Mechanism.}
The properties of retransmission events
enable a systematic approach to compensate for
retransmission-induced delays.
We follow a three-step process.
First, we identify the temporal boundaries of each retransmission event
using RAN information from \tool,
as illustrated in Figure~\ref{fig:retx_revert_b}.
Second, we identify affected packets based on their characteristics:
a large gap from previous packets due to the retransmission-induced idle period,
followed by near-simultaneous delivery of multiple packets at the event's end.
Finally, we compensate for the delay by
subtracting $T_M$ (MAC) or $T_R$ (RLC) from the error packet
and adjusting all simultaneously reported packets (a burst) to match this corrected delay.
This approach effectively restores packet timing to its baseline state.
Further details about obtaining retransmission delays ($T_M$ and $T_R$)
and tracking retransmission event boundaries using \tool
are provided in Appendix~\ref{appendix:retx_delay} and \ref{appendix:track_retx},
respectively.

%We observe that two reTx event could couple together 
%when a new error happens before the retransmission of the last error is finished.
\vspace{0.08cm}
\noindent
\textit{Handling uplink retransmissions.}
For downlink traffic, correlating retransmission events with affected packets is straightforward
since both RAN information and packet data are available at the UE.
However, uplink presents a unique challenge:
while RAN information is collected at the UE,
packet delays are measured at the remote receiver.

To address this challenge, we transmit retransmission event information
collected using \tool to the remote receiver and synchronize it with packet delay measurements.
The synchronization exploits the unique timing patterns of retransmissions:
packets affected by MAC layer retransmission exhibit an idle period of $T_M$
followed by a burst of packets arriving almost simultaneously,
with the first packet showing a delay increase of $T_M$,
just as shown in Figure~\ref{fig:retx_revert_a}.
Using these timing characteristics,
we can identify potential retransmission events from packet measurements 
alone\footnote{While packet-based identification may miss some retransmission events 
or identify false positives, the high frequency of MAC layer retransmissions 
ensures sufficient events for accurate synchronization.}.
We then search for a time shift that maximizes the overlap between
packet-identified events and \tool-reported events.
Specifically, we compute the ratio of matched events for different time shifts
and select the shift that yields the highest matching ratio.
This synchronized information enables accurate delay compensation
even when RAN information and packet data are collected at different locations.

\subsubsection{Uplink Buffer induced Delay}\label{sec:uplink}
Uplink buffer delay differs fundamentally from retransmission delay.
While retransmissions affect packets independently and discretely,
uplink buffering impacts almost every packet with varying delays.
This continuous nature makes it impossible to identify and eliminate
individual delay events as we did with retransmissions.

Instead, we treat the end-to-end delay (with retransmission delay removed)
as a superposition of two components:
uplink buffer delay and network delay
(including propagation delay and network buffer queuing).
By isolating these components,
we can extract the true network delay signal
for use by higher-layer protocols and applications.
This separation is crucial because uplink buffer occupancy
does not indicate network congestion and should not trigger
congestion control mechanisms in end-to-end applications.

\paragraph{Periodic Buffer Behavior.}
We observe that the uplink buffer exhibits periodic filling and clearing
due to the periodic nature of BSR reporting.
Consequently, the uplink buffer-induced delay manifests as
a periodic signal with frequency determined by the BSR interval $T_{bsr}$:
\begin{equation}
f_{u} = \frac{1}{T_{bsr}}
\end{equation}
The base station configures and communicates $T_{bsr}$ to the UE,
with possible values ranging from five milliseconds to several seconds.
Through \tool, we can continuously track the value of $T_{bsr}$ in real-time 
(detailed in Appendix~\ref{appendix:bsr}).
Our measurements show that $T_{bsr}$ is typically set to five milliseconds
for both LTE and 5G networks.
%though it may increase to tens of milliseconds
%during periods of limited data exchange.

\paragraph{RAN-aware Delay Filtering.}
We leverage this periodic behavior to isolate network delay
from uplink buffer delay using signal processing techniques.
After removing retransmission delays,
we apply FFT to the end-to-end delay measurements,
then use a low-pass filter to remove the higher-frequency
buffer-induced delay component ($f_u$) while preserving
the lower-frequency network delay variations.
Network delay variations typically occur at lower frequencies
as they stem from congestion buildup and dissipation,
which happen over longer time scales
(hundreds of milliseconds to seconds)
compared to the BSR-driven buffer dynamics
(typically every 5-10 milliseconds).
An inverse FFT then recovers the filtered delay signal,
\textit{i.e.}, the network delay.

\begin{figure}[tb]
    \setlength{\abovecaptionskip}{2pt}
    \setlength{\belowcaptionskip}{-2pt}
    \centering
    \begin{subfigure}[b]{0.31\linewidth}
        \centering
        \includegraphics[width=\textwidth]{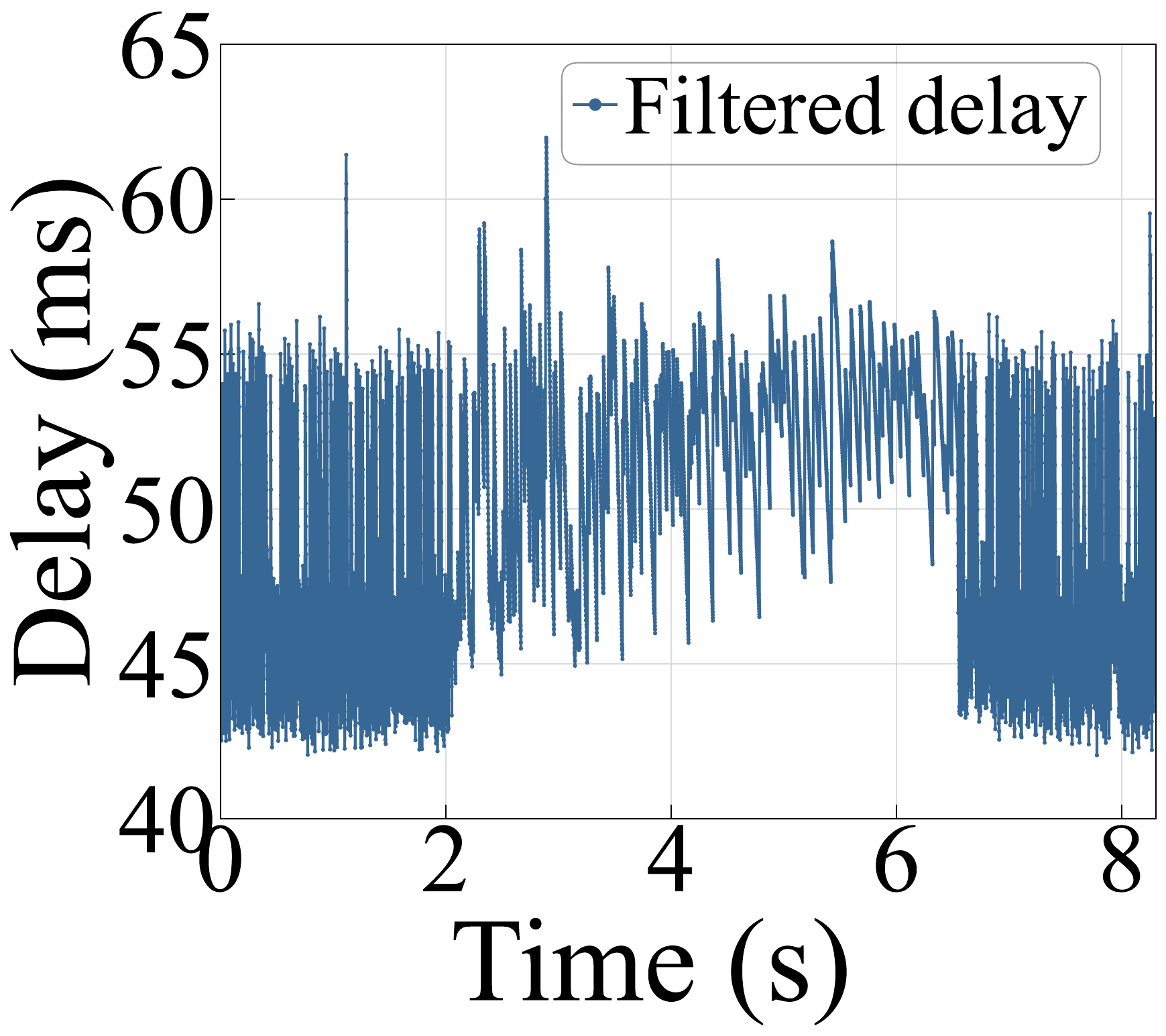}
        \caption{The raw delay}
        \label{fig:filter_raw_delay}
    \end{subfigure}
    \hfill
    \begin{subfigure}[b]{0.31\linewidth}
        \centering
        \includegraphics[width=\textwidth]{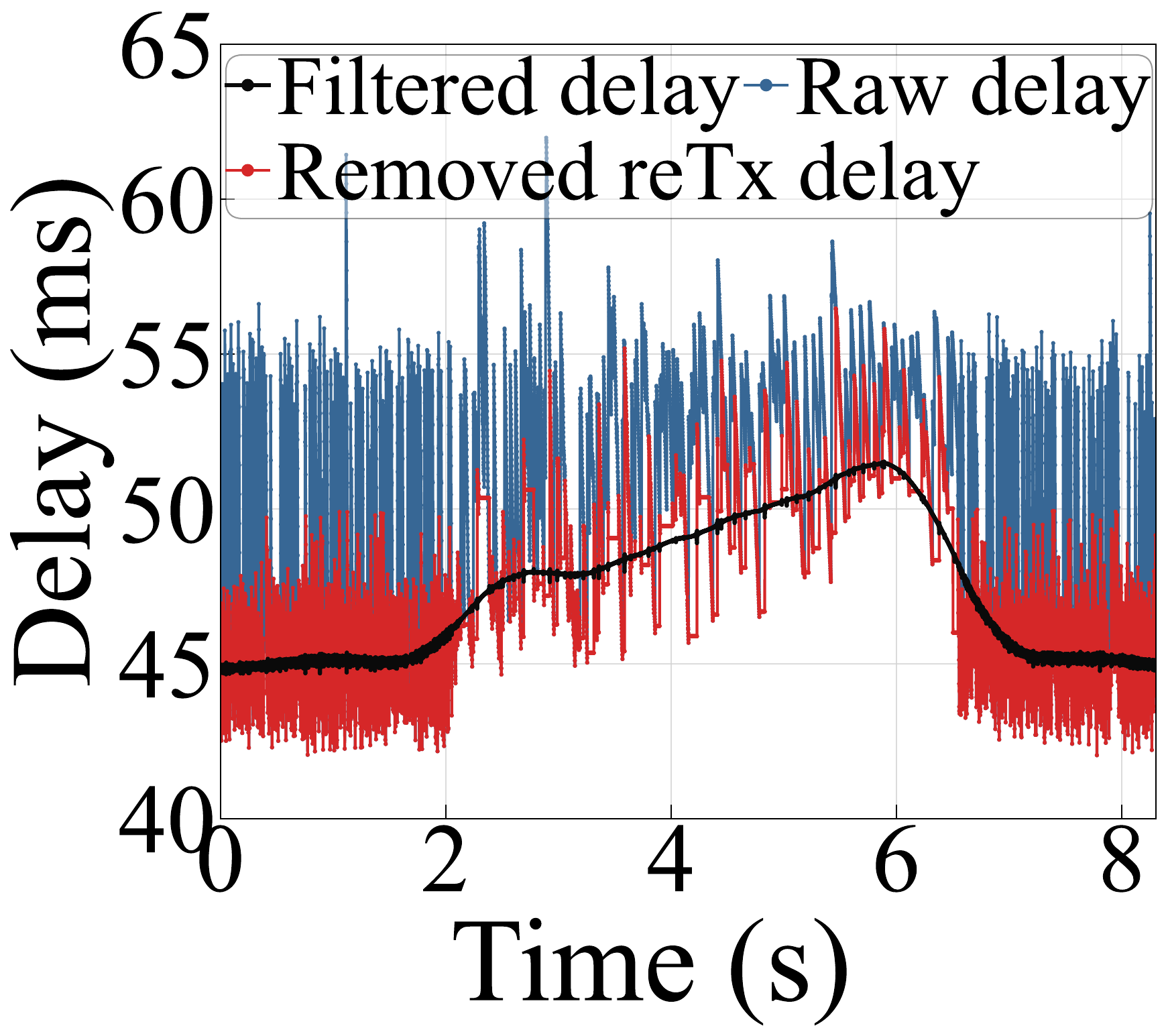}
        \caption{Comparision}
        \label{fig:filter_3type_cmp}
    \end{subfigure}
    \hfill
    \begin{subfigure}[b]{0.34\linewidth}
        \centering
        \includegraphics[width=\textwidth]{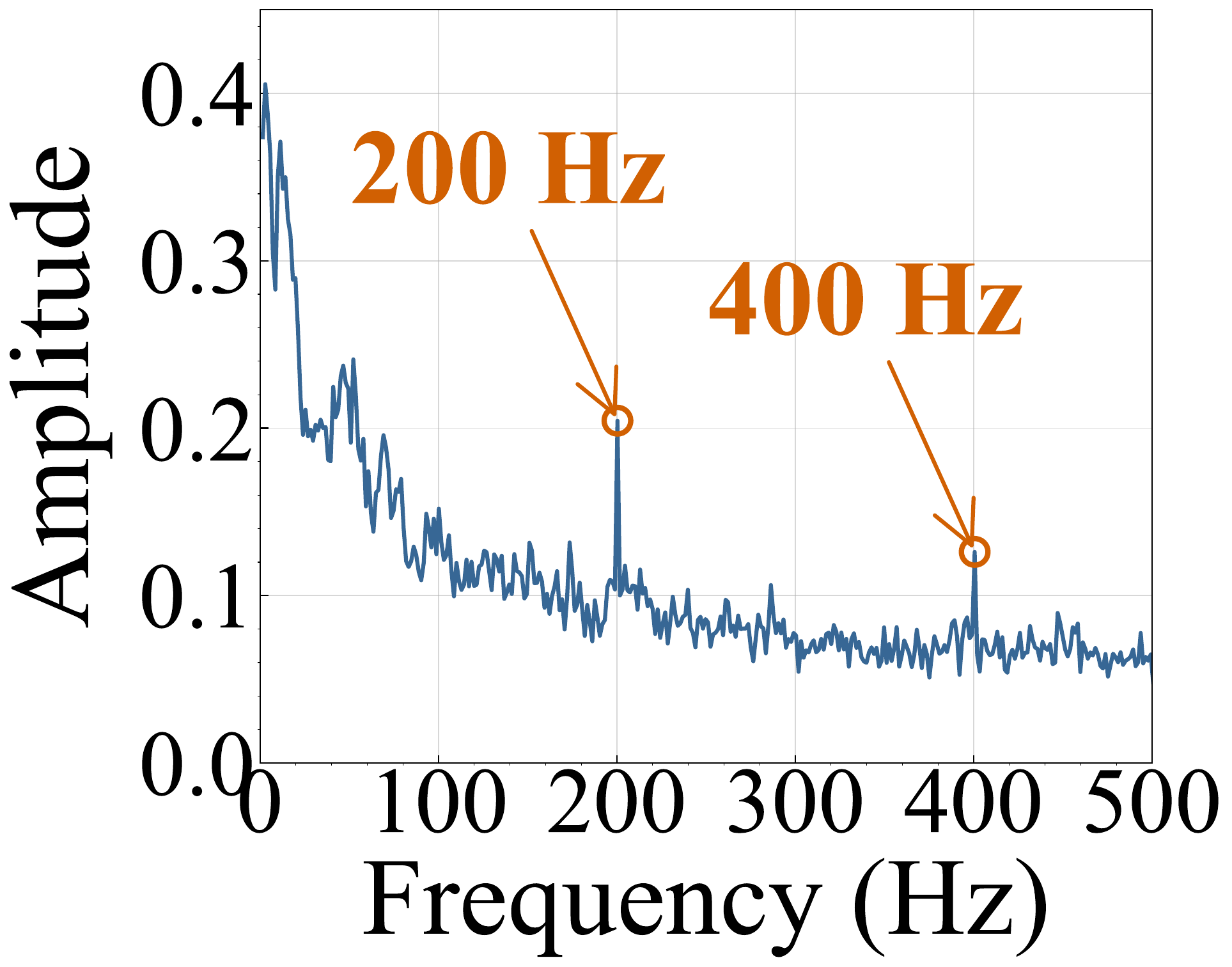}
        \caption{Spectrum after FFT}
        \label{fig:filter_lomb_spectrum_3cmp}
    \end{subfigure}
    \caption{\textbf{Processing steps to isolate network delay:} 
    (a) raw network delay measurements, 
    (b) comparison of raw delay, retransmission-removed delay, and filtered delay, 
    and (c) frequency spectrum after FFT showing RAN-specific components at 200Hz and 400Hz.}
    \label{fig:delay_isolation}
\end{figure}

\subsubsection{Delay Filtering in Practice}
To demonstrate how \sys removes RAN-induced delays in practice, 
we show the step-by-step delay filtering process in Figure~\ref{fig:delay_isolation}.
Figure~\ref{fig:filter_raw_delay} presents the raw network delay measurements.
In Figure~\ref{fig:filter_3type_cmp}, we show three delay signals: the raw delay (blue), 
the delay after removing retransmission effects (red), and the final filtered delay (black) 
that represents true network congestion. 
The difference between raw and retransmission-removed delays reveals the impact of sporadic RAN retransmissions. 
Finally, Figure~\ref{fig:filter_lomb_spectrum_3cmp} displays the frequency spectrum of the delay signal after FFT analysis, where we observe two distinct peaks at 200Hz and 400Hz corresponding to periodic RAN buffer behaviors. By identifying and removing these RAN-specific frequency components, \sys successfully extracts the underlying network delay that accurately reflects actual network conditions.

\section{\sys for Enhanced Protocol and App}
We demonstrate how \sys's delay compensation mechanisms improve performance
of three delay-sensitive applications:
Copa~\cite{arun_copa_2018} and PCC~\cite{dong_pcc_2018} congestion control algorithms that rely on delay measurements,
and WebRTC~\cite{blum_webrtc_2021} real-time communication that requires accurate delay estimation.

\subsection{Congestion Control: COPA}
\label{s:copa_impact}

COPA~\cite{arun_copa_2018} aims at simultaneously maximizing throughput 
and minimizing the delay.
COPA theoretically proves that such a goal can be achieved by setting the 
sending rate $R_{s}$ to:
\begin{equation}
   R_s = 1/d_q
   \label{eqn:copa_rate}
\end{equation}
where $d_q$ is the mean per-packet queuing delay.

\begin{figure}
    \includegraphics[width=0.98\linewidth]{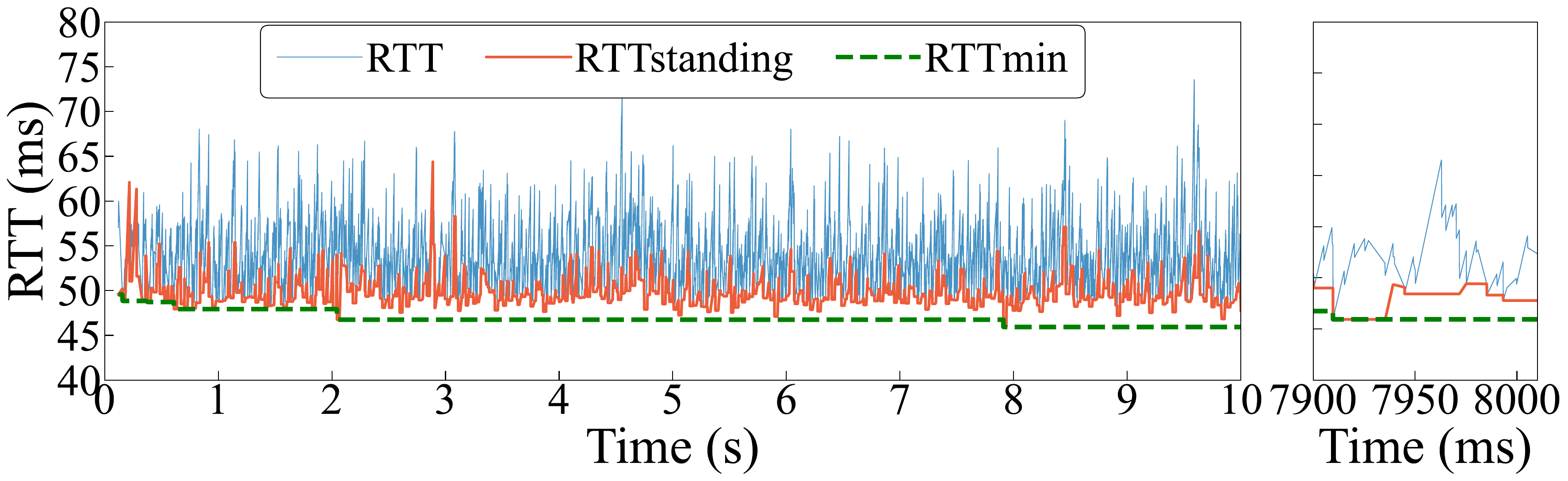}
    \caption{The workflow of COPA's queuing delay estimation. 
    COPA estimates the queuing delay by calculating the difference 
    between $\mathrm{RTT}_{standing}$ and $\mathrm{RTT}_{min}$.}
    \label{fig:rtt_standing}
\end{figure}

\noindent
\textbf{Misinterpreting RAN Queuing.} 
COPA's queuing delay measurement captures both network buffer queuing and
RAN-internal buffer queuing indiscriminately, 
treating both of them as signals of congestion.
To estimate queuing delay, COPA uses: 
\begin{equation}
    d_q = \mathrm{RTT}_{\mathrm{standing}} - \mathrm{RTT}_{\mathrm{min}} 
\end{equation}
where $\mathrm{RTT}_{\mathrm{min}}$ and $\mathrm{RTT}_{\mathrm{standing}}$ are the 
minimal RTT observed within a longer (10 seconds) and shorter (one RTT) window, respectively. 
In cellular networks, $\mathrm{RTT}_{\mathrm{min}}$ typically captures the true minimum delay
during rare moments when uplink buffers are empty or near empty (BSR approaching zero),
as we describe in \S\ref{sec:buffer_clear},
while $\mathrm{RTT}_{\mathrm{standing}}$
reflects delays with normal uplink buffer occupancy,
which is typically non-empty as we demonstrate in \S\ref{sec:buffer_acu}.
Figure~\ref{fig:rtt_standing} shows an example where COPA underutilizes the channel at 8Mbps:
$\mathrm{RTT}_{\mathrm{standing}}$ (red line) consistently stays above $\mathrm{RTT}_{\mathrm{min}}$ (green dashed line),
causing COPA to persistently estimate heavy queuing in the network, despite the absence of actual network congestion.

\paragraph{Misinterpreted Queuing Triggers Rate Reduction.} 
This misinterpretation of RAN queuing directly triggers COPA's congestion control response. 
When COPA mistakes normal RAN buffer occupancy for network congestion, 
it unnecessarily reduces its sending rate to clear what it perceives as harmful queues,
according to Eqn~\ref{eqn:copa_rate}.
However, since RAN buffers maintains a certain occupancy level due to normal cellular operation, 
these rate reductions are not only unnecessary but counterproductive. 
Furthermore, since RAN buffers are persistently present by design, 
COPA constantly sees congestion signals and remains stuck at low data rates, 
severely underutilizing the available network capacity,
just as evident in Figure~\ref{fig:rtt_standing}.

\begin{figure*}
    \begin{minipage}[t]{0.77\linewidth}
        \centering
        \begin{subfigure}[b]{0.24\linewidth}
            \centering
            \includegraphics[width=\textwidth]{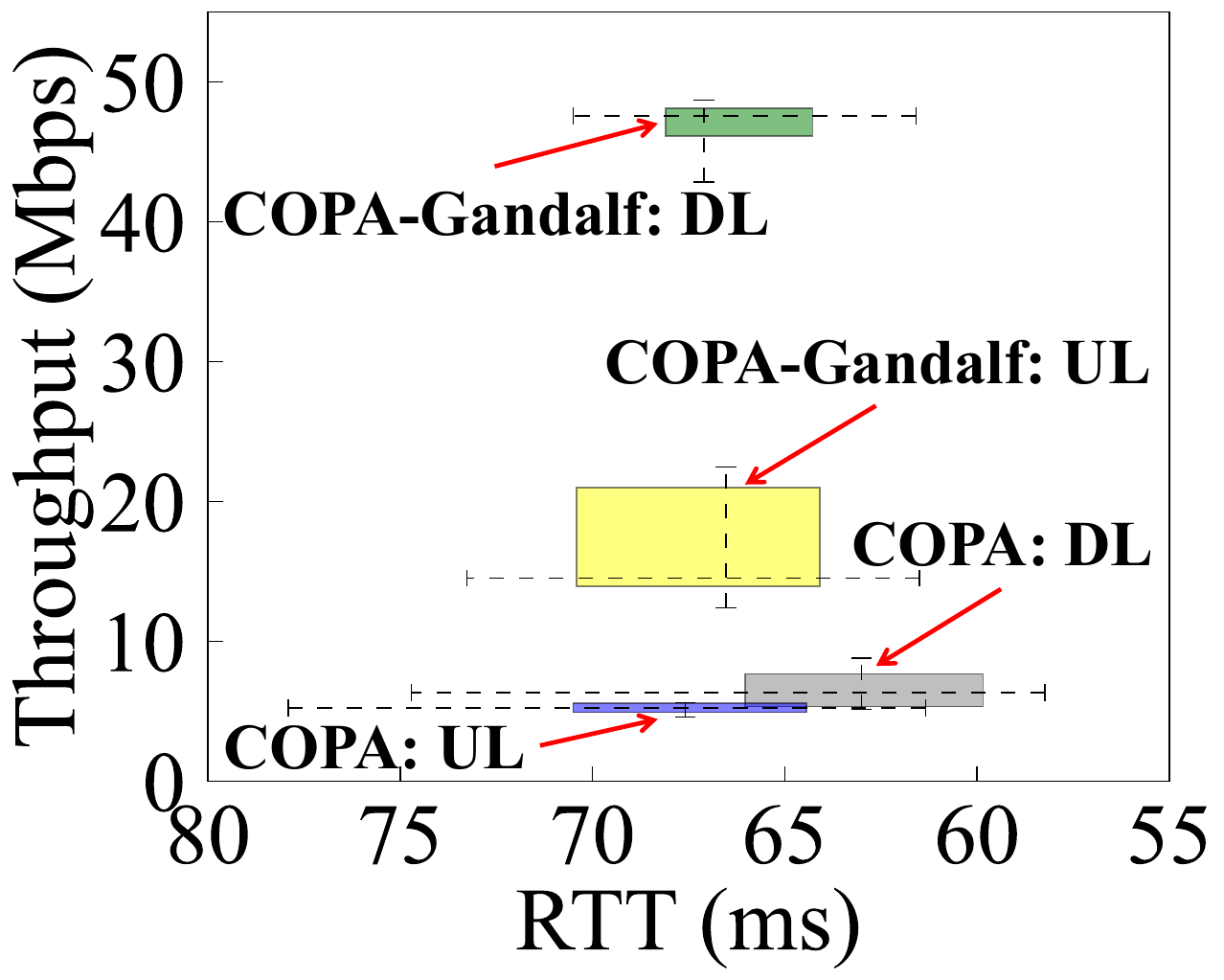}
            \caption{5G SA stationary link}
            \label{fig:copa_5g}
        \end{subfigure}
        \hfill
        \begin{subfigure}[b]{0.24\linewidth}
            \centering
            \includegraphics[width=\textwidth]{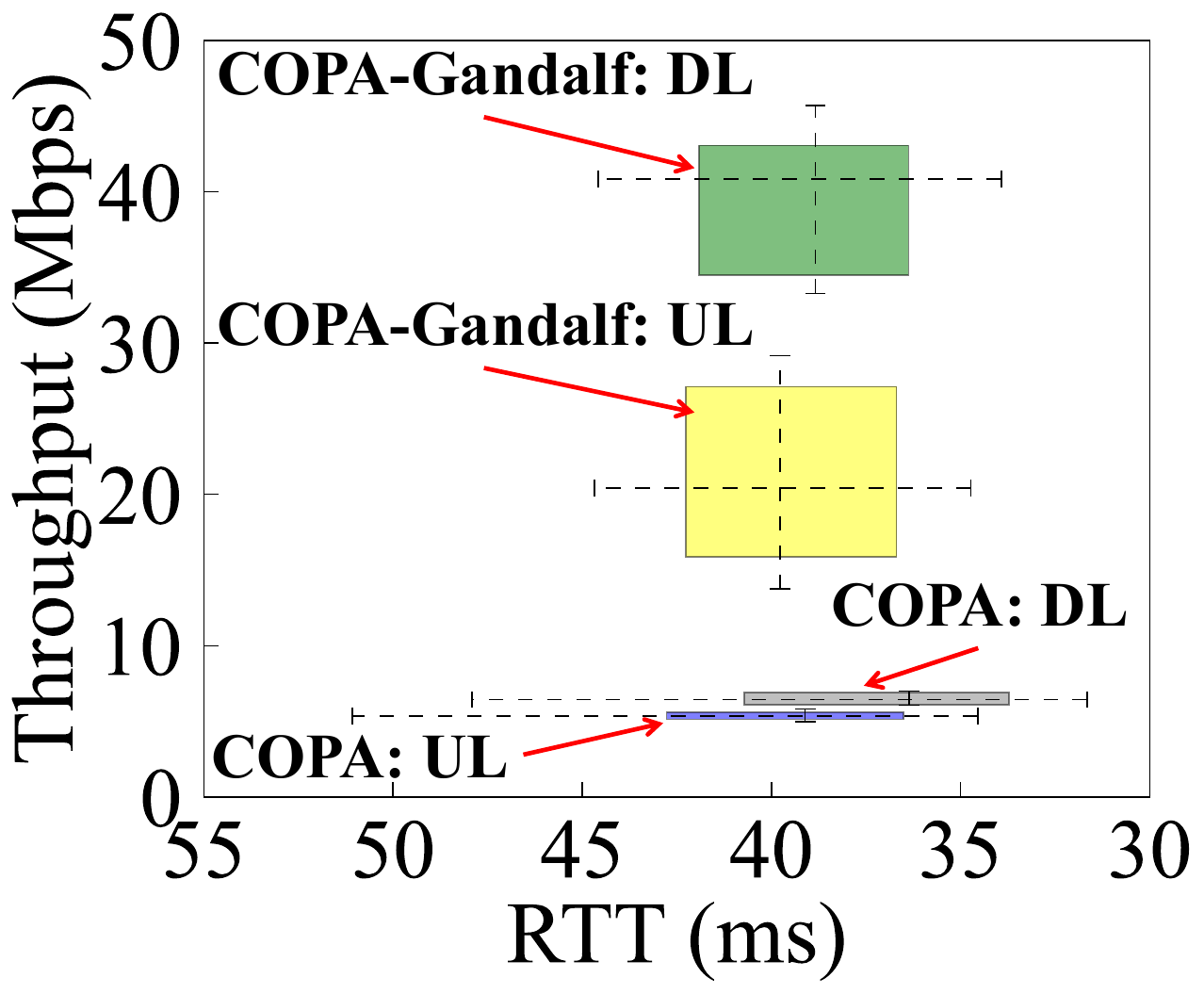}
            \caption{LTE stationary link}
            \label{fig:copa_lte}
        \end{subfigure}
        \hfill
        \begin{subfigure}[b]{0.24\linewidth}
            \centering
            \includegraphics[width=\textwidth]{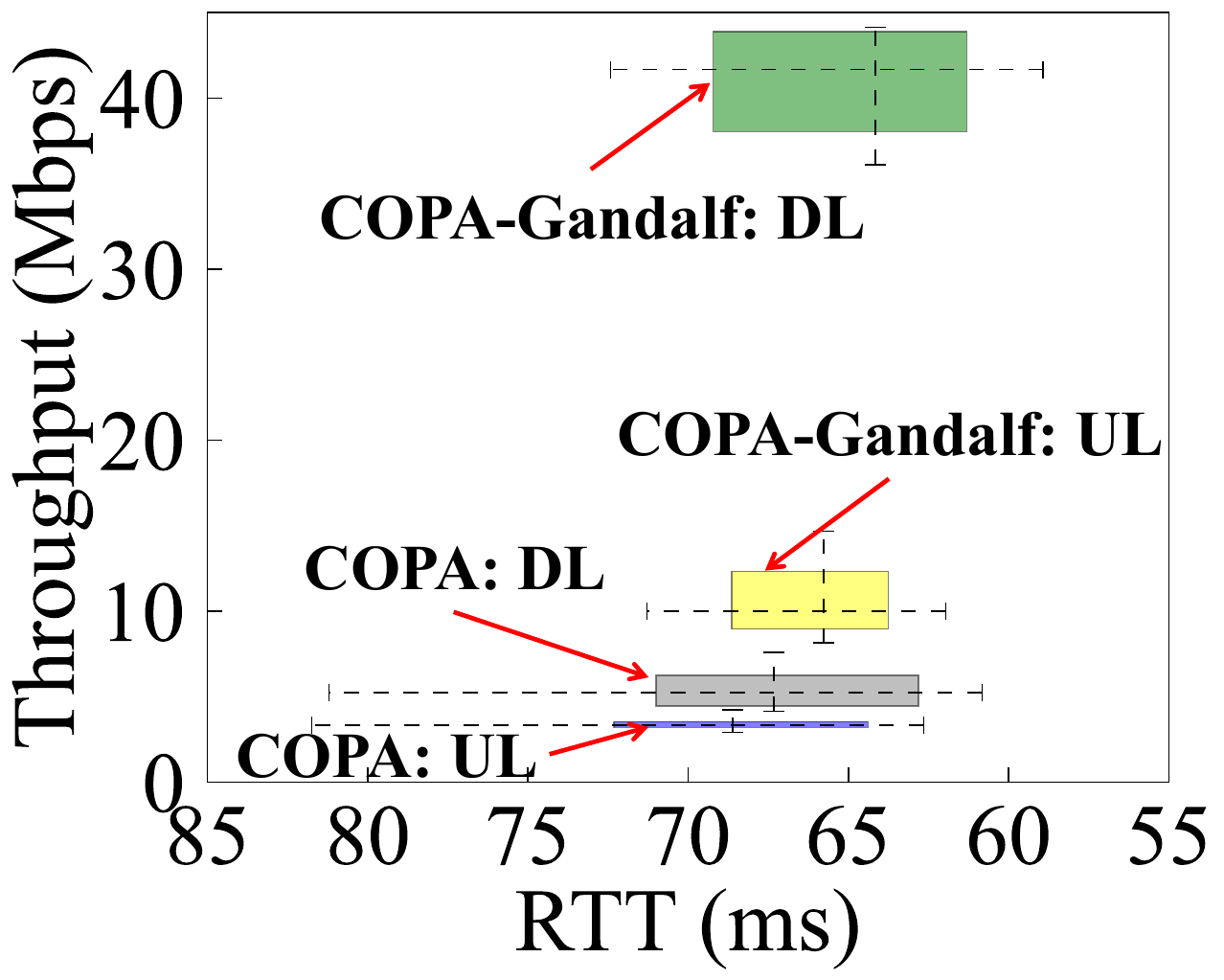}
            \caption{5G SA under mobility}
            \label{fig:copa_5g_mobile}
        \end{subfigure}
        \hfill
        \begin{subfigure}[b]{0.24\linewidth}
            \centering
            \includegraphics[width=\textwidth]{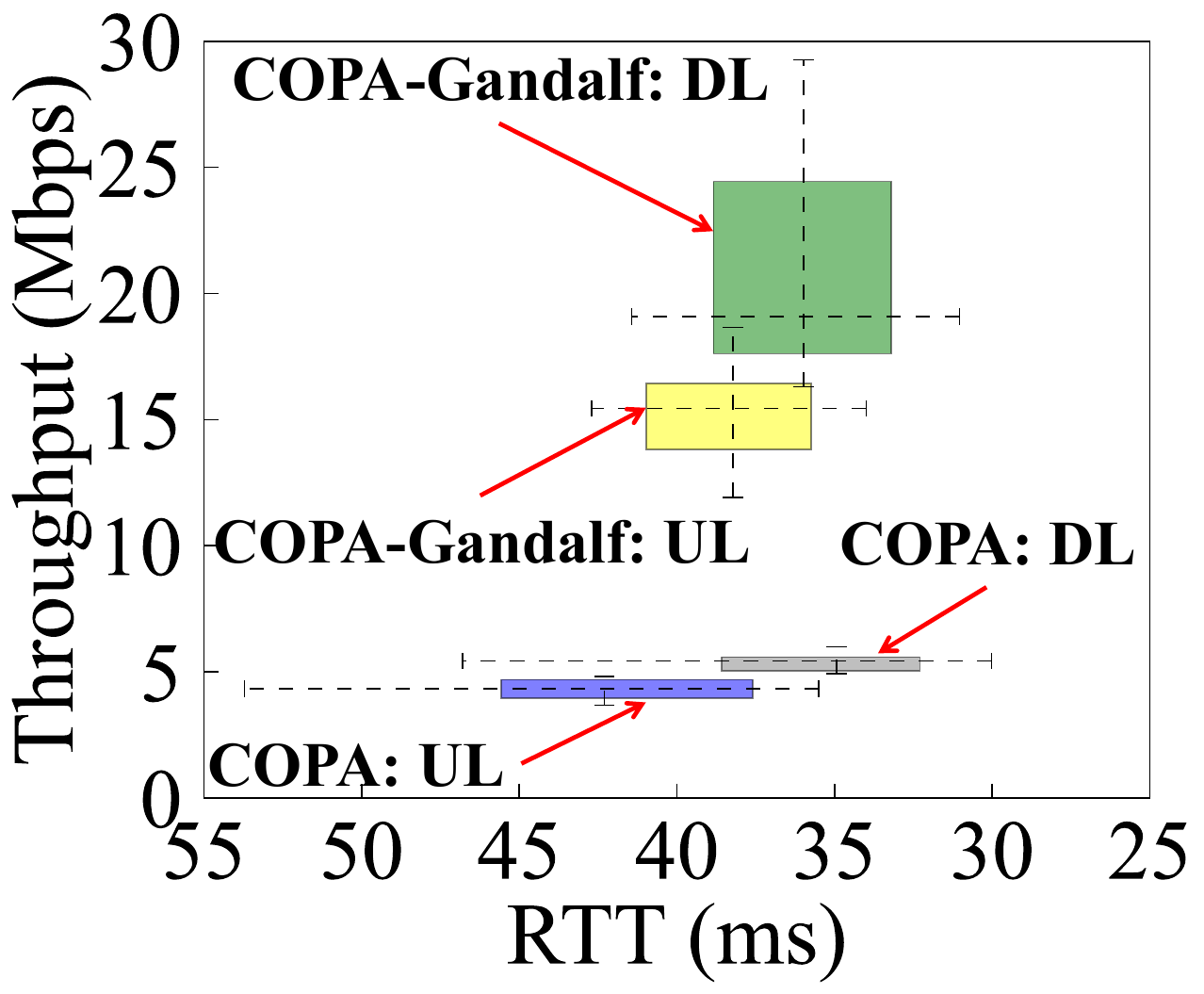}
            \caption{LTE under mobility}
            \label{fig:copa_lte_mobile}
        \end{subfigure}
        \caption{Performance comparison between COPA and COPA-\sys across different scenarios: static user in 5G SA~(\textbf{a}) and
        LTE~(\textbf{b}), mobile user in 5G SA~(\textbf{c}) and LTE~(\textbf{d}),
        with both uplink (UL) and downlink (DL) results shown in each subplot.
        Each box plot shows throughput vs. RTT, where box edges represent 25th and 75th percentiles, 
        and whiskers extend to 10th and 90th percentiles. }
        \label{fig:copa_perf}          
    \end{minipage}
    \hfill
    \begin{minipage}[t]{0.22\linewidth}
        \centering 
        \includegraphics[width=0.98\textwidth]{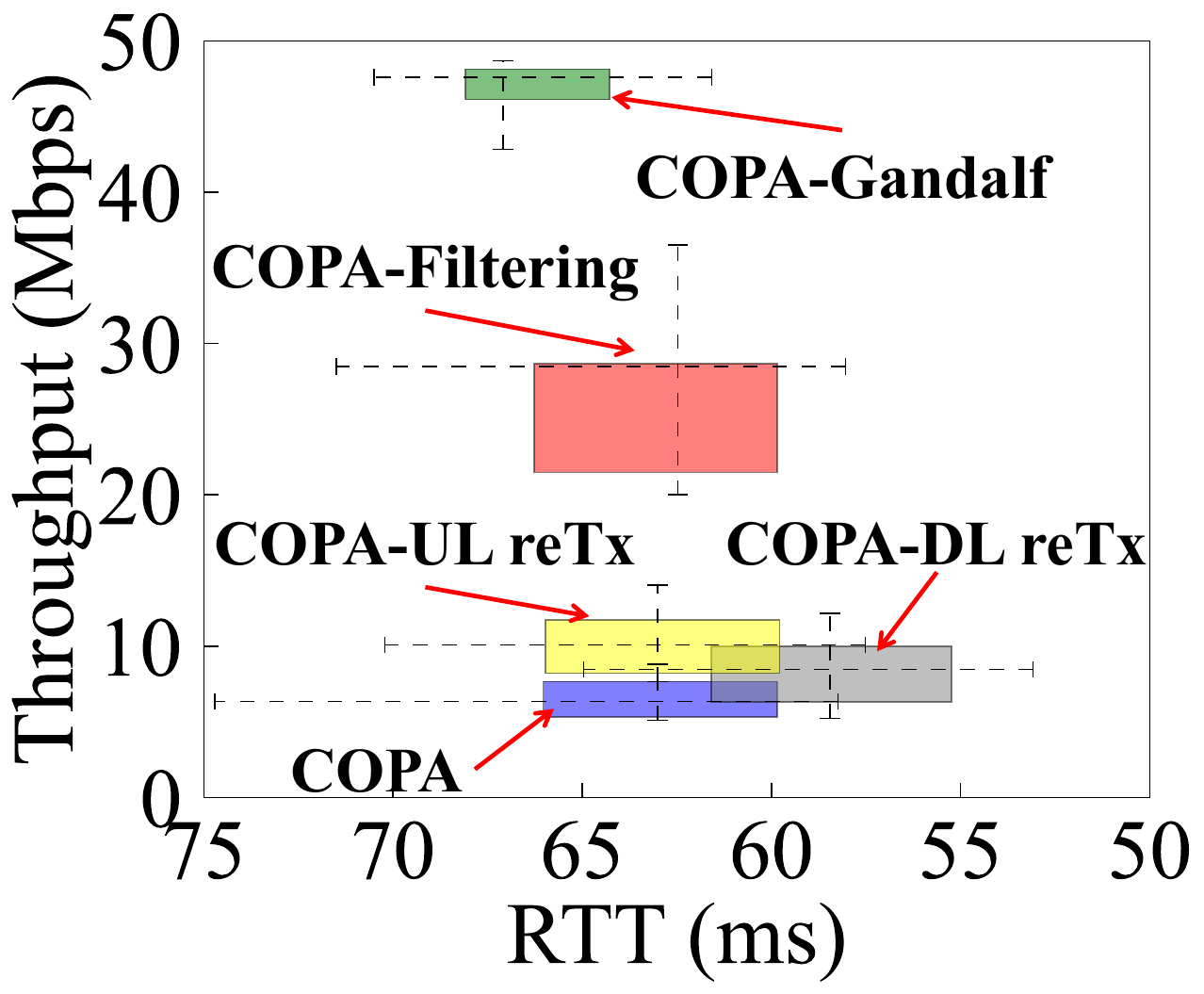}
        \caption{Ablation study comparing performance of COPA variants with different \sys components.}
        \label{fig:copa_5g_compare}
    \end{minipage}
\end{figure*}
\subsubsection{\sys-enhanced COPA}\label{sec:copa_eval}
Since COPA's performance degradation stems from 
misinterpreting RAN-internal queuing as network congestion, 
we can improve its performance by using \sys to remove RAN-induced delays from its congestion signals. 
\sys allows COPA to focus solely on actual network queuing, 
enabling it to maintain appropriately high sending rates 
while still responding to true network congestion.
We denote the new system as COPA-\sys.

\paragraph{Experimental Methodology.}
To evaluate the performance in a realistic scenario, 
we adopted a common network usage configuration: 
a mobile client (Xiaomi 10 Lite) communicating with an AWS server~\cite{rochman_comprehensive_2024, narayanan_first_2020}. 
For each test, we established a COPA connection between the AWS server 
and the mobile client for 30 seconds. 
We repeated each test five times to account for potential network fluctuations.
We also investigate both uplink and downlink performances in each test
and conduct experiments inside both LTE and 5G SA networks.
We test the performance with both static and mobile users.
We plot the results in Figure~\ref{fig:copa_perf}.

\paragraph{Performance.}
COPA-\sys dramatically outperforms standard COPA across all tested scenarios
while maintaining similar or lower RTT.
In stationary scenarios,
COPA-\sys improves throughput by 7.49$\times$ for downlink and 2.77$\times$ for uplink in 5G SA,
and achieves gains in LTE with 6.34$\times$ downlink and 3.82$\times$ uplink improvement.
These significant performance gains persist under mobility:
COPA-\sys maintains throughput improvements of 7.95$\times$ downlink and 3.01$\times$ uplink in 5G SA,
and 3.51$\times$ downlink and 3.56$\times$ uplink in LTE networks.
These substantial improvements demonstrate that
by removing RAN-induced delays,
COPA can better utilize available network capacity
without sacrificing latency performance.

\paragraph{Ablation Study.}
To understand which RAN delays most significantly impact COPA's performance,
we evaluate COPA working with three versions of \sys:
\sys-DL-reTx (downlink retransmission compensation only),
\sys-UL-reTx (uplink retransmission compensation only),
and \sys-filter (RAN-aware filtering only)
in 5G SA networks.
Figure~\ref{fig:copa_5g_compare} shows that RAN-aware filtering
provides the most significant improvement,
with COPA working with \sys-filter achieving performance close to full COPA-\sys.
This result aligns with our analysis:
COPA's queuing delay estimation primarily captures
the persistent packet queuing in the uplink buffer.
By filtering out this RAN-induced component,
we effectively improve COPA's ability to detect true network congestion,
instead of buffering inside RAN.

\subsection{Congestion Control: PCC Vivace}

\begin{figure}[t]
    \centering
    \begin{subfigure}[b]{0.41\linewidth}
        \centering
        \includegraphics[width=0.99\textwidth]{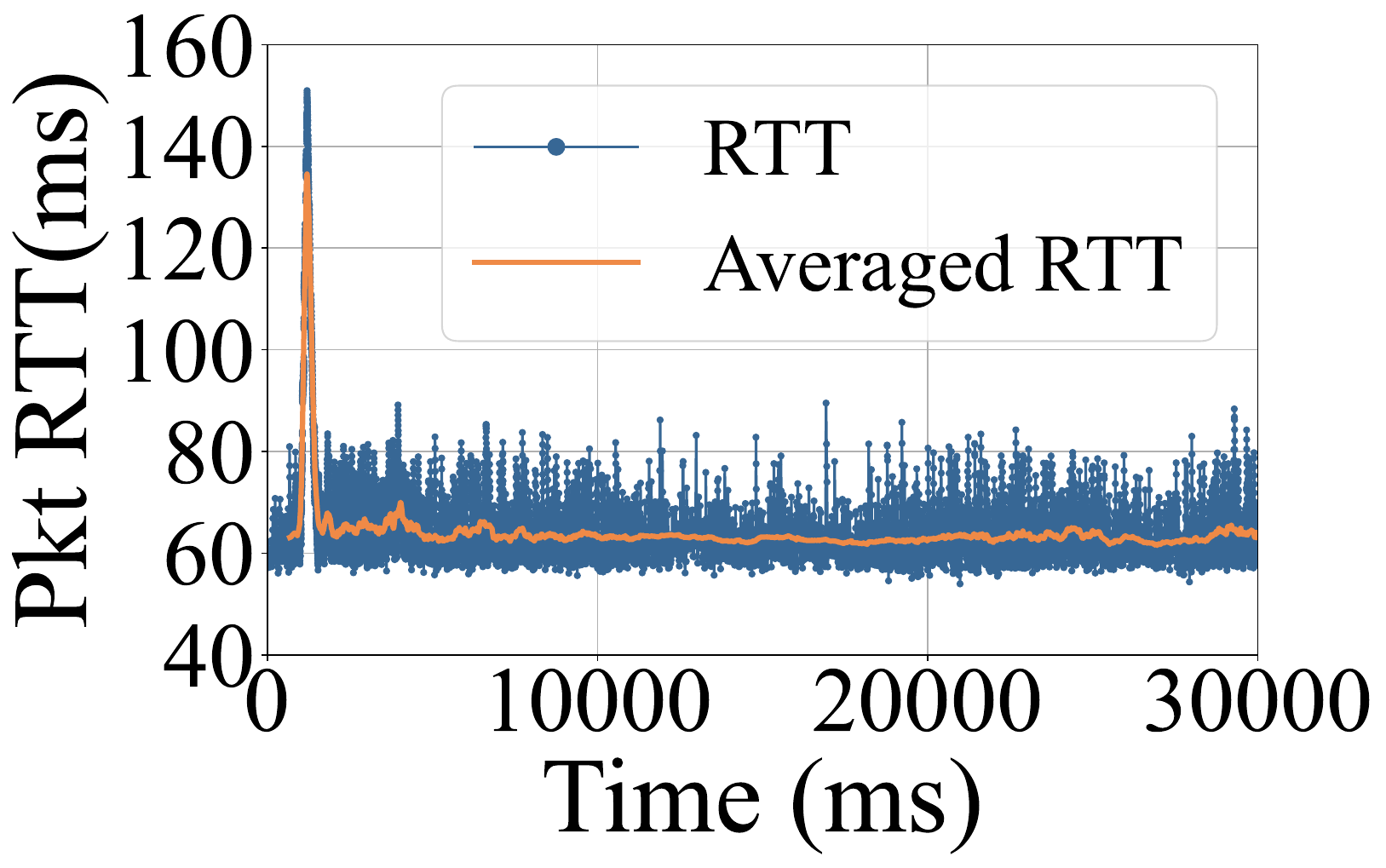}
        \caption{The measured RTT}
        \label{fig:pcc_analysis_rtt}
    \end{subfigure}
    \hfill
    \begin{subfigure}[b]{0.58\linewidth}
        \centering
        \includegraphics[width=0.99\textwidth]{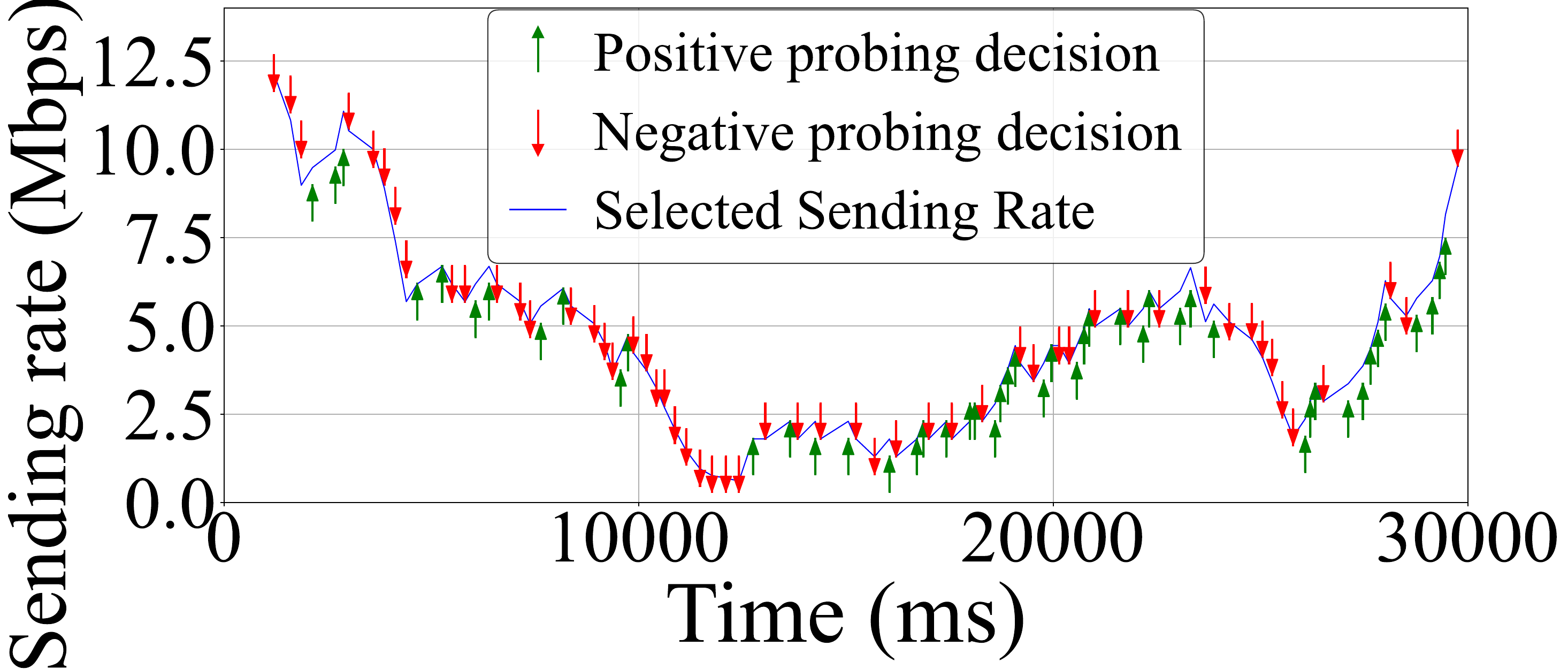}
        \caption{Rate probing process of PCC}
        \label{fig:pcc_analysis_rate}
    \end{subfigure}
    \caption{PCC misinterprets RAN-induced RTT variations as network congestion: 
    despite stable network conditions shown in (\textbf{a}), 
    PCC makes erroneous probing decisions in (\textbf{b}), resulting in unnecessarily low sending rates.}
    \label{fig:pcc_probe}
\end{figure}

\begin{figure*}
    \begin{minipage}[t]{0.76\linewidth}
        \centering
        \begin{subfigure}[b]{0.24\linewidth}
            \centering
            \includegraphics[width=\textwidth]{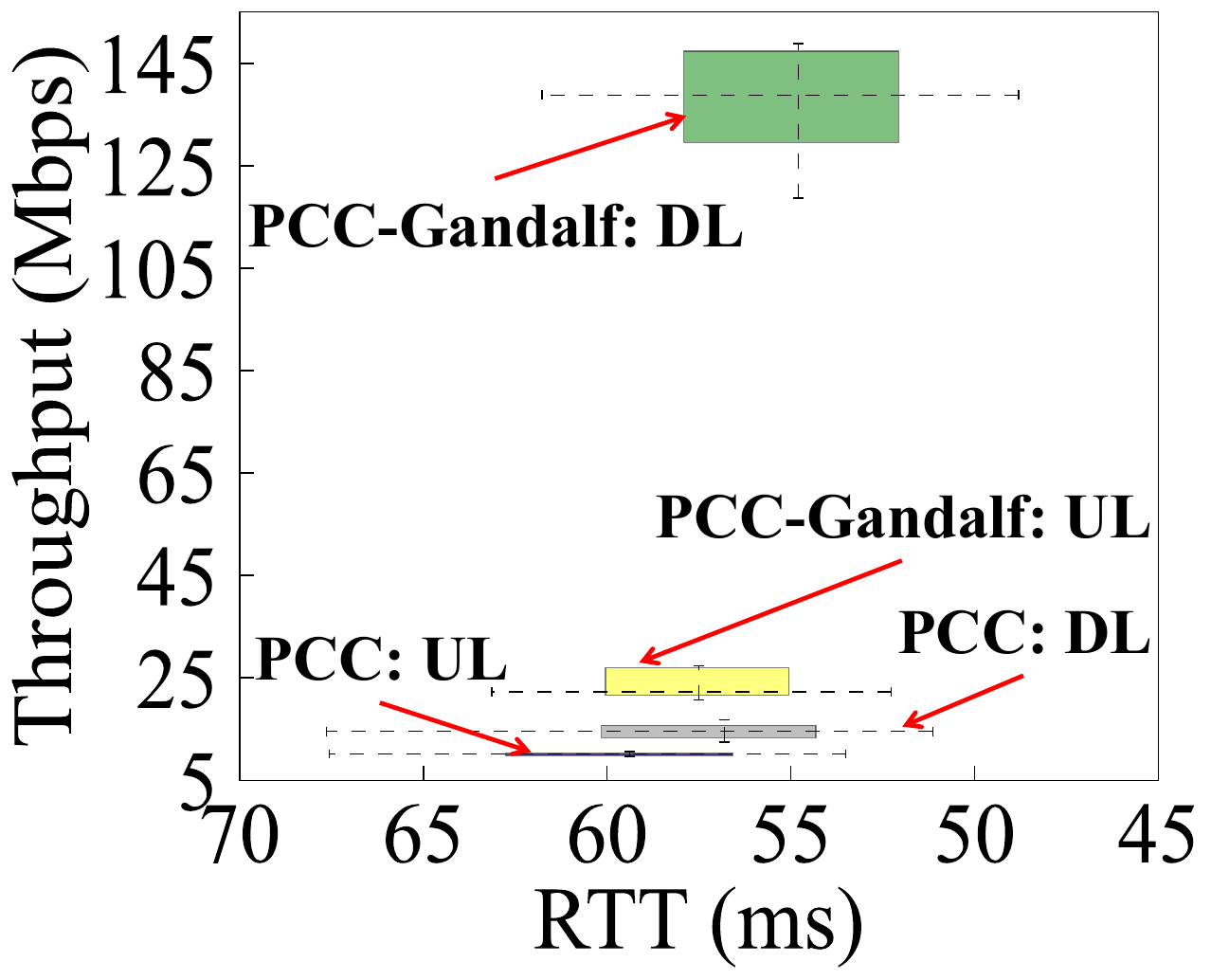}
            \caption{5G SA stationary link}
            \label{fig:pcc_5g}
        \end{subfigure}
        \hfill
        \begin{subfigure}[b]{0.24\linewidth}
            \centering
            \includegraphics[width=\textwidth]{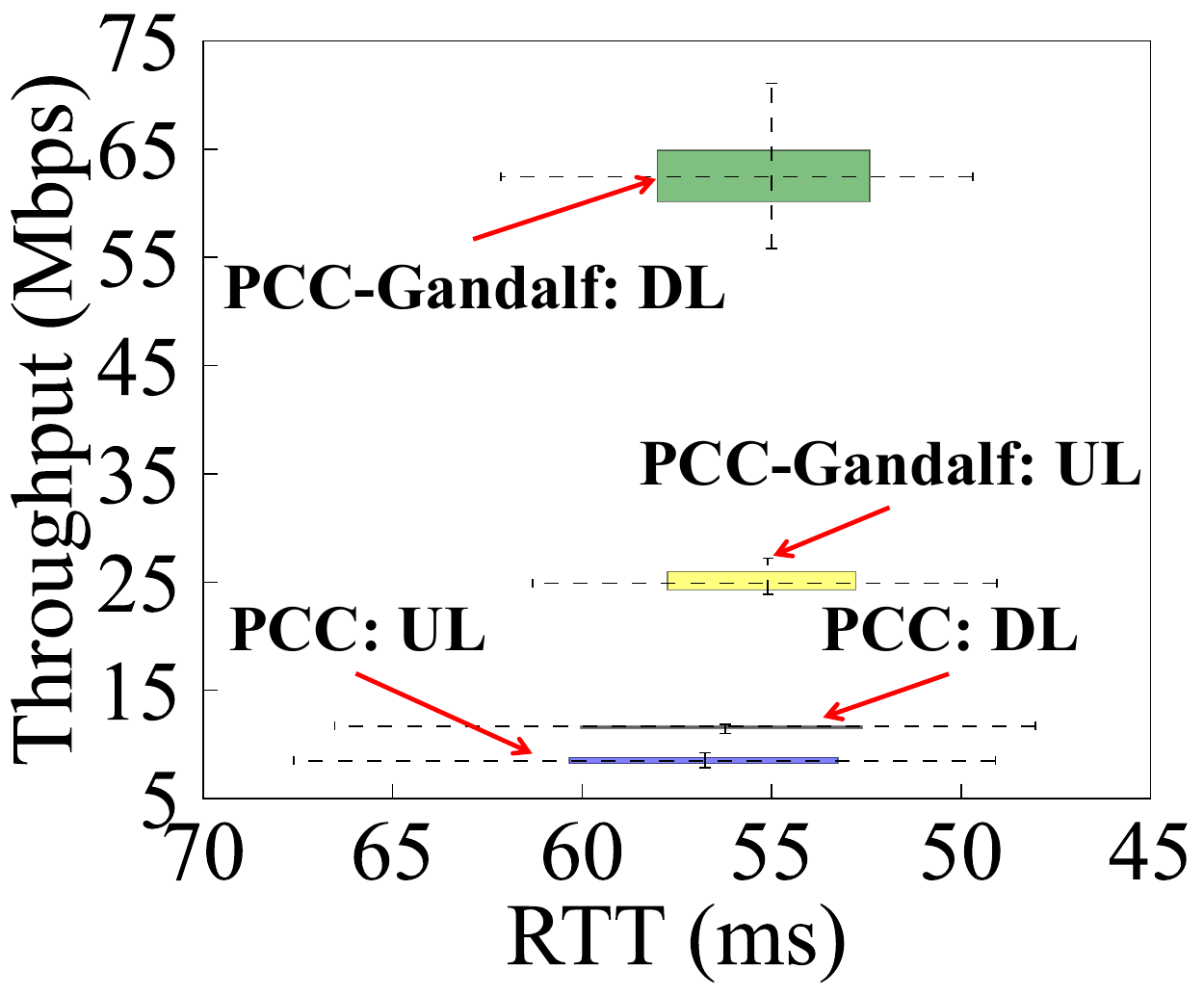}
            \caption{LTE stationary link}
            \label{fig:pcc_lte}
        \end{subfigure}
        \hfill
        \begin{subfigure}[b]{0.24\linewidth}
            \centering
            \includegraphics[width=\textwidth]{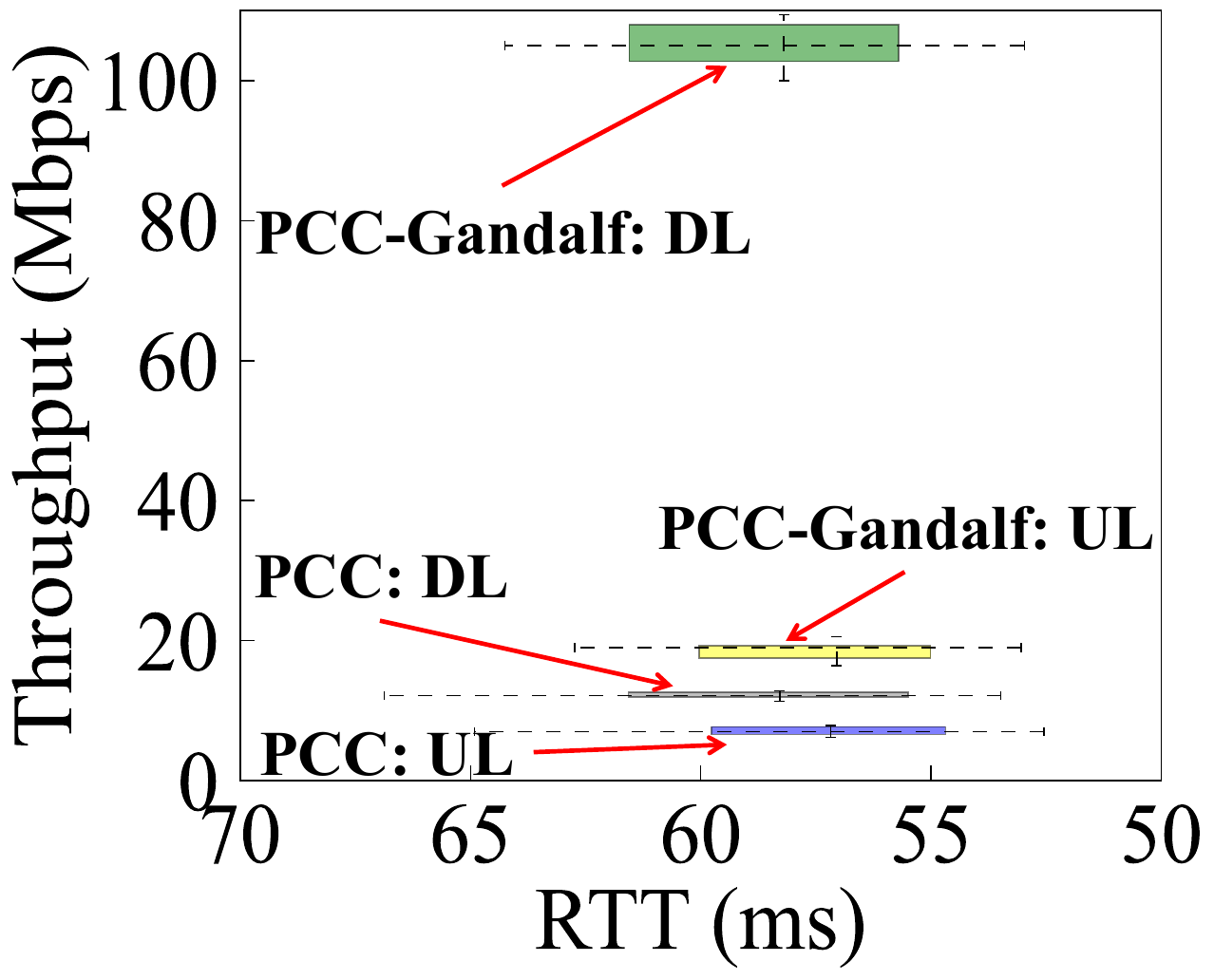}
            \caption{5G SA under mobility}
            \label{fig:pcc_5g_mobile}
        \end{subfigure}
        \hfill
        \begin{subfigure}[b]{0.239\linewidth}
            \centering
            \includegraphics[width=\textwidth]{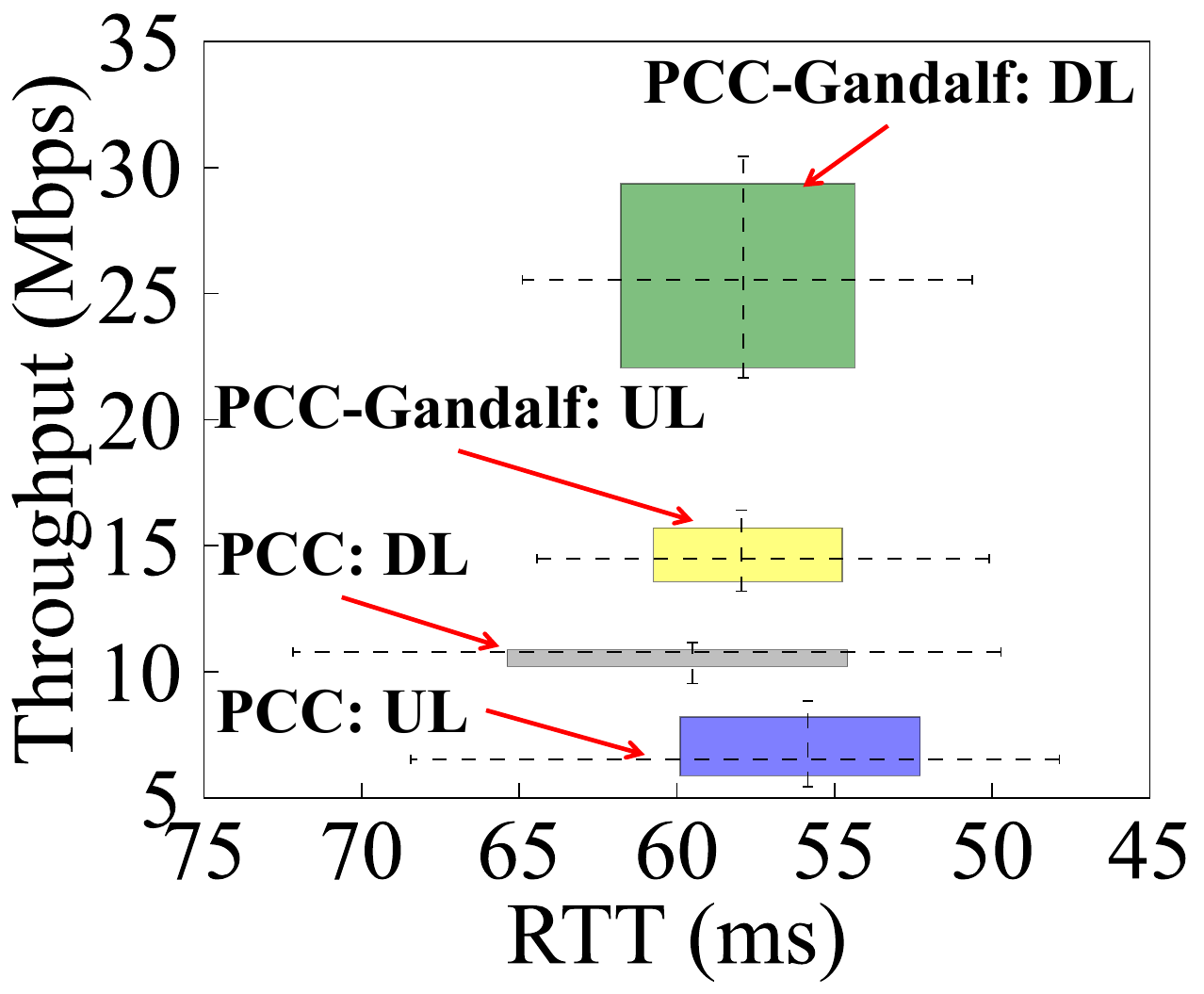}
            \caption{LTE under mobility}
            \label{fig:pcc_lte_mobile}
        \end{subfigure}
        \caption{Performance comparison between PCC and PCC-\sys across different scenarios: static user in 5G SA~(\textbf{a}) and
        LTE~(\textbf{b}), mobile user in 5G SA~(\textbf{c}) and LTE~(\textbf{d}),
        with both uplink (UL) and downlink (DL) results shown in each subplot.}
        \label{fig:pcc_perf}          
    \end{minipage}
    \hfill
    \begin{minipage}[t]{0.225\linewidth}
        \centering 
        \includegraphics[width=0.98\textwidth]{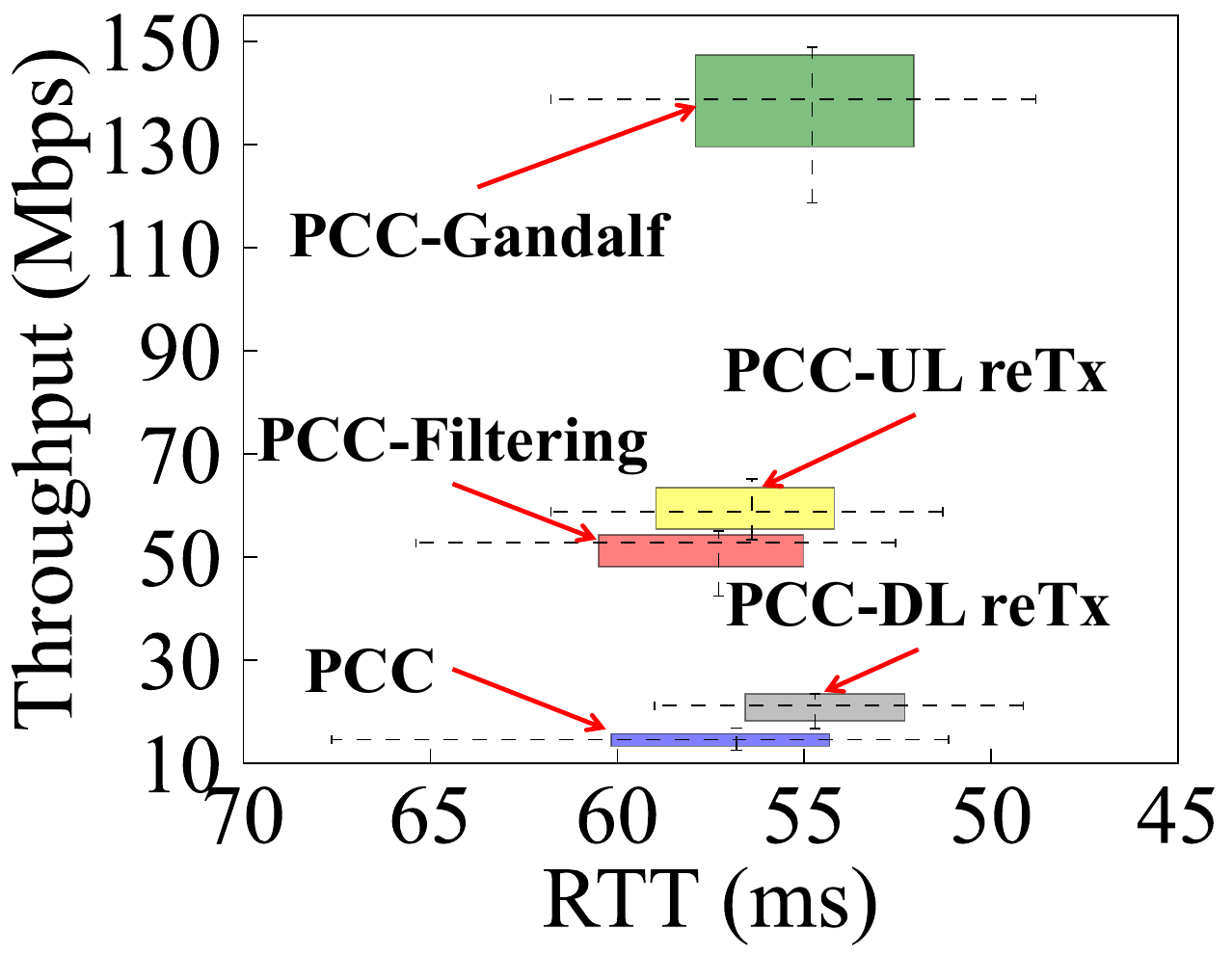}
        \caption{Ablation study comparing performance of different PCC variants.}
        \label{fig:pcc_ablation}
    \end{minipage}
\end{figure*}
\paragraph{PCC's Rate Probing Mechanism.}
PCC~\cite{dong_pcc_2018} continuously probes network capacity by creating \textit{monitor intervals} (MIs)
to test different sending rates.
During these probing phases, PCC tests two rates:
one slightly higher and one slightly lower 
than the current rate.
To determine whether to increase or decrease its rate,
PCC primarily examines RTT gradient changes during these tests,
along with throughput and packet loss measurements.
This approach aims to find sending rates that maintain high throughput
without causing excessive queuing delays.
Since RTT gradient serves as a key indicator of potential congestion,
accurate delay measurements are crucial for PCC's rate probing effectiveness.

\paragraph{RAN-induced RTT Gradient Errors.}
In cellular networks, RTT measurements inherently include
multiple sources of RAN-induced delays:
uplink buffering delay and retransmission delays in both uplink and downlink paths.
As our analysis in previous sections shows,
these delays collectively introduce substantial RTT variations,
often 40-50ms even in uncongested networks.
Such significant delay variations fundamentally disrupt PCC's RTT gradient calculations.
When probing with higher rates, PCC observes RTT gradient changes
that appear to signal congestion, but actually stem from
normal RAN buffering and retransmission behaviors.
Unable to distinguish between RAN-induced variations and true congestion,
PCC frequently reduces its sending rate unnecessarily,
leading to suboptimal performance despite available network capacity.

% \begin{figure}[t]
%     \centering
%     \begin{subfigure}[b]{0.41\linewidth}
%         \centering
%         \includegraphics[width=0.99\textwidth]{figure/pcc_analysis_rtt.pdf}
%         \caption{The measured RTT.}
%         \label{fig:pcc_analysis_rtt}
%     \end{subfigure}
%     \hfill
%     \begin{subfigure}[b]{0.58\linewidth}
%         \centering
%         \includegraphics[width=0.99\textwidth]{figure/pcc_analysis_rate.pdf}
%         \caption{Rate probing process of PCC.}
%         \label{fig:pcc_analysis_rate}
%     \end{subfigure}
%     \caption{PCC misinterprets RAN-induced RTT variations as network congestion: 
%     despite stable network conditions shown in (\textbf{a}), 
%     PCC makes erroneous probing decisions in (\textbf{b}), resulting in unnecessarily low sending rates.}
%     \label{fig:pcc_probe}
% \end{figure}

\noindent
\textit{Experimental Demonstration.}
We experimentally demonstrate this behavior by having a UE upload data
to an AWS server using PCC.
%Figure~\ref{fig:pcc_probe} shows the results.
%the left subplot presents PCC's rate selections and probing decisions,
%while the right subplot shows the measured RTT values.
The RTT measurements in Figure~\ref{fig:pcc_analysis_rtt} show only RAN-induced variations 
without any significant network queuing,
indicating an uncongested network.
However, PCC repeatedly misinterprets these RAN-induced RTT variations
as congestion signals during its rate probing.
As shown in Figure~\ref{fig:pcc_analysis_rate}, 
despite available network capacity,
PCC's sending rate remains stuck at low levels (below 7.5 Mbit/s)
due to these false congestion signals.
Notably, while PCC employs low-pass filtering on RTT measurements
this filtering fails to effectively remove RAN-induced variations
as it lacks the RAN internal information needed to distinguish
between RAN delays and actual network congestion.

% \begin{figure}[t]
%     \setlength{\abovecaptionskip}{2pt}
%     \setlength{\belowcaptionskip}{-2pt}
%     \centering 
%     \includegraphics[width=1\linewidth]{figure/pcc_analysis.pdf}
%     \caption{The calculated utilization, utilization gradient, and selected sending rate of PCC Vivace is shown in~(\textbf{a});
%     the measured RTT is shown in~(\textbf{b}).}
%     \label{fig:pcc_util}
% \end{figure}
% \begin{figure}[t]
%     \setlength{\abovecaptionskip}{2pt}
%     \setlength{\belowcaptionskip}{-2pt}
%     \centering 
%     \includegraphics[width=1\linewidth]{figure/pcc_analysis.pdf}
%     \caption{The calculated utilization, utilization gradient, and selected sending rate of PCC Vivace is shown in~(\textbf{a});
%     the measured RTT is shown in~(\textbf{b}).}
%     \label{fig:pcc_util}
% \end{figure}

\subsubsection{\sys-enhanced PCC}
Since PCC's performance degradation stems from
misinterpreting RAN-induced delay variations as network congestion,
we can improve its performance by using \sys to remove these variations from its RTT gradient calculations.
\sys allows PCC to compute RTT gradients based on true network queuing delays,
enabling it to probe network capacity effectively while still maintaining
sensitivity to actual congestion.
We denote the new system as PCC-\sys.

\paragraph{Experimental Methodology.}
Following the same experimental setup as in \S\ref{sec:copa_eval},
we evaluate PCC and PCC-\sys under both static and mobile scenarios
in LTE and 5G SA.

\paragraph{Performance.}
PCC-\sys similarly achieves substantial throughput gains across all scenarios
while maintaining comparable or lower RTT.
In stationary environments,
PCC-\sys improves throughput by 9.53$\times$ for downlink and 2.19$\times$ for uplink in 5G SA,
and shows gains in LTE with 5.34$\times$ downlink and 2.93$\times$ uplink improvement.
The performance advantages continue under mobility:
PCC-\sys maintains throughput improvements of 8.65$\times$ downlink and 2.72$\times$ uplink in 5G SA,
and 2.37$\times$ downlink and 2.22$\times$ uplink in LTE networks.
These consistent improvements demonstrate that
by providing accurate delay signals without RAN-induced variations,
PCC can better optimize its sending rates
while maintaining low latency.

\paragraph{Ablation Study.}
To understand which RAN delays most significantly impact PCC's performance,
we evaluate PCC working with three versions of \sys:
\sys-DL-reTx, \sys-UL-reTx, and \sys-filter in 5G SA networks.
Figure~\ref{fig:pcc_ablation} shows that all delay compensation mechanisms
contribute significantly to performance improvement,
except for downlink retransmission compensation in this specific experiment
where the downlink channel quality happened to be better than uplink,
resulting in fewer retransmission events.
This result aligns with our analysis:
PCC relies heavily on RTT gradient measurements for rate probing,
making it sensitive to all sources of delay variation.
Both retransmission delays and uplink buffering create RTT variations
that disrupt PCC's gradient calculations.
By removing these RAN-induced variations,
each component of \sys helps PCC better estimate true network conditions,
leading to more accurate rate adjustments.

\subsection{Real-Time Communication: WebRTC}
\paragraph{WebRTC Framework.}
WebRTC~\cite{blum_webrtc_2021} is an open-source project defining 
the standard for real-time video communication over the internet. 
It implements a Google Congestion Control (GCC)~\cite{carlucci_analysis_2016} 
to estimate the available network bandwidth,
which is fed back to the video encoder within the WebRTC framework.
The encoder dynamically adjusts the encoded video frame rate and frame resolution (both width and height)
to ensure that the encoded video bitrate 
aligns with the network's capacity.

%firstly, the encoded video frame rate; 
%secondly, the frame resolution (both width and height).
%and thirdly, the quantization parameter, which represents 
%the balance between video quality and compression level. 
%Lower quantization parameters yield higher quality but result in larger file sizes.
\begin{figure*}[htb]
    \setlength{\abovecaptionskip}{2pt}
    \setlength{\belowcaptionskip}{-2pt}
     
    \begin{minipage}[t]{0.32\linewidth} 
        \centering
        \includegraphics[width=0.99\linewidth]{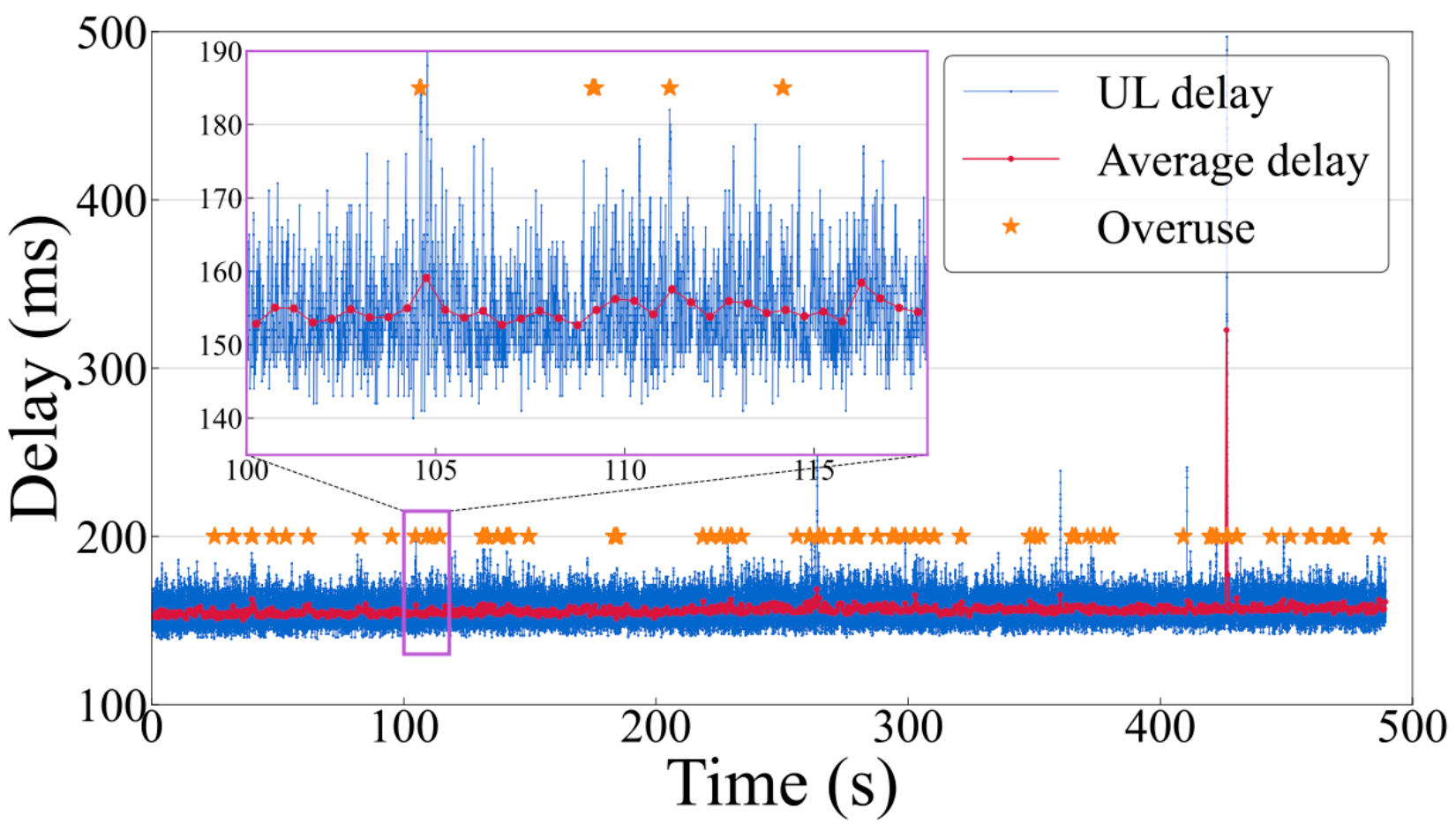}
        \caption{The original and mean uplink oneway delay measured when a UE uploads its video to an SFU server using a WebRTC connection.}
        \label{fig:webrtc_delay}
    \end{minipage}
    \hfill
    \begin{minipage}[t]{0.32\linewidth} 
        \centering
        \includegraphics[width=0.99\linewidth]{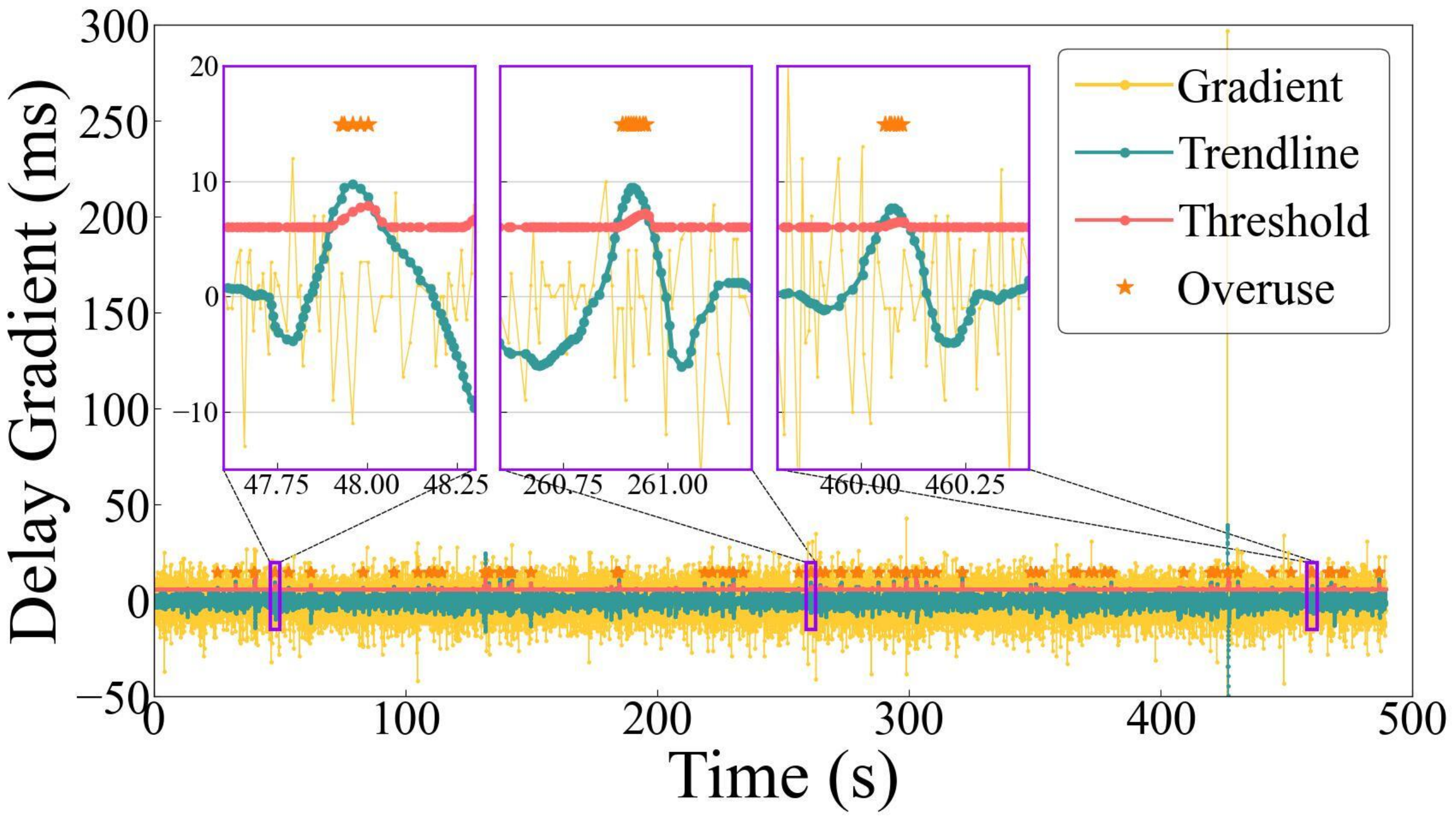}
        \caption{The process of detecting an overuse. 
        GCC calculates the delay gradient, filters it using a Trendline filter 
        and compares it with a threshold.}
        \label{fig:webrtc_gradient}
    \end{minipage}
    \hfill
    \begin{minipage}[t]{0.32\linewidth} 
        \centering
        \raisebox{-1.6pt}{\includegraphics[width=\textwidth]{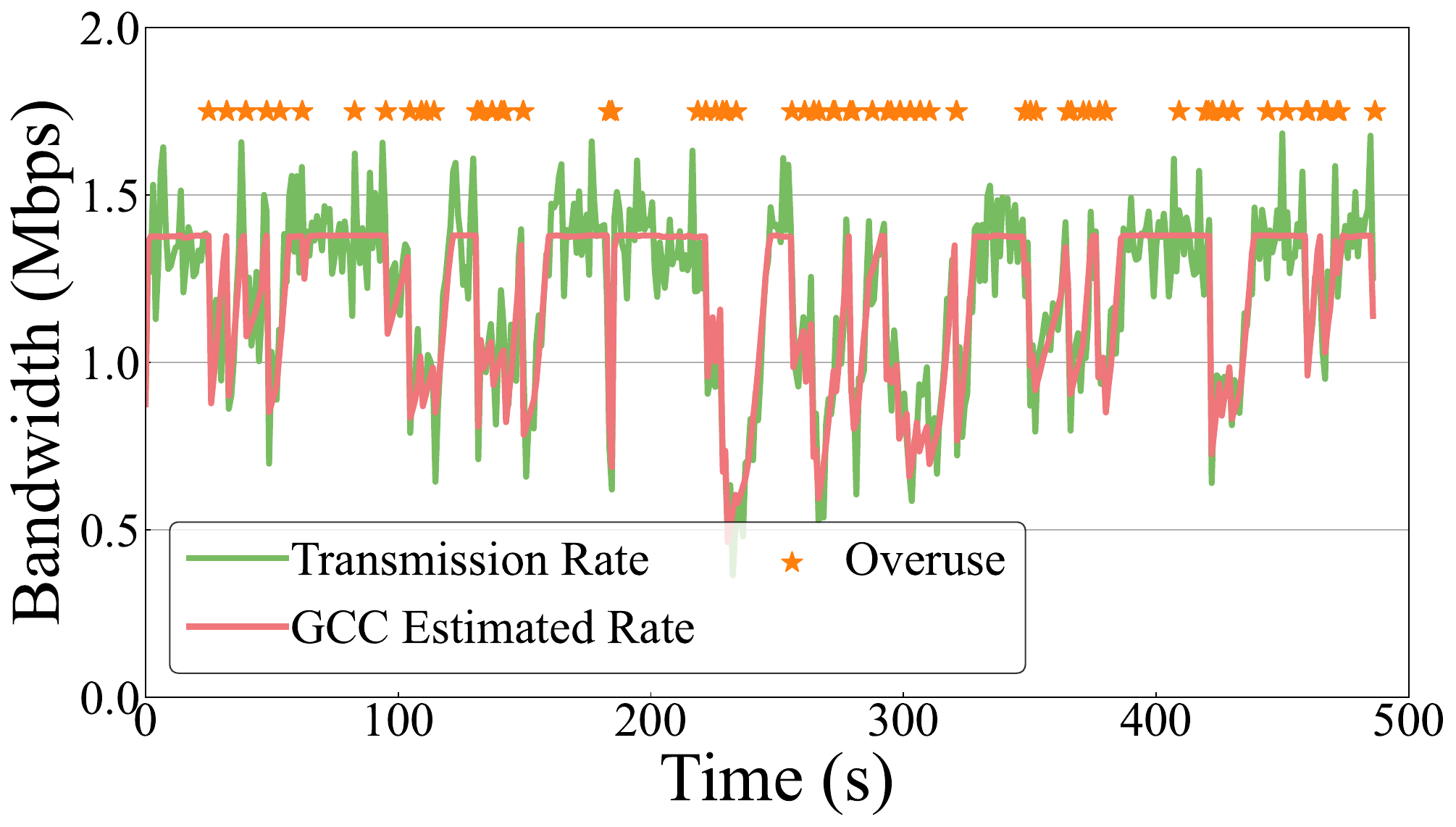}}
        \caption{The rate estimated by GCC and the final video streaming rate generated by UE. 
        Frequently detected overuse severely affects GCC's rate estimation. }
        \label{fig:gcc_rate}
    \end{minipage}
    \vspace{-0.3cm}
\end{figure*}
\begin{figure}[!t]
    \setlength{\abovecaptionskip}{2pt}
    \setlength{\belowcaptionskip}{-2pt}
    \centering
    \begin{subfigure}[b]{0.49\linewidth}
        \centering
        \includegraphics[width=\textwidth]{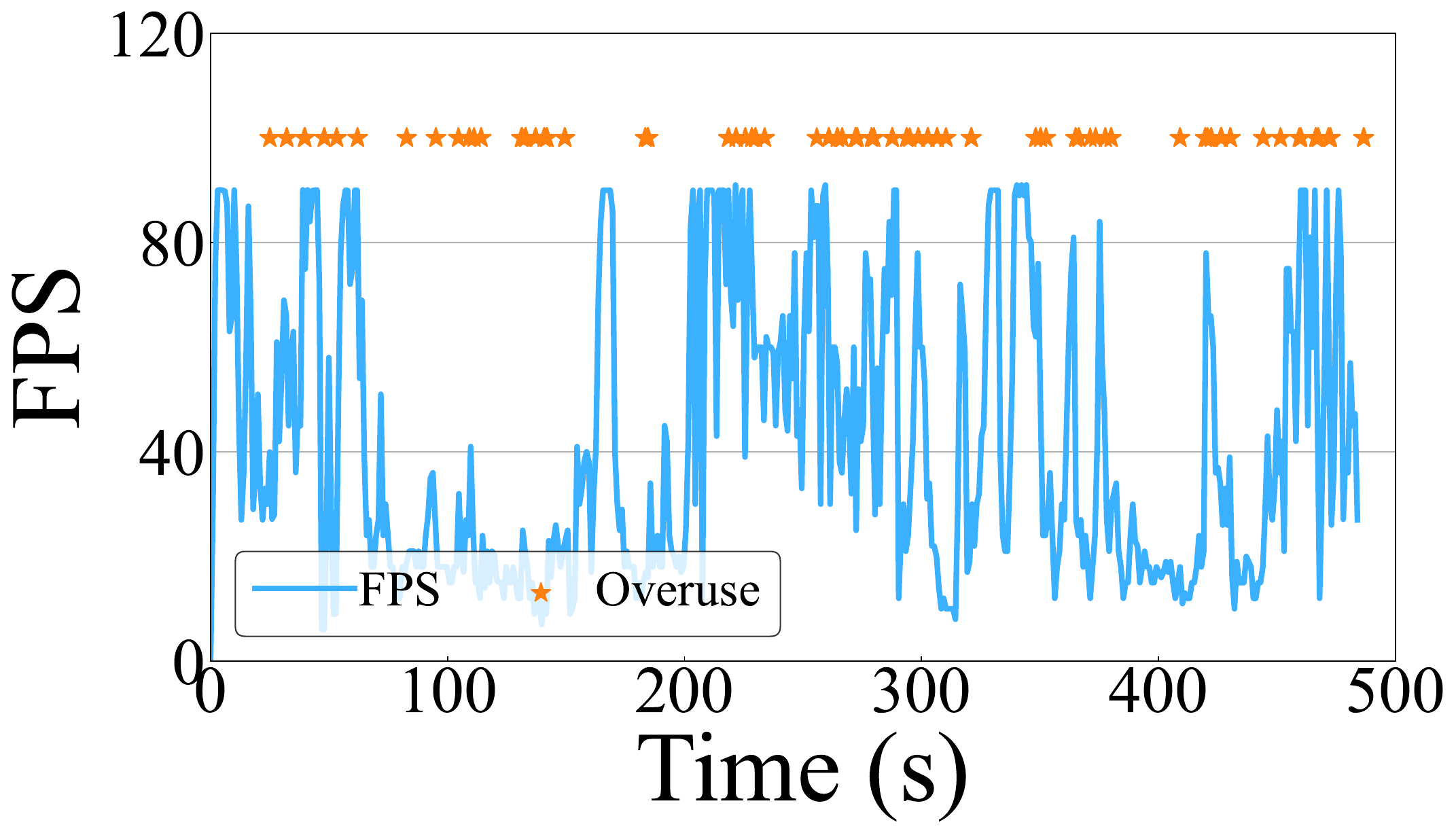}
        \caption{Frames per second (FPS)}
        \label{fig:webrtc_fps_trend}
    \end{subfigure}
    \hfill
    \begin{subfigure}[b]{0.49\linewidth}
        \centering
        \includegraphics[width=\textwidth]{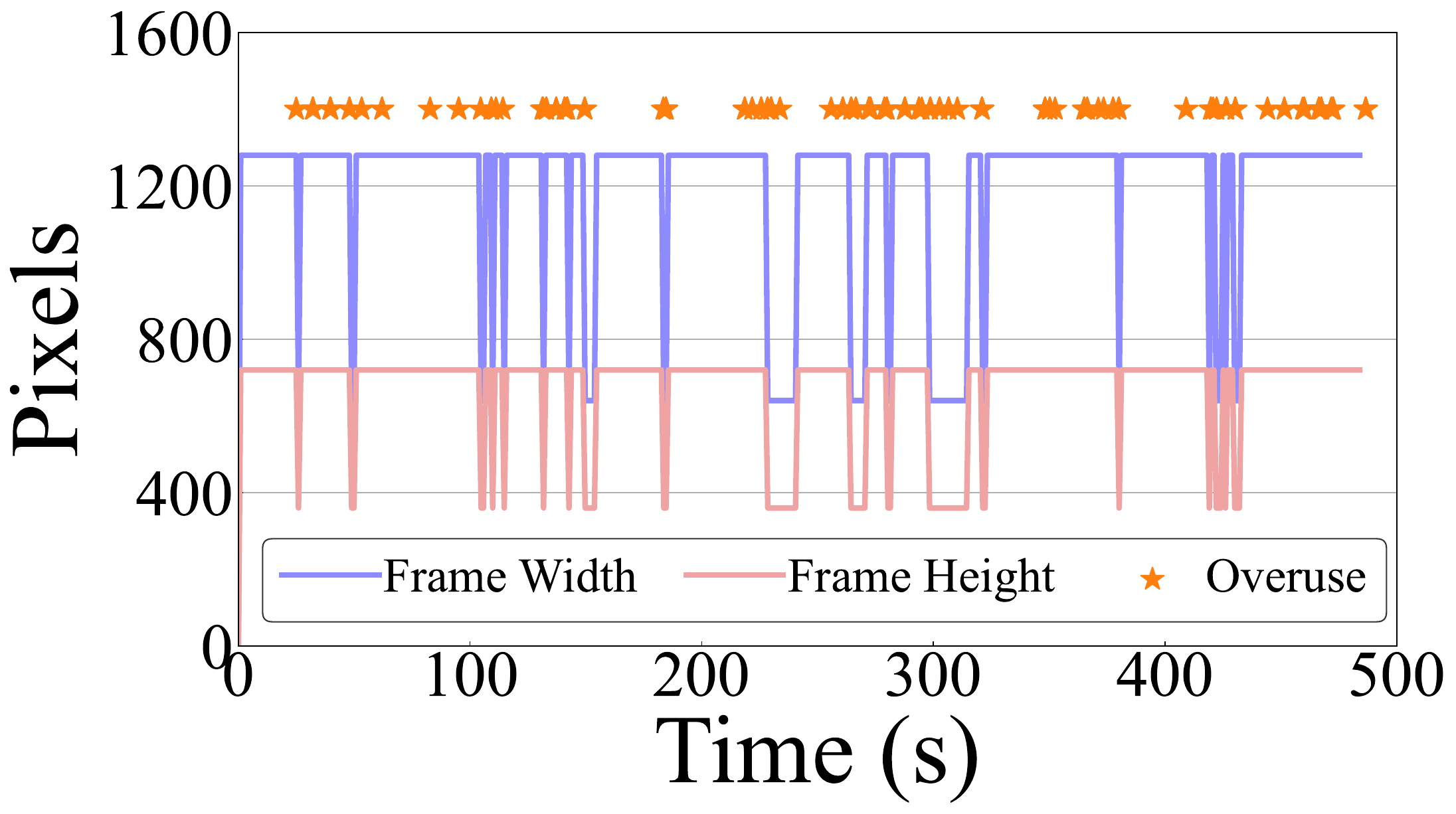}
        \caption{Frame resolution}
        \label{fig:webrtc_resolution_trend}
    \end{subfigure}
    \caption{The GCC estimated rate and video quality parameters such as FPS, Quantization parameter, and Resolution are affected by the Overuse signal.}
    \vspace{-0.0cm}
\end{figure}

\paragraph{Core Idea of GCC.}
WebRTC~\cite{blum_webrtc_2021} leverages the oneway delay gradient $d_m$ to signal channel congestion. 
GCC builds upon this principle but enhances robustness by 
applying a Trendline filter to the raw gradient, 
reducing the impact of random noise. 
It then compares the filtered delay gradient $d_t$
to a threshold $\gamma$ to guide its rate estimation $A_r$.
Specifically, if $d_t$ surpasses the threshold, 
GCC identifies the network as \textit{overused} and 
adjusts its rate estimation $A_r$ downwards. 
Conversely, if $d_t$ falls below the negative threshold $-\gamma$, 
indicating \textit{underuse}, GCC increases its rate estimation $A_r$ accordingly. 
In cases where $d_t$ falls within the range of $[-\gamma, \gamma]$, 
GCC recognizes the network as operating normally and 
maintains its current estimation of $A_r$. 
Moreover, GCC dynamically adapts its threshold $\gamma$ in response to evolving network conditions.

\paragraph{Uplink Delay Variance Misleads GCC.}
RAN induces significant uplink delay jitter, 
as we have demonstrated in Section~\S\ref{sec:ran_delay},
which could mislead the GCC algorithm. 
To demonstrate that, we let a UE stream its video to a selective forwarding unit (SFU) server 
set up on AWS using WebRTC for 200 seconds. 
We log the oneway delay of all the packets,
calculate the average oneway delay every 500 ms 
and plot the data together with the overuse detected by GCC,
in Figure~\ref{fig:webrtc_delay}.
We could observe the mean delay remains very stable, indicating no congestion. 
On the other hand, significant network jitters are introduced by RAN operation,
which misleads the GCC to generate a substantial amount of overuse.
To further verify that, we also plot the raw delay gradient $d_m$, the filtered gradient $d_t$, 
the threshold $\gamma$ in Figure~\ref{fig:webrtc_gradient}.
We see that the network jitter results in nonzero highly dynamic delay gradients,
which cannot be filtered out by the Trendline filter.
Consequently, we observe frequent cases where the gradient exceeds the thresholds,
causing GCC to mistakenly identify the network is being overused.

Figure~\ref{fig:gcc_rate} illustrates how GCC's estimated data rate $A_r$ 
sharply decreases upon detecting overuse. 
While this conservative behavior is intended to prevent congestion collapse (which is positive), 
it can also lead to excessive network underutilization in scenarios like ours, 
where overuse detection is triggered mostly by network jitter.
The lowered capacity estimate received by the encoder leads to a reduction in video quality. 
As shown in Figures~\ref{fig:webrtc_fps_trend} and \ref{fig:webrtc_resolution_trend}, 
the encoder adjusts the video frame rate and resolution downwards, 
even in the absence of actual congestion. 
These unnecessary fluctuations in video quality negatively impact the user experience.

\section{Related Work}
\label{s:related}
Prior research has explored cellular network latency in specific scenarios,
including handovers during mobility~\cite{ni_polycorn_nodate,wang_active-passive_2019,hassan_vivisecting_2022},
discontinuous reception (DRX) during power-saving mode~\cite{brunstrom_-depth_2023},
and Radio Resource Control (RRC) connection states~\cite{rosen_discovering_2014}.
In contrast, this paper examines RAN-induced delay jitters at the subframe level
in a common scenario: steady data download without handovers or DRX.
By focusing on this frequent use case, our findings have significant implications
for understanding and mitigating delay jitter in everyday cellular usage.

Most related to our work, LRP~\cite{tan_device-based_2021} addresses scheduling request (SR) delay
using a prefetcher mechanism: it sends a dummy packet $T_{sr}+4$ subframes before
the predicted packet generation time to trigger uplink grant allocation.
However, LRP only works when packet generation can be accurately predicted,
and its benefits are limited to reducing initial SR delay for infrequent small packet transmissions.

\section{Conclusion}
This paper presents \tool and \sys to address the challenge of 
RAN-induced delays in cellular networks. 
By providing real-time visibility into RAN operations 
and systematically handling RAN-induced delays, 
our solution enables delay-based protocols to achieve their full 
potential in cellular networks without modifying their core logic.
This work does not raise any ethical issues. 

\clearpage
{\footnotesize \bibliographystyle{unsrt}
\bibliography{references}}
\begin{appendices}
\section{Comparison with Existing Tools}
\label{appendix:tool_comparison}
Table~\ref{tab:tool_comparison} presents a detailed comparison between CellNinja and existing cellular network monitoring tools. Here we explain each evaluation metric:
\begin{itemize}
    \item \textbf{Real-Time:} Whether the tool provides instantaneous message reporting with minimal delay. 
    \tool implements buffer draining mechanisms 
    to achieve true real-time reporting.

    \item \textbf{5G SA:} Support for standalone 5G networks. 
    \tool has been extensively tested across multiple commercial 5G SA networks, 
    while other tools either lack 5G SA support or have limited validation.
    
    \item \textbf{NO. of LTE/5G msg:} Number of unique message types decoded by each tool. 
    NG-Scope and NR-Scope focus solely on DCI messages transmitted over control channels, 
    thus supporting only one message type. In contrast, 
    \tool and MobileInsight decode a comprehensive range of cellular messages.
    
    \item \textbf{Mobile-Ready:} Whether the tool can run directly on mobile devices. 
    NG-Scope and NR-Scope require external USRP (Universal Software Radio Peripheral) hardware for sniffing cellular signals.
    
    \item \textbf{Granularity:} The temporal resolution of network monitoring. 
    All tools achieve TTI (Transmission Time Interval) level granularity.
\end{itemize}

This comparison highlights CellNinja's advantages: 
comprehensive message coverage, true real-time operation through buffer draining, 
validated 5G SA support across commercial networks, and 
direct mobile device deployment without additional hardware requirements.

\begin{table*}[t]
    \centering
    \begin{tabular}{c|ccccccc}
    \toprule
    Tools               & Real-Time      &5G SA      &NO. of LTE msg     &NO. of 5G msg     &Mobile-Ready     & Granularity \\      
    \hline
    NG-Scope~\cite{xie_ng-scope_2022}       
                        &\ding{51}      &\ding{55}   &1           &0              &\ding{55} Sniffing with USRP  & TTI\\
    \hline
    NR-Scope~\cite{wan_nr-scope_2024}       
                        &\ding{51}      &\ding{51}   &0           &1              &\ding{55} Sniffing with USRP & TTI\\
    \hline
    MobileInsight~\cite{li_mobileinsight_2016}       
                        &\ding{55}      & \ding{55}  &57           &8            &\ding{51}     & TTI\\
    \hline
    \rowcolor{myLightGreen}
    CellNinjia    
                        &\ding{51}      &\ding{51}   &99    &60     &\ding{51} & TTI\\
    \bottomrule
    \end{tabular}
    \caption{Comparison of cellular network monitoring tools: CellNinja enables real-time monitoring of both LTE and 5G SA networks with comprehensive message coverage, while maintaining direct mobile device deployment.}
    \label{tab:tool_comparison}
\end{table*}

\section{Monitoring RAN with \tool }
In this section, we present how to extract critical RAN parameters
and monitoring RAN behaviors using \tool.

\begin{table}[H]
    \centering
    \begin{tabular}{c|ccccccc}
    \toprule
    Config              &0       &1     &2     &3     &4     &5       &6      \\ \hline
    $N_{slot}$        &10      &20    &40    &80    &160   &320     &640    \\ \hline 
    $T_{slot}$ ($\mu s$) &1000    &500   &250   &125   &62.5  &31.25   &15.625 \\ 
    \bottomrule
    \end{tabular}
    \caption{\textbf{Subcarrier spacing (SCS) configurations in 5G}: 
    relationship between spacing parameters, 
    number of slots per system frame ($N_{slot}$), and slot duration ($T_{slot}$).}
    \label{tab:numerology}
\end{table}

\subsection{Extracting Subcarrier Spacing}
The 3GPP standard~\cite{3gpp_3gpp_nodate} defines seven subcarrier spacing configurations,
each of which determines the temporal structure of a system frame.
These configurations affect two key parameters:
the number of slots per system frame and the duration of each slot,
as detailed in Table~\ref{tab:numerology}. This configuration parameter is represented by the \textit{subcarrierSpacing} field within the diagnostic message \textsf{NR5G\_SIB1}.

\subsection{Extracting BSR Interval $T_{bsr}$}\label{appendix:bsr}
% Explain how to extract the value of BSR interval and 
% how to track the change of BSR interval (if the BSR changes)
There are two approaches to extracting the BSR interval. 
The first method is to directly obtain the \textit{periodic BSR timer} value 
from the LTE diag messages \textsf{LTE\_RRC\_Connection\_Setup} and \textsf{LTE\_RRC\_Connection\_Reconfiguration}, the 5G diag messages \textsf{NR5G\_RRC\_Setup} and \textsf{NR5G\_RRC\_Reconfigurat-ion}. 
The \textit{periodic BSR timer} defines the BSR interval $T_{bsr}$, 
which is configured and may be reconfigured by the RRC layer. 
When an RRC reconfiguration occurs, 
any modifications to the BSR interval can be tracked by monitoring newly received RRC setup and RRC reconfiguration messages. 
The second method determines the BSR interval by 
calculating the time interval of TTI index between two BSRs.
The LTE diag message \textsf{LTE\_ML1\_UL\_Transport\_Block} and 
the 5G diag message \textsf{NR5G\_L2\_UL\_TB} contains detailed information of each BSR report. 
This approach enables continuous tracking of $T_{bsr}$, 
ensuring that updates are incorporated with each newly received BSR event.

\subsection{Extracting Retransmission Delay}
\label{appendix:retx_delay}
This section details how we leverage \tool's diagnostic message 
to extract retransmission delay values for both MAC $T_M$ and RLC $T_R$ layers.
The accurate measurement of these delays is crucial for \sys
as it needs to compensate for their impact on end-to-end performance.

\subsubsection{MAC Layer Retransmission Delay $T_M$}\label{app:t_m}

\paragraph{LTE.}
For MAC layer retransmission in LTE networks,
timing parameters remain constant.
In downlink transmission, $K_0=0$, $K_1=4$, and $K_d=4$,
resulting in a fixed retransmission delay $T^{d}_M=8$ milliseconds.
Similarly, in uplink transmission, $K_2=4$ and $K_u=4$,
yielding a fixed retransmission delay $T^{u}_M=8$ milliseconds.

\paragraph{5G downlink.}
In the 5G downlink, we extract retransmission timing parameters 
from multiple diagnostic messages.
The parameter $K_1$ comes directly from message \textsf{NR5G\_MAC\_\\PDSCH\_Status},
while obtaining $K_0$ requires a two-step process:
first, we extract the \textit{time resource assignment} field from message \textsf{NR5G\_MAC\_DCI\_Info},
then use this value to look up $K_0$ in the \textit{PDSCH time domain allocation list}
provided by \textsf{NR5G\_RRC\_Reconfiguration}.
We derive the parameter $K_d$ by computing the time interval between the NACK and the corresponding MAC PDU.
The ACK/NACK feedback for each downlink transmission is captured through the \textit{Num HARQ ACK bits} field in 
\textsf{NR5G\_MAC\_UL\_PHY\_Channel\_Schedule\_\\Report}. 
A value of \textit{0} indicates an ACK, while a nonzero value signifies a NACK.
We then identify a retransmission MAC PDU by checking the \textit{redundancy version} in \textsf{NR5G\_MAC\_PD-SCH\_Status}.

\paragraph{5G uplink.}
In the 5G uplink, the extraction of $K_2$ begins with obtaining \textit{symbol allocation index} field from \textsf{NR5G\_MAC\_DCI\\\_Info}. This \textit{symbol allocation index} is then used as a lookup key within the \textit{PUSCH time domain allocation list} of \textsf{NR5G\_R-RC\_Reconfiguration} to retrieve the corresponding $K_2$ value.

Unlike in the downlink, explicit ACK/NACK feedback is not required in the uplink transmission. Instead, this information is inferred from the \textit{NDI} field in \textsf{NR5G\_MAC\_DCI\_Info}, which signals a new uplink grant. A change in the \textit{NDI} value indicates that the previous transmission has been successfully acknowledged, allowing a new transmission. Conversely, if the \textit{NDI} remains unchanged, it implies a NACK. Similar to the 5G downlink, the \textit{redundancy version} field in \textsf{NR5G\_MAC\_UL\_PHY\_Channel\_Schedule\_Report} determines the end boundary of a retransmission event. Finally, the $K_u$ is calculated as the time interval between the occurrence of NACK and the identified end boundary of the retransmission event.

\subsubsection{RLC Layer Retransmission Delay $T_R$}\label{subsubsec:rlc}

\paragraph{Uplink.} 
In the 5G uplink, the \textit{NACK SN} field in the \textsf{NR5G\_RL-C\_DL\_Status\_PDU} specifies the sequence number of the PDU that requires retransmission. By leveraging the \textit{NACK SN}, both the initial transmission and the retransmitted PDU can be traced within \textsf{NR5G\_L2\_UL\_Data\_PDU} using the \textit{transmission type} field. The \textit{transmission type} indicates whether a given PDU is a new transmission or a retransmission. If the \textit{transmission type} value is \textit{new transmission}, it denotes a new transmission, whereas a value of \textit{retransmission} signifies a retransmission. The $T_R$ is computed as the time interval between the initial transmission and its corresponding retransmission.

In the LTE uplink, the \textit{NACK SN} is also essential and it is extracted from \textsf{LTE\_RLC\_DL\_AM\_Control\_PDU}. Once the \textit{NACK SN} is obtained, the corresponding \textit{SN} field in \textsf{LTE\_RLC\_UL\_AM\_All\_PDU} determines whether the given PDU is an initial transmission or a retransmission. If an \textit{SN} appears for the first time, it indicates a new transmission, whereas if the same \textit{SN} reappears, it signifies that the PDU has been retransmitted. The $T_R$ is then determined by calculating the time interval between the initial transmission and its corresponding retransmission.

% \begin{figure}[h]
%     \centering
%     \includegraphics[width=0.98\linewidth]{figure/ul_dl_config.pdf}
%     \caption{Two downlink and uplink configurations we measured from commercial 5G SA networks.}
%     \label{fig:5g_frame_uldl_config}
% \end{figure}

% \paragraph{TDD Uplink Downlink Configuration.}
% 5G networks primarily operate in TDD mode,
% where slots within each SFN are designated for either
% uplink, downlink, or flexible use.
% The pattern of uplink and downlink slot allocation is highly configurable in 5G.
% Figure~\ref{fig:5g_frame_uldl_config} shows
% two distinct configurations observed in our measurements from commercial 5G providers,
% each using 0.5ms slot duration.

\paragraph{Downlink.} 
Similarly, to monitor 5G downlink retransmission events, the \textit{NACK SN} and \textit{SN} fields are extracted from \textsf{NR5G\_RLC\_UL\_Status\_PDU} and \textsf{NR5G\_L2\_DL\_Data\_PDU}, respectively. For LTE downlink retransmission tracking, the \textit{NACK SN} and \textit{SN} fields are extracted from \textsf{LTE\_RLC\_UL\_AM\\\_Control\_PDU} and \textsf{LTE\_RLC\_DL\_AM\_All\_PDU}, respectively. In both cases, the $T_R$ is computed as the time interval between the initial transmission and its corresponding retransmission.

\subsubsection{Extracted MAC Delay $T_R$ of Commercial 5G SA}
\begin{figure}[h]
    \centering
    \includegraphics[width=0.98\linewidth]{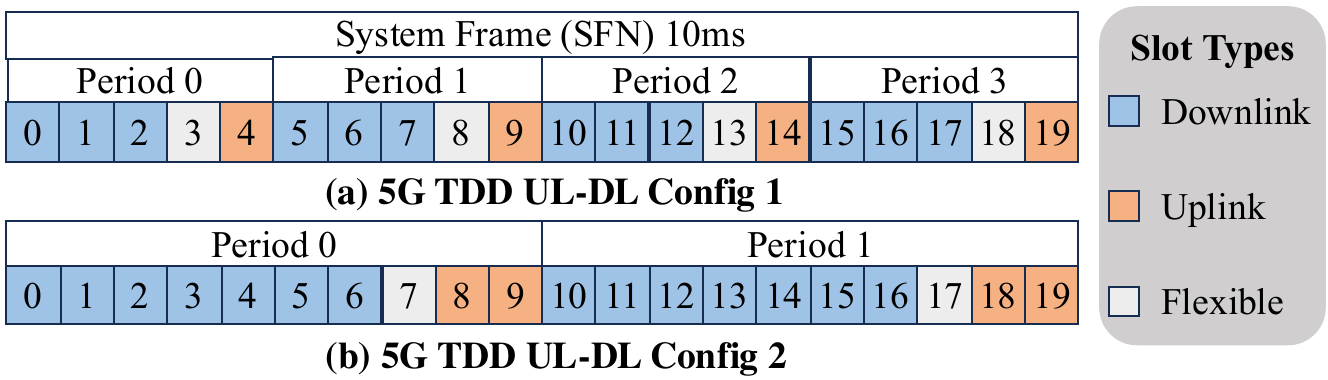}
    \caption{Two downlink and uplink configurations we measured from commercial 5G SA networks.}
    \label{fig:5g_frame_uldl_config}
\end{figure}

We present the extracted MAC layer retransmission delay $T_M$ from commercial 5G SA networks 
with two distinct configurations in Figure~\ref{fig:5g_frame_uldl_config}.

\paragraph{TDD Uplink Downlink Configuration.}
5G networks primarily operate in TDD mode~\cite{ye_dissecting_2024, k_fezeu_unveiling_2024},
where slots within each SFN are designated for either
uplink, downlink, or flexible use.
The pattern of uplink and downlink slot allocation is highly configurable in 5G.
Figure~\ref{fig:5g_frame_uldl_config} shows
two distinct configurations observed in our measurements from commercial 5G providers,
each using 0.5ms slot duration.

\paragraph{Downlink Retransmission Delay.}
The downlink retransmission delay $T^d_{M}$
is determined by three parameters:
$K_0$, $K_1$, and $K_d$.
Our measurements show that in 5G SA, $K_0$ maintains a constant value of zero,
while both $K_1$ and $K_d$ vary with the slot index $i$ ($i \in [0, 19]$)
within a system frame.
Tables~\ref{tab:dl_reTx_config1} and~\ref{tab:dl_reTx_config2}
provide the complete values of $K_1$, $K_d$, and the resulting $T_{M}^d$
for the first and second TDD UL-DL configurations illustrated in
Figure~\ref{fig:5g_frame_uldl_config}, respectively.

\begin{table}[htb]
    \centering
    \begin{tabular}{c|ccccccc}
    \toprule
    Slot index $i$    &0/10      &1/11     &2/12     &3/13     &5/15     &6/16    &7/17 \\      
    \hline
    $K_{1}$       &4      &8     &7     &6     &4    &8      &7 \\
    \hline
    $K_{d}$       &6      &6     &6     &6     &6    &6      &6\\
     \hline
    \rowcolor{myLightGreen}
    $T^d_{M}$    &10     &14    &13    &12    &10   &14     &13  \\ 
    \bottomrule
    \end{tabular}
    \caption{Downlink retransmission timing parameters ($K_1$, $K_d$) and resulting delay ($T_{M}^d$) for each slot index under the TDD UL-DL configuration shown in Figure~\ref{fig:5g_frame_uldl_config}(a).}
    \label{tab:dl_reTx_config1}
\end{table}

\begin{table}[htb]
    \centering
    \begin{tabular}{c|cccccccc}
    \toprule
    %Slot index    &0    &1    &2    &3    &4    &5   &6   &7 \\      
    $i$     &0/10      &1/11     &2/12     &3/13     &4/14 &5/15     &6/16    &7/17 \\      
    \hline
    $K_{1}$      &8    &7    &7    &6    &5    &4   &12  &11 \\
    \hline
    $K_{d}$      &5    &5    &5    &5    &5    &5   &5  &5 \\
    \hline
    \rowcolor{myLightGreen}
    $T^d_{M}$   &13   &12   &12   &11   &10   &9   &17  &16 \\ 
    \bottomrule
    \end{tabular}
    \caption{Downlink retransmission timing parameters ($K_1$, $K_d$) and resulting delay ($T_{M}^d$) for each slot index under the TDD UL-DL configuration shown in Figure~\ref{fig:5g_frame_uldl_config}(b).}
    \label{tab:dl_reTx_config2}
\end{table}

\begin{figure}[h]
    \centering
    \includegraphics[width=0.99\linewidth]{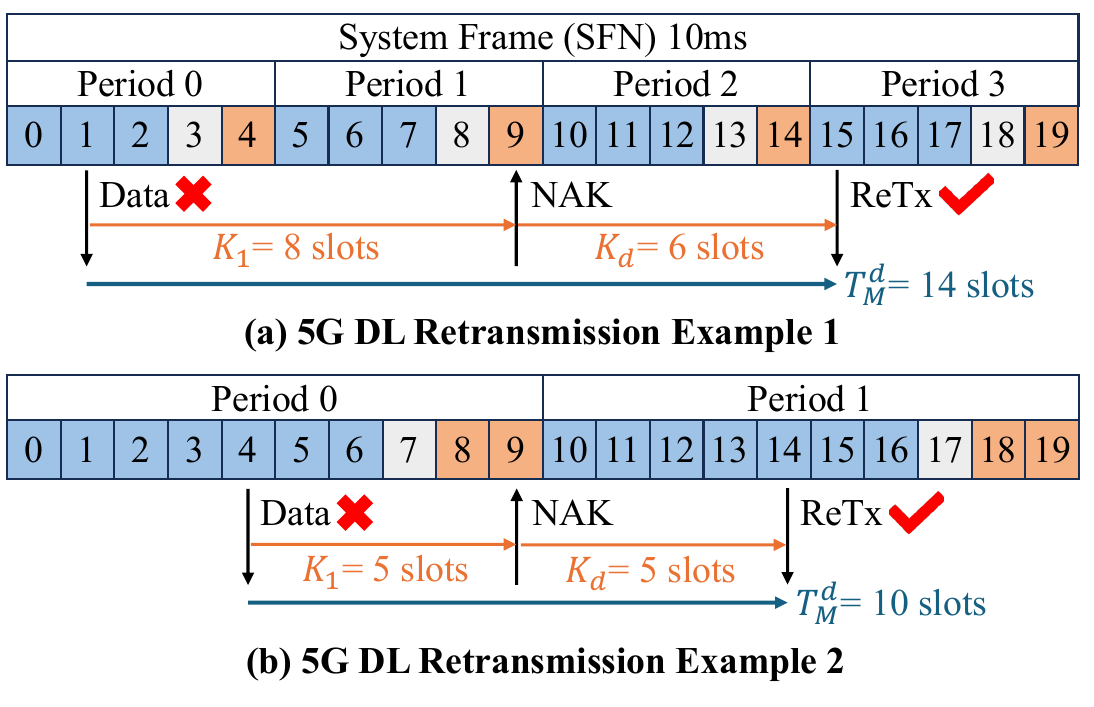}
    \caption{Examples of downlink retransmission timing parameters under two TDD UL-DL configurations shown in Table~\ref{tab:dl_reTx_config1} and Table~\ref{tab:dl_reTx_config2}, respectively.}
    \label{fig:5g_frame_uldl_config_e1}
\end{figure}

\paragraph{Uplink Retransmission Delay.}
The uplink retransmission delay $T^u_{M}$
is determined by two parameters:
$K_2$ and $K_u$.
Our measurements show that in 5G SA,
both $K_2$ and $K_u$ vary with the slot index $i$ ($i \in [0, 19]$)
within a system frame.
Tables~\ref{tab:ul_reTx_config}
provide the complete values of $K_2$, $K_u$, and the resulting $T_{M}^u$
for two TDD UL-DL configurations illustrated in
Figure~\ref{fig:5g_frame_uldl_config}.

\begin{table}[htb]
    \centering
    \begin{subtable}{0.49\linewidth}
        \centering
        \begin{tabular}{ccc}
        \hline
        Slot index $i$  &4/14       &9/19      \\ \hline
        $K_2$           &4          &3           \\ \hline
        $K_u$           &6          &7           \\ \hline
        \rowcolor{myLightGreen}
        $T^u_{M}$      &10         &10           \\
        \bottomrule
        \end{tabular}
        \caption{UL-DL configuration 1}
        \label{tab:ul_reTx_config1}
    \end{subtable}%
    \hfill
    \begin{subtable}{0.49\linewidth}
        \centering
        \begin{tabular}{cccc}
        \hline
        Slot index $i$  &8/18       &9/19      \\ \hline
        $K_2$           &3          &3          \\ \hline
        $K_u$           &7          &6          \\ \hline
        \rowcolor{myLightGreen}
        $T^u_{M}$      &10         &9           \\
        \bottomrule
        \end{tabular}
        \caption{UL-DL configuration 2}
        \label{tab:ul_reTx_config2}
    \end{subtable}
    \caption{Uplink retransmission timing parameters ($K_2$, $K_u$) 
    and resulting delay ($T_{M}^u$) for each slot index 
    under two TDD UL-DL configuration shown in Figure~\ref{fig:5g_frame_uldl_config}.}
    \label{tab:ul_reTx_config}
\end{table}

Figure \ref{fig:5g_frame_uldl_config_e1} illustrates two examples of downlink retransmission process under two distinct TDD UL-DL configurations, respectively. In Figure \ref{fig:5g_frame_uldl_config_e1}(\textbf{a}), the retransmission timing is characterized by $K_1=8$ slots and $K_d=6$ slots,
resulting in a total downlink retransmission delay $T^{d}_M=14$ slots, under the configuration in Table~\ref{tab:dl_reTx_config1}. Conversely, Figure \ref{fig:5g_frame_uldl_config_e1}(\textbf{b}) presents a case with $K_1=5$ slots and $K_d=5$ slots, leading to a total delay of $T^{d}_M=10$ slots. This case aligns with the TDD configuration outlined in Table~\ref{tab:dl_reTx_config2}.

\subsection{Tracking the Retransmission Events}\label{appendix:track_retx}
% explain how to monitoring and tracking the retransmission events

\paragraph{Tracking MAC layer Retransmissions}.
To track the downlink MAC layer retransmission, it is crucial to obtain the key fields \textit{redundancy version} and \textit{HARQ process id} from diagnostic messages. In the LTE downlink, these information is retrieved from \textsf{LTE\_ML1\_PDSCH\_Stat\_Indication}, while in the 5G downlink, it is obtained from \textsf{NR5G\_MAC\_PDSCH\_S-tatus}. The \textit{redundancy version} determines which version of a coded transport block is transmitted or retransmitted, making it a key indicator for identifying the end boundary of a retransmission event. Using the \textit{HARQ process id} of the end boundary, the corresponding initial transmission can be traced, establishing the start boundary of the retransmission process.
Similar to the downlink retransmissions, the uplink retransmissions are monitored with extracted key fields \textit{redundancy version} and \textit{HARQ process id}. However, for the uplink, these values are obtained from \textsf{LTE\_ML1\_PDSCH\_St-at\_Indication} in LTE and \textsf{NR5G\_MAC\_UL\_PHY\_Channel\_S-chedule\_Report} or \textsf{NR5G\_MAC\_DCI\_Info} in 5G.

\paragraph{Tracking RLC layer Retransmissions}.
As we mentioned in Section~\ref{subsubsec:rlc}, the \textit{NACK SN} and \textit{SN} fields are leveraged to trace initial transmissions and corresponding retransmissions. In the 5G uplink, these information is extracted from \textsf{NR5G\_RLC\_DL\_Status\_PDU} and \textsf{NR5G\_L2\_UL\_Data\_PDU}, with the \textit{transmission type} field distinguishing between new transmissions \textit{new transmission} and retransmissions \textit{retransmission}. Similarly, LTE uplink follows the same principle, using \textsf{LTE\_RLC\_DL\_AM\_Control\_PDU} and \textsf{LTE\_RLC\_UL\_AM\\\_All\_PDU}.
For the downlink retransmission tracking, the same methodology applies: 5G utilizes \textsf{NR5G\_RLC\_UL\_Stat-us\_PDU} and \textsf{NR5G\_L2\_DL\_Data\_PDU}, while LTE extracts relevant fields from \textsf{LTE\_RLC\_UL\_AM\_Control\_PDU} and \textsf{LTE\_RLC\_DL\_AM\_All\_PDU}.

\begin{table}[H]
    \centering
    \begin{tabular}{p{7.5cm}}
        \toprule
        \multicolumn{1}{c}{\textbf{5G Message Type}}\\ 
        \toprule
        \textsf{NR5G\_MAC\_PDSCH\_Status}\\
        \hline
        \textsf{NR5G\_MAC\_DCI\_Info} \\
        \hline
        \textsf{NR5G\_RLC\_DL\_Status\_PDU} \\
        \hline
        \textsf{NR5G\_L2\_UL\_Data\_PDU} \\
        \hline
        \textsf{NR5G\_RLC\_UL\_Status\_PDU}\\
        \hline
        \textsf{NR5G\_L2\_DL\_Data\_PDU} \\
        \hline
        \textsf{NR5G\_L2\_UL\_TB}\\
        \hline
        \textsf{NR5G\_L2\_UL\_BSR}\\
        \hline
        \textsf{NR5G\_SIB1}\\
        \hline
        \textsf{NR5G\_RRC\_Setup}\\
        \hline
        \textsf{NR5G\_RRC\_Reconfiguration}\\
        \hline 
        \textsf{NR5G\_MAC\_UL\_PHY\_Channel\_Schedule\_Report}\\
        \bottomrule
    \end{tabular}
    \caption{Essential 5G message types utilized in {\sys}.}
    \label{tab:msg_tti_5g}
\end{table}

\begin{table}[H]
    \centering
    \begin{tabular}{p{7.5cm}}
        \toprule
        \multicolumn{1}{c}{\textbf{LTE Message Type}} \\ 
        \toprule
        \textsf{LTE\_ML1\_PDSCH\_Stat\_Indication} \\
        \hline 
        \textsf{LTE\_LL1\_PUSCH\_Tx\_Report} \\
        \hline
        \textsf{LTE\_ML1\_DCI\_Information\_Report} \\
        \hline
        \textsf{LTE\_RLC\_DL\_AM\_Control\_PDU} \\
        \hline
        \textsf{LTE\_RLC\_UL\_AM\_All\_PDU} \\
        \hline
        \textsf{LTE\_RLC\_UL\_AM\_Control\_PDU} \\
        \hline
        \textsf{LTE\_RLC\_DL\_AM\_All\_PDU} \\
        \hline
        \textsf{LTE\_ML1\_UL\_Transport\_Block} \\
        \hline
        \textsf{LTE\_ML1\_GM\_Tx\_Report} \\
        \hline
        \textsf{LTE\_LL1\_PDCCH\_Decoding\_Results} \\
        \hline
        \textsf{LTE\_RRC\_Connection\_Setup} \\
        \hline
        \textsf{LTE\_RRC\_Connection\_Reconfiguration} \\
        \bottomrule
    \end{tabular}
    \caption{Essential LTE message types utilized in {\sys}.}
    \label{tab:msg_tti_lte}
\end{table}

\section{Summary of Messages used in \sys }\label{app:msg_list}
As shown in \autoref{tab:msg_tti_5g} and \autoref{tab:msg_tti_lte}, 
we outline the key message types collected in 5G and LTE for analyzing both uplink and downlink transmissions.

\end{appendices}
% \clearpage
% \input{chapter/Appendix}
\end{document}